\documentclass {udthesis}

\usepackage{booktabs}
\usepackage{graphicx}
\usepackage{subcaption}
\usepackage{amsmath}
\usepackage{booktabs}
\usepackage{subcaption}
\usepackage{float}
\usepackage{xcolor}

\begin{document}
%
%

\title[]{On Fin Based Propulsion and Maneuvering\\ for uncrewed underwater vehicles}
\author{Parker Thomas Grobe}
\type{dissertation}
\degree{Doctor of Philosophy}
\majorfieldtrue\majorfield{Mechanical Engineering}
\degreedate{Spring 2026}

\begin{front} 

\maketitlepage 

\maketocloflot

\prefacesectiontoc{Abstract}
Bio-inspired propulsion systems based on oscillating fins have attracted significant interest due to their potential to achieve high propulsive thrust, efficiency, and maneuverability. Many aquatic organisms generate thrust through coordinated oscillations of fins or flukes, and understanding the underlying hydrodynamic mechanisms can inform the design of efficient underwater vehicles. This dissertation investigates force production and performance characteristics of oscillating fin propulsion systems using computational simulations.

A computational framework is developed to simulate oscillating fins in a uniform freestream using a NACA 0020 profile undergoing prescribed heaving and pitching motions about the leading edge. The simulations are performed using a two-dimensional viscous incompressible flow solver based on the Boundary Data Immersion Method (BDIM), as implemented in WaterLily, which resolves the fluid–structure interaction on a fixed Cartesian grid. The kinematics are characterized by parameters including oscillation frequency, heave amplitude, pitch amplitude, and phase offset. These parameters are non-dimensionalized using the Strouhal number, which governs the relationship between fin motion and the surrounding flow field. The Reynolds number is held constant across all simulations to isolate kinematic effects, while an additional nondimensional parameter incorporating phase offset is introduced to characterize multi-fin interactions. Numerical simulations are performed to examine the resulting flow field structures, force production, and interactions between structures.

The study first examines the behavior of a single oscillating fin to establish baseline force generation mechanisms and to validate the numerical approach. A reduced order dynamical model is then developed to describe the motion of a passively pitching fin with a leading edge torsional spring, enabling the pitch response to arise from hydrodynamic force balancing rather than being fully prescribed. This formulation provides insight into how passive pitching influences force production and system performance. We then investigate asymmetric actuation of a single fin, examining how non-symmetric force generation produces a net lateral force that enables maneuvering. Specifically we study asymmetric heave speeds, added pitch bias, and asymmetric stiffness can create these net lateral forces. 

Building on the single fin results, the work explores the hydrodynamic interactions that occur in multi-fin propulsion systems. We study configurations with two or more fins arranged in-line. Meaning one fin directly in front of the other in the streamwise direction, the downstream fin interacts with vortices shed by upstream fins; similar to the benefits birds see when flying in a v-shaped formation. The simulations demonstrate that appropriate tuning of phase offsets and spacing between fins can significantly enhance thrust production by allowing downstream fins to extract energy from the upstream wake. The timing of these interactions can amplify thrust or, in unfavorable configurations, reduce overall performance. As the parameter space for multi-fin systems grows rapidly with the number of fins, Bayesian optimization is employed to identify high performance configurations. The optimization framework searches combinations of phase offsets and spacing to maximize thrust or efficiency. The results reveal coordinated motion patterns that produce favorable vortex interactions and improved propulsion compared to single-fin systems.

Together, the findings provide new insight into the hydrodynamic mechanisms governing oscillating fin propulsion and the role of vortex interactions in multi-fin systems. The combination of computational simulations, scaling equations, and optimization offers an architecture for understanding and designing bio-inspired propulsion systems for underwater vehicles.

\end{front}

\chapter{Introduction, methodology, and validation}
\newcommand{\ptg}[1]{{{\color{green}#1}}}

\section{Basics of oscillating propulsion}
Traditional marine vessel design relies on continuous rotation mechanisms for forward propulsion. These designs then depend on additional control surfaces and devices such as rudders, bow planes, thrusters, and other auxiliary systems to provide maneuvering control. As a result, the vessel relies on multiple independent systems, increasing the overall complexity as well as the cost of manufacturing, operation, and repair. Reorienting the lens through which we approach vessel design may therefore lead to architectures that offer distinct advantages over traditional configurations. One such lens is biological inspiration, or more simply bio-inspired design. In recent years this approach has gained popularity, largely driven by the rise of uncrewed underwater vehicles (UUVs). As UUVs become more prevalent in both the commercial sector and the modern battlefield, the demand for higher performance and lower system cost continues to grow. This pressure has motivated a renewed effort to rethink how underwater vehicles are designed. With this motivation, we turn to a different design theory, bio-inspired propulsion. Numerous studies have examined biological swimming as a model for vehicle propulsion, highlighting advantages in maneuverability, efficiency, and control authority compared to conventional propellers \cite{Fish2020,Tangorra2007,Wen2013,Bale2014}. Fish and swimming mammals exhibit remarkable capabilities including rapid acceleration, exceptional maneuverability, superb top speeds, and lastly high cruising ranges.  We organize these as three main metrics for our discussion, maneuverability, thrust, and efficiency. In many cases, the performance of these animals rival that of traditional vessels. These capabilities raise a fundamental question: why not attempt to replicate what nature already does so effectively? 

We begin by looking at how nature operates and the mechanisms by which forces can be generated from basic oscillatory motions. Then we examine the history of this line of research and where it leaves us today.

\subsection{Animal locomotion}

Aquatic animals have evolved a wide range of propulsion strategies that allow them to move efficiently through water \cite{van2020bioinspired, Lighthill1970}. Unlike traditional vehicles relying on continuously rotating propellers, biological swimmers typically generate propulsion through periodic body or fin motions. These motions interact with the surrounding fluid to produce forces that propel the animal forward while simultaneously allowing for maneuvering and control \cite{Triantafyllou2000}. 

Several broad categories of biological swimming have been identified based on the dominant mechanism used to generate thrust. Undulatory swimmers, such as eels and lampreys, produce propulsion through traveling waves that propagate along the length of the body. As the body bends laterally, the wave motion pushes fluid backward, generating forward thrust. Other animals, such as rays, produce similar undulatory waves along extended pectoral fins. Pulsatile propulsion represents another class of swimming in which animals periodically take in and then expel water to produce a jet, as seen in jellyfish and squid. A third mechanism is drag-based propulsion, where rigid appendages are pushed through the water to generate thrust by reaction forces. This mechanism is commonly observed in animals such as turtles, ducks, and it is how humans swim. The fourth among these different propulsion strategies, is our primary focus, oscillatory propulsion. This mode has received significant attention in both biological and engineering studies due to its high performance above drag, jet, or undulatory swimmers. In oscillatory swimming, propulsion is produced primarily by the periodic pitching and heaving motion of a fin or fluke. This mechanism is characteristic of high performance swimmers such as tuna, salmon, and dolphins. In these animals the body remains relatively rigid while a caudal fin or fluke oscillates laterally, generating forces that are redirected into forward thrust through the unsteady interaction with the surrounding flow. The resulting wake consists of a structured pattern of vortices that transfer momentum to the fluid and produce net thrust.

Due to their high propulsive efficiency and strong thrust generation capabilities, oscillatory propulsion systems have become a primary focus of bio-inspired propulsion research. Simplified models such as pitching and heaving foils have been widely used to investigate the underlying fluid dynamics and to develop propulsion systems for underwater vehicles \cite{Scherer1968, Read2003, Floryan2017}. These reduced order models capture the essential physics of oscillatory propulsion while remaining simple enough to allow systematic experimental and computational study. As a result, oscillating foils have become a canonical model for studying biologically inspired propulsion. In the following sections we focus on the history of oscillatory propulsion research and then examine mechanisms by which periodic fin motion generates thrust and efficiency in aquatic propulsion systems.

\subsection{Historical biological studies}
Early investigations into fish locomotion established the fundamental physical mechanisms underlying aquatic propulsion. Lighthill’s elongated body theory showed that thrust is generated through reactive forces arising from the acceleration of surrounding fluid mass as a traveling wave propagates along the body \cite{Lighthill1970}. This framework demonstrated that propulsion can be achieved without relying on drag-based mechanisms, instead depending on the interaction between body kinematics and the induced fluid motion. This distinction between reactive and resistive force generation provides a foundation for understanding oscillatory propulsion systems.

Further work on oscillating foils clarified the relationship between kinematics, vortex formation, and efficiency. Experimental studies demonstrated that the timing and strength of vortex shedding are strongly influenced by the prescribed motion of the foil \cite{Koochesfahani1989}. Later studies showed that when the oscillation frequency and amplitude are tuned such that the foil motion resonates with the natural vortex shedding process, propulsive efficiency can be significantly enhanced \cite{Moored2012,Dewey2013}. This highlights the importance of properly selecting kinematic parameters to control fluid--structure interactions.

Observations of biological swimmers also revealed that many animals operate within a narrow range of kinematic motion called the Strouhal number. The Strouhal number, which is discussed further in later chapters, relates the oscillation frequency and amplitude of the propulsor to the forward swimming velocity. Taylor et al. \cite{Taylor2003} demonstrated that swimming and flying animals typically cruise within a characteristic Strouhal number range.  Taylor found that efficient swimmers typically operate within a range of approximately $0.2 < St < 0.5$, corresponding to conditions associated with high propulsive efficiency. This discovery suggested that the relationship between oscillation kinematics and forward velocity plays a fundamental role in determining performance.

Further work by Gazzola et al. examined scaling laws across a wide range of aquatic species and identified consistent relationships between body size, swimming speed, and propulsion mechanics \cite{Gazzola2014}. Their analysis demonstrated that swimming performance across species can often be described by a small set of non-dimensional parameters. These findings have helped guide the development of simplified models and numerical simulations aimed at reproducing the performance of biological swimmers.

Together, these studies establish the physical and biological foundation for oscillatory propulsion research. They demonstrate that efficient aquatic propulsion arises from the interaction between periodic fin motion and the surrounding fluid, producing organized vortex generation that generates thrust. Building on this understanding, modern computational and experimental studies now seek to further explore the performance limits of oscillatory propulsion systems and apply these principles to the design of next-generation underwater vehicles.

\subsection{Historical hydrodynamic studies}
To better understand the mechanisms underlying biological propulsion, researchers began studying simplified models of oscillating foils. These models replicate the key motions observed in biological fins and flukes while allowing controlled experimental and numerical investigation. By reducing the complexity of biological swimmers to a pitching and heaving foil, researchers were able to isolate the hydrodynamic mechanisms responsible for thrust generation and efficiency.

One of the earliest theoretical investigations into unsteady forces acting on oscillating lifting surfaces was conducted by Theodorsen and Garrick in their study of aeroelastic flutter \cite{Theodorsen1935}. Their work sought to understand the interaction between structural motion and aerodynamic forces on wings undergoing oscillatory motion. In particular, they examined how pitching and heaving motions of an airfoil generate time dependent aerodynamic loads that can either dampen or amplify structural oscillations. The primary motivation of this work was to predict, and ultimately prevent, destructive flutter in aircraft wings, the analysis also provided important insight into the forces generated by oscillating lifting surfaces in a fluid. 

This theoretical framework has proven highly relevant to studies of oscillatory propulsion. The same pitching and heaving motions examined in flutter analysis are commonly used to model biological propulsion systems and oscillating foil propulsors. As a result, Theodorsen’s formulation remains a cornerstone of modern oscillating foil theory and provides a basis for understanding lift-based thrust generation in bio-inspired propulsion systems.

Building on early unsteady aerodynamic theory, later work began to examine oscillating foils specifically as propulsion devices rather than purely aeroelastic systems. Scherer conducted one of the early investigations into oscillating foil propulsion at large amplitudes \cite{Scherer1968}, combining both analytical modeling and experimental measurements to examine the forces generated by rigid foils undergoing periodic motion. In this work, the foil was subjected to combinations of pitching, heaving, and surging oscillations in order to better represent the motions associated with biological propulsion systems. Scherer developed an analytical framework to predict the forces and moments acting on a finite-span foil undergoing these large amplitude oscillations. Unlike earlier small amplitude theories, the analysis incorporated several additional physical effects that become important at larger motions, including flow separation and stall as well as the influence of the induced slipstream generated by the oscillating foil. By accounting for these non-linear effects, the model aimed to better represent the hydrodynamic conditions experienced by practical propulsion systems. In addition to examining the fundamental physics of oscillating foil propulsion, Scherer also explored the potential application of such systems for marine propulsion. Performance predictions were presented for a conceptual oscillating foil propulsor designed for use on a small shallow draft vessel capable of operating at speeds of approximately 15 knots. This work represented an early attempt to translate oscillating foil theory into practical propulsion system design and helped establish the feasibility of oscillatory propulsion systems for marine vehicles.

Following earlier work on oscillating foil propulsion, researchers began to examine the hydrodynamic mechanisms responsible for thrust generation in greater detail. Later experimental and theoretical studies \cite{KatzWeihs1978, Triantafyllou1993, anderson1997concept, Czarnowski1997} focused on understanding how oscillating foils generate propulsive forces and how these forces depend on the motion kinematics of the foil. These investigations helped establish oscillating foils as a useful simplified model for studying biologically inspired propulsion.

While much of the early work on oscillating foils focused on forward propulsion, later studies began to explore their potential for maneuvering and control. By modifying the kinematics of the motion, particularly through pitch bias, asymmetric flapping, or transient impulses, the same lifting surface used for propulsion can generate large lateral forces and turning moments in addition to thrust.

This focus of maneuverability has been studied in regards to aquatic animals, with analyses examining how fish generate turning forces and maintain stability during swimming \cite{Bandyopadhyay2002,Webb2004,Webb2005}. Read et al. examined the forces generated by oscillating foils used for both propulsion and maneuvering \cite{Read2003}. Their experiments demonstrated that properly phased heave and pitch motions produce strong thrust performance, while introducing a pitch bias can generate significant lateral force coefficients. These results highlight the ability of an oscillating foil to function simultaneously as both a propulsor and a control surface. Rather than acting purely as a thrust-generating device, the oscillating foil can therefore provide integrated propulsion and maneuvering authority.

This concept represents a significant departure from traditional marine propulsion systems. Conventional vessels typically separate propulsion and control, relying on propellers for thrust and additional control surfaces such as rudders or bow planes for maneuvering. Biological swimmers, however, routinely combine these functions using coordinated fin and body motion. The oscillating foil literature suggests that similar multifunctional capabilities can be engineered in bio-inspired propulsion systems. By exploiting the unsteady vortex formation associated with oscillatory motion, large lateral forces and turning moments can be generated, enabling efficient maneuvering and control using a single propulsive surface \cite{Read2003}.

Beyond maneuvering, a large body of work has focused on understanding the flow field structures and scaling relationships that govern oscillatory propulsion. Early studies primarily examined the wake structures generated by oscillating foils, identifying thrust producing vortex patterns and connecting them to propulsion performance. More recent work has expanded this understanding by developing predictive scaling laws that relate foil kinematics to thrust, power, and efficiency.

A comprehensive review of oscillatory and undulatory swimming was presented by Smits \cite{Smits2019}. This work highlighted that propulsion in oscillatory swimmers such as tuna and dolphins cannot be explained solely by added-mass forces. Instead, both added mass and circulatory lift contributions play important roles in determining thrust generation and propulsive efficiency. The review also emphasized that the characteristic lateral velocity of the trailing edge often serves as a more relevant velocity scale than the forward swimming speed itself.

Scaling laws for rigid heaving or pitching foils were developed \cite{Floryan2017}. Their work demonstrated that thrust, power, and efficiency can be collapsed using non-dimensional parameters based on Strouhal number, reduced frequency, and physically motivated lift and added mass contributions. Importantly, they found that when viscous drag is not dominant, thrust scales approximately linearly with reduced frequency. This work marked a shift in the field from qualitative descriptions of wake structures toward predictive scaling relations capable of describing propulsion performance across a wide range of kinematic conditions.

Subsequent studies continued to refine these scaling relationships. Examining the influence of flow speed on oscillating foil propulsion and showed that forward velocity has comparatively little effect on thrust generation relative to foil kinematics \cite{VanBuren2018}. Their findings further support the view that the trailing-edge velocity and motion amplitude are dominant parameters controlling propulsion performance. Additional work examined the role viscous drag plays in defining an ideal Strouhal range for efficient cruising \cite{Floryan2019}, and how the forces generated in oscillating foils can be scaled effectively \cite{van2019scaling}.

Flow visualization studies have also remained important for connecting these scaling laws to the underlying fluid dynamics. Earlier research established that oscillating foils producing thrust typically generate a reverse K\'arm\'an vortex street. However, later work demonstrated that similar propulsion efficiencies can arise from a range of wake structures depending on foil geometry, motion amplitude, and flexibility \cite{Smits2019}. Studies of batoid-inspired oscillating fins further showed that traveling wave fin motions produce distinct wake topologies, with thrust increasing with non-dimensional frequency and optimal gaits emerging from the balance between wake formation and input power \cite{Clark2017}.

Taken together, these studies illustrate the progression of oscillatory propulsion research. Early investigations identified the vortex production associated with thrust generation. Later work connected these motions to force production and maneuvering capability, and more recent studies have developed predictive scaling laws that relate propulsion performance directly to foil kinematics. This progression provides the foundation for modern computational and experimental studies of bio-inspired propulsion systems.

While early oscillating foil studies primarily focused on rigid lifting surfaces, many biological propulsors exhibit significant flexibility. Fish fins, flukes, and other appendages deform under hydrodynamic loading, which can alter the effective kinematics of the propulsor and influence both thrust production and efficiency. As a result, a growing body of research has investigated how structural flexibility affects oscillatory propulsion. Both experimental and theoretical studies have shown that flexibility can enhance propulsion performance \cite{Heathcote2007,Michelin2009,Quinn2014, yudin2023propulsive}.

Early theoretical work examined how flexibility could improve propulsive performance by modifying the phase relationship between foil motion and fluid forces. In particular, studies of flapping appendages in inviscid flow demonstrated that there exists an optimal level of structural flexibility that maximizes thrust and propulsive efficiency \cite{Alben2008}. In these models, the deformation of the foil allows the trailing edge motion to adjust dynamically to the surrounding flow, producing favorable vortex formation and improved momentum transfer to the wake.

More recent computational and experimental studies have further explored how flexibility alters the interaction between the foil and its wake. Quinn et al. investigated flexible flapping foils using simulations and showed that structural deformation can significantly modify wake topology and thrust production \cite{Quinn2017}. Their results demonstrated that flexible foils can enhance propulsion by allowing the propulsor to passively adapt its shape during each oscillation cycle, effectively tuning the motion to the surrounding flow.

Flexibility can also play an important role in maneuvering. Deformation changes the instantaneous angle of attack and effective trailing-edge velocity, flexible foils can produce asymmetric force distributions when operated under biased or transient motions. This allows flexible propulsors to generate both thrust and lateral forces, expanding their potential role in maneuvering and control of bio-inspired vehicles. Recent work has begun to explore these effects in greater detail, examining how structural compliance influences both propulsion efficiency and maneuvering authority in oscillating fin systems \cite{yudin2024generating}.

Taken together, these studies demonstrate that structural flexibility introduces an additional design parameter in oscillatory propulsion systems. By coupling structural deformation with fluid dynamics, flexible foils can passively modify the effective kinematics of the propulsor, leading to changes in wake formation, thrust production, and maneuvering capability.

While much of the oscillating foil literature has focused on single propulsors, many biological propulsion systems involve multiple interacting fins. Fish often use paired fins or multiple propulsors along the body, and schooling behavior introduces additional hydrodynamic interactions between swimmers. These observations have motivated studies examining how multiple oscillating foils interact through their wakes. The hydrodynamic benefits of interacting swimmers were first explored in studies of fish schooling, which showed that vortices shed by upstream swimmers can reduce energetic cost for downstream individuals \cite{Weihs1973}. More recent work has examined these interactions in engineered oscillating foil systems, including tandem propulsors and schooling inspired propulsion arrangements \cite{Muscutt2017,Saadat2021,SeoMittal2022, Zhou2024, Guo2023, peng2009energy}.

One such configuration examined in the literature is the tandem arrangement of two in-line oscillating foils. In these systems, the downstream foil interacts directly with the vortex wake generated by the upstream foil. Studies have shown that when the spacing and phase relationship between the foils are properly tuned, the downstream foil can extract energy from the upstream wake and produce enhanced thrust \cite{Muscutt2017}. 

Related studies have also drawn connections between oscillating foil interactions and the hydrodynamics of fish schooling. In schooling configurations, swimmers may take advantage of the vortices generated by neighbors to reduce energy expenditure or enhance thrust production \cite{liu2017computational}. These collective hydrodynamic effects provide insight into how biological systems coordinate multiple propulsors and suggest potential strategies for multi-fin propulsion systems in engineered vehicles.

Multi-foil systems have also been investigated in the context of energy harvesting, where oscillating foils are arranged to extract energy from incoming vortex wakes. These studies highlight the broader relevance of wake interaction physics and demonstrate that the same mechanisms governing propulsion can also be exploited for energy extraction \cite{peng2009energy}. Understanding these interactions is therefore critical for both propulsion and energy systems involving oscillatory foils.

Together, these studies show that oscillating foil propulsion can be further benefited from studying the timing between a fins and the upstream structures they may encounter. When multiple propulsors are present, wake interactions introduce additional dynamics that can strongly influence thrust generation, efficiency, and maneuvering capability. Investigating these multi-foil interactions provides important insight into both biological propulsion systems and the design of multi-fin bio-inspired vehicles.

\section{Numerical methodology and validation}
\label{chap:validation}
Before discussing our studies and results, we begin by presenting our computational methods and subsequent validations we ran. We simulate oscillating foils in a uniform freestream with velocity $U_\infty$. The foil geometry is a NACA 0020 profile with chord length $c$. The prescribed kinematics consist of sinusoidal heaving and pitching about the leading edge,
\begin{equation}
h(t) = h_0 \sin(2\pi f t + \phi), \qquad
\theta(t) = \theta_0 \sin(2\pi f t + \phi + \Phi),
\label{eq:kinematics}
\end{equation}
\noindent where $h_0$ is the heave amplitude, $\Phi = 270^\circ$ as per previous research determined it to result in the highest efficiency \cite{van2019scaling}, $\theta_0$ is the pitch amplitude, $f$ is the frequency, and $\phi$ is a phase offset that we can adjust for $N$-fin systems. Adjusting the phase offset for downstream fins can introduce a time lag between upstream and downstream fins. 

Kinematics are non-dimensionalized using the Strouhal number and reduced frequency, 

\begin{equation}
St = \frac{2 f a_0}{U_\infty}
\qquad
f^* = \frac{f c}{U_\infty}
\label{eq:St_fstar}
\end{equation}

\noindent where $a_0$ is the trailing edge amplitude. The significance of the Strouhal number is that it relates the motion of the foil to that of the freestream. It characterizes the relationship between oscillation frequency, motion amplitude, and freestream velocity, and governs the wake structure from oscillating fins. In nature it has been found that cruising, or highly efficient, animals tend to exist in the range of $St \approx 0.2 -0.5 $, for our studies here we operate at $St = 0.36$ roughly in the center of this range. In this flapping regime efficient shedding of vortices forms and momentum is directed downstream most effectively. The importance of the reduced frequency, $f^*$, is that it compares the oscillation time scale of the fin to the convective time scale of the flow over the chord. It assists to quantify how strongly unsteady effects influence the fluid structure interaction. For combined heave and pitch, the trailing edge displacement is
\begin{equation}
y_{TE}(t) = h(t) + c\sin\theta(t),
\end{equation}
and the corresponding amplitude $a_0$ is computed from the maximum extension of $y_{TE}(t)$ over an oscillation cycle, this is primarily changed by the pitching amplitude $\theta_0$. To quantify the inertia to viscous forces in our flow we use the Reynolds number. The Reynolds number being defined as
\begin{equation}
Re = \frac{\rho U_\infty c}{\nu},
\label{eq:Re}
\end{equation}
with $\rho$ being fluid density, $c$ being chord length, and $\nu$ is the kinematic viscosity. The Reynolds governs the relative influence of viscosity, determining boundary layer behavior, and separation characteristics into the downstream. Unless otherwise noted, all simulations are performed at $Re = 5000$. This Reynolds number is selected in part due to limitations of the computational solver, which does not fully resolve turbulent flow regimes. However, turbulence is not expected to play a primary role in the present study, as the dominant forces are governed by large-scale flow separation and coherent vortex dynamics.

Fin performance is quantified using the coefficient of thrust, coefficient of power, and the Froude efficiency,
\begin{equation}
C_T = \frac{F_x}{\frac{1}{2}\rho U_\infty^2 s c}, \qquad
C_P = \frac{F_y \dot{h} + M_z \dot{\theta}}{\frac{1}{2}\rho U_\infty^3 s c}, \qquad
\eta = \frac{C_T}{C_P},
\label{eq:CT_CP_eta}
\end{equation}
\noindent where $F_x$ and $F_y$ are the streamwise and lateral forces, respectively, and $M_z$ is the spanwise moment about the pitching axis. Here, $\dot{h}$ represents the heave velocity and $\dot{\theta}$ is the angular velocity. The freestream velocity is used for normalization to maintain consistency with common convention; however, the lateral velocity scale is more physically representative of the foil kinematics and resulting force generation. Although these quantities vary over time, results are reported using cycle averaged values unless otherwise specified.

\section{Solver framework and methodology}
The core computational tool used in these studies is WaterLily. WaterLily is a free open-source computational fluid dynamics (CFD) solver written in the Julia programming language \cite{weymouth2024}. Additionally, the solver is back-end agnostic meaning it can easily be run on serial or multi-thread central processing units (CPUs). The solver has been shown to accurately match direct numerical simulation (DNS) and large eddy simulation (LES) models for flow over a Taylor-green vortex, oscillating cylinders, and oscillating foils \cite{weymouth2024}. 

WaterLily solves the incompressible viscous Navier-Stokes equations across a uniform cartesian grid. 
\begin{equation}
\frac{\partial \mathbf{u}}{\partial t}
+
(\mathbf{u} \cdot \nabla)\mathbf{u}
=
-\frac{1}{\rho}\nabla p
+
\nu \nabla^2 \mathbf{u}
+
\rho a_i
\label{eq:NS_vector}
\end{equation}

\noindent where $\mathbf{u} = (u,v,w)$ is the velocity vector, $p$ is pressure,
$\rho$ is density, $\nabla$ being spatial derivatives, and $\nu$ is the kinematic viscosity.

\begin{equation}
\nabla \cdot \mathbf{\vec{V}} = 0
\label{eq:continuity}
\end{equation}

Equations \ref{eq:NS_vector} and \ref{eq:continuity} represent conservation of momentum and mass, respectively governing the motion for Newtonian fluids assuming a constant density. This assumption is forced in the solver by way of divergence free velocity field preventing fluid expansion or compression. The momentum equation is advanced forward in time using a two step method that first predicts the velocity and then corrects it. After each time step, the velocity field is adjusted so that it remains incompressible. This is done by solving an additional equation for pressure. Since the presence of immersed boundaries (like a moving fin) changes the flow field, the pressure equation varies in space and must be solved carefully. A geometric multi-grid method is used to solve this equation efficiently. The time step size is automatically adjusted to keep the simulation numerically stable according to the Courant–Friedrichs–Lewy (CFL) condition. The CFL number is defined as

\begin{equation}
\mathrm{CFL} =
\frac{u_{\max}\,\Delta t}{\Delta x}
\label{eq:CFL}
\end{equation}

\noindent where $u_{\max}$ is the maximum velocity magnitude in the domain,
$\Delta t$ is the time step size, and $\Delta x$ is the grid spacing. For stability, the CFL number must satisfy \cite{courant1928}

\begin{equation}
\mathrm{CFL} < 1
\label{eq:CFL_condition}
\end{equation}

\noindent Physically, this condition ensures that fluid information does not travel farther than one grid cell within a single time step. If this constraint were to be violated, numerical instability may occur. 

WaterLily simulates static or dynamic bodies using the immersed boundary method (IBM) \cite{mittal2005immersed}. Rather than forcing the computational grid to conform to the geometry, IBM modifies the governing equations within a narrow transition region surrounding the immersed surface. Within this blending region, the fluid and solid equations are combined using a localized weighting procedure. Grid cells entirely within the fluid use the standard Navier–Stokes equations, while cells fully inside the solid follow the prescribed body motion. For cells near the interface, the equations are smoothly blended based on proximity to the surface, creating a gradual transition between fluid and solid behavior.

This blending enforces the no-slip boundary condition at the body surface, meaning that the fluid velocity matches the velocity of the solid boundary. For a stationary body, the fluid velocity at the surface is zero. For a moving or oscillating fin, the local fluid velocity matches the instantaneous velocity of the surface as it heaves or pitches. By embedding this transition directly into the governing equations, IBM maintains physically consistent boundary behavior while preserving numerical stability.

Since the mesh remains fixed, dynamic re-meshing is unnecessary when bodies move. This significantly reduces computational overhead and allows complex time dependent geometries to be simulated efficiently on structured Cartesian grids. As a result, large amplitude motions, multi-body interactions, and extensive parameter sweeps can be performed without the traditional cost associated with body fitted mesh approaches.

Lastly, WaterLily does not employ a core turbulence model, instead utilizing an implicit large eddy system (iLES) framework. This model does not directly model small scale turbulent structures but rather directly resolving large turbulent systems. While small scale structures are in theory then covered by relying on numerical dissipation between cells to act as a subgrid for small scale turbulence. Inherently then anything smaller than one cell is not resolved on the cartesian grid. This iLES resolution cutoff is discussed further in this chapter in section \ref{sec:mesh_comparison}.

\section{Coarse and fine setting theory }
The studies explored here focus on oscillating hydro-foils, referred to as fins or $N$-fin systems from here on. For $N \ge 2$ systems, the computational time was directly correlated with the number of fins. Ensuring that we had a balance of well resolved simulations that were not too computationally costly was key then to explore these greater $N$-fin systems. 

We balance well resolved resolutions for final data measurements and a coarser resolution to traverse and explore large parameter spaces from a topographical perspective. The reasoning here was when we looked at multi-fin systems across performance maps we planned to use the coarse cell refinement to locate zones of interest. After locating these zones, we would run specific cases at the high fidelity settings in order to analyze forces and animations of the flow fields. To achieve this we performed kinematic, temporal, and mesh refinement studies to determine ideal settings. In these refinement studies we analyzed the number of cyclic oscillations, amount of data points saved per cycle, and lastly cells per chord length.

\section{Cycle convergence study}

\begin{figure}[h!]
\centering
\includegraphics[width=0.9\textwidth]{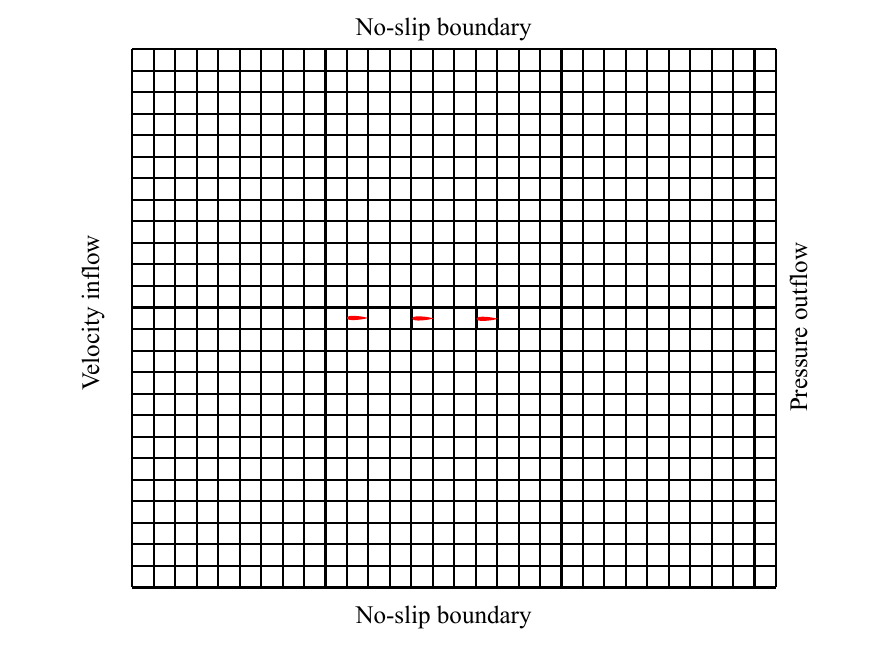}
\caption{Computational domain schematic of our standard $N=3$ fin system. Domain boundaries are 30 chords in $x$ by 25 chords in $y$, with boundary conditions indicated in the schematic. Each of these boxes would be 32x32 cells (coarse) or 64x64 (fine).}
\label{fig:ch1_domain}
\end{figure}

\begin{figure}[h!]
\centering
\includegraphics[width=1.0\textwidth]{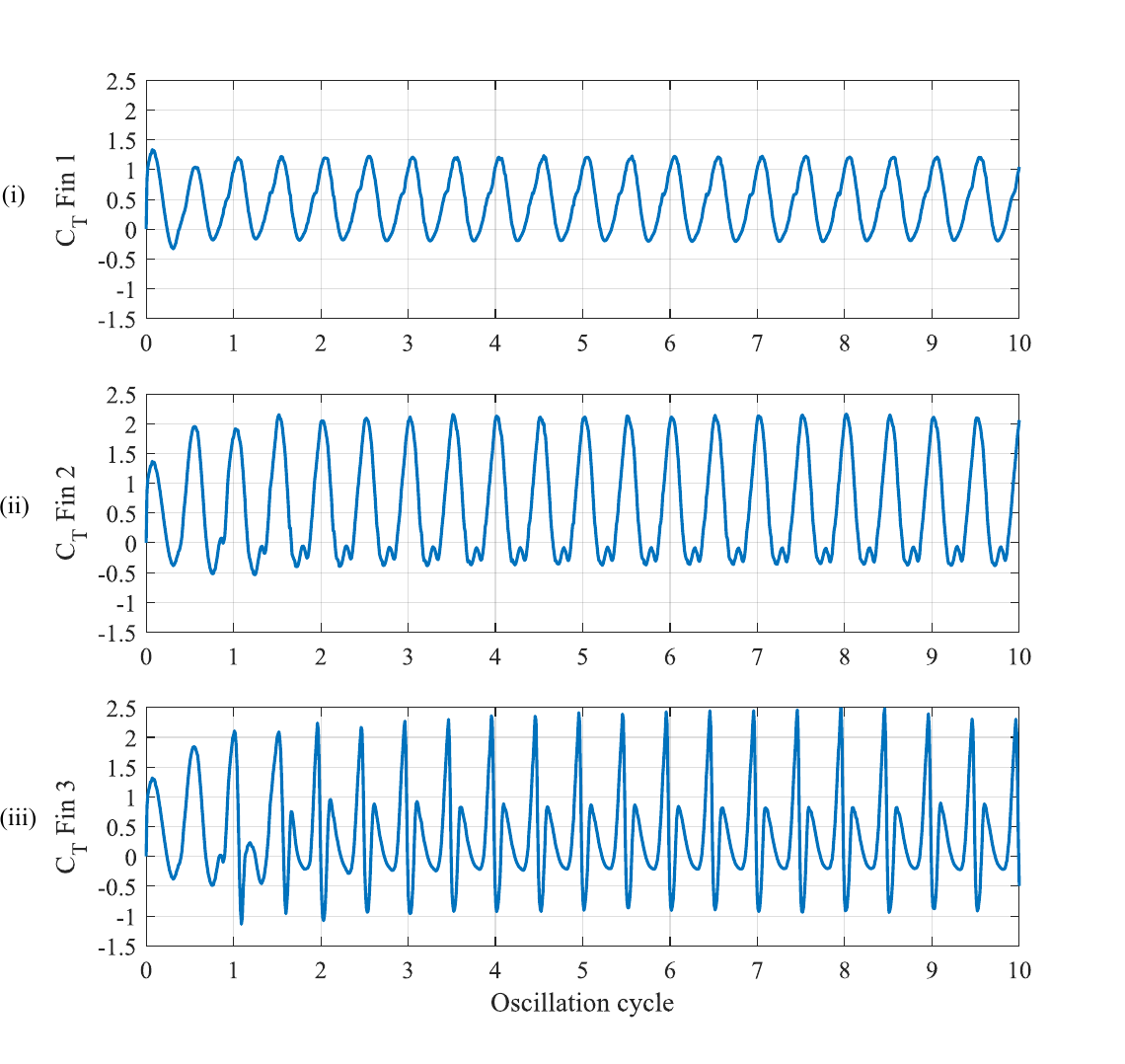}
\caption{Instantaneous coefficient of thrust, $C_T$, for each fin in a $N=3$ fin system over 10 oscillation cycles.}
\label{fig:ch1_cycle}
\end{figure}

We begin with a kinematic refinement study using the base configuration of three in-line fins spaced two chord lengths apart, and all in phase with one another, i.e. no phase or time lag offset is present in their heaving and pitching motions. The computational domain extends 30 chord lengths in the streamwise direction and 25 chord lengths in the lateral direction. Each fin has a NACA 0020 profile. The flow field setup of our general case is presented in figure \ref{fig:ch1_domain}. The full kinematic definitions of the prescribed heave and pitch motions presented earlier in this chapter, are again presented here for continuity but are later expanded upon in Chapter 2 for a more refined explanation. For a brief summary here, each fin undergoes a sinusoidal heave of one chord amplitude in both the positive and negative lateral directions while simultaneously pitching between $\pm 30$ degrees. The foils complete 10 full oscillation cycles, with 100 data points being saved per oscillatory cycle. We plot the instantaneous coefficient of thrust $C_T$ for each fin in figure \ref{fig:ch1_cycle}, and record the average coefficient of thrust for each fin at the third cycle, tenth (final) cycle, and across the last five cycles in table \ref{tab:cycle_convergence}.

To assess convergence, we compare the average thrust coefficient ($C_T$) computed during the third oscillation cycle to values obtained at later times in the simulation. Specifically, we use two reference metrics: (1) the average $C_T$ from the tenth cycle, and (2) the average $C_T$ over the final five cycles of the simulation.

For fins 1 and 2, the average $C_T$ in the third cycle differs by less than $1.00\%$ from both reference values, indicating that these fins have effectively reached a periodic steady state by the third cycle. In contrast, fin 3 exhibits a larger discrepancy: its third-cycle average $C_T$ is approximately $1.56\%$ lower than the tenth cycle and $5.41\%$ lower than the average over the final five cycles. This suggests that fin 3 converges more slowly, likely due to its position downstream and its dependence on upstream wake development. We deemed that for a coarse and fine search taking the third cycle of data would be sufficient in predicting system performance. However, for $N \ge 3$ we would add a cycle for each $N^{th}$ fin and measure the forces across the final $N^{th}$ cycle. 

\begin{table}[h!]
\caption{Relative percent difference in mean thrust coefficient, $\overline{C_T}$, computed using different cycle averaging windows. Cycle 3 (C3), Cycle 10 (C10), and the mean of the final five cycles (L5) are compared to assess convergence.}
\label{tab:cycle_convergence}
\centering
\small
\begin{tabular}{lccccc}
\toprule
 & $\overline{C_T}$ (C3) & $\overline{C_T}$ (C10) & $\overline{C_T}$ (L5) & \% Diff (C3 vs.\ C10) & \% Diff (C3 vs.\ L5) \\
\midrule
Fin 1 & 0.477 & 0.474 & 0.472 & 0.63\% & 1.06\% \\
Fin 2 & 0.622 & 0.622 & 0.626 & 0.00\% & 0.64\% \\
Fin 3 & 0.315 & 0.320 & 0.333 & 1.56\% & 5.41\% \\
\bottomrule
\end{tabular}
\end{table}

\section{Temporal Resolution Study}
The next refinement considered the number of temporal data points saved per oscillation cycle. For baseline simulations, force data was recorded at 100 intervals per cycle. However, large parameter sweeps substantially increase storage requirements and post processing time. To evaluate whether a reduced sampling frequency would preserve the essential force dynamics and characteristics, we compared simulations saved at 100 points per cycle to those saved at 99 points over three cycles, corresponding to 33 points per cycle when analyzed on a per cycle basis.

\begin{figure}[h!]
\centering
\includegraphics[width=1.0\textwidth]{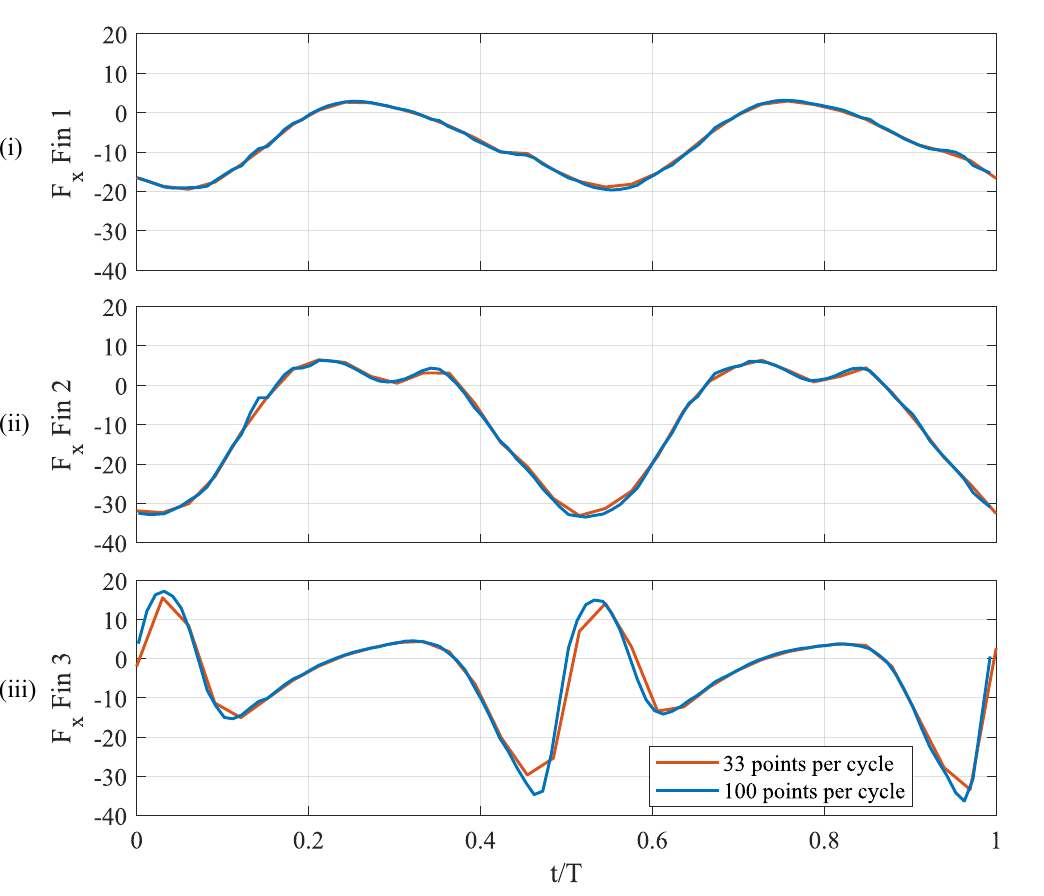}
\caption{Instantaneous force in the $x$ direction for each fin in a 3-fin system over the third oscillation cycle ($t/T \in [0,1]$). Temporal resolution comparison between 33 saved points per cycle and 100 saved points per cycle.}
\label{fig:ch1_temporal}
\end{figure}

It is important to note that the solver time step remained unchanged; only the frequency at which data was saved and then written to the file was altered. Thus, the underlying flow solution retained identical temporal accuracy, and only the output resolution was modified. Figure \ref{fig:ch1_temporal} compares the time-varying forces for all three fins over a representative oscillation cycle. The waveform shape, peak and trough magnitudes, and phase alignment are preserved between the two sampling strategies, with negligible differences in both instantaneous and cycle-averaged quantities for the first two fins. The primary difference seen in wave shape is across the third fin, emphasizing that when specific cases are rerun, 100 points will be used to capture temporal accuracy. Due to the fact that the reduced sampling captures the full oscillatory structure of the force signal while significantly decreasing file size and post-processing cost, 33 data points per cycle were adopted for coarse searches in large parameter studies. 

Table \ref{tab:temporal_comparison} compares the cycle averaged coefficient of thrust, $C_T$, for each individual fin as well as the system averaged value using both 33 and 100 data points per cycle. Across all cases, the differences between sampling strategies are small. The largest deviation occurs for fin-2, with a percent difference of $4.61\%$. All other fins and the system averaged thrust exhibit smaller discrepancies.

Importantly, the overall thrust hierarchy and relative performance trends remain unchanged between sampling resolutions. The observed differences fall within acceptable numerical tolerance for time averaged quantities and do not alter the physical interpretation of the system behavior. These results further support the adequacy of reduced temporal sampling for coarse searches in large parameter sweeps.

\begin{table}[h!]
\caption{Temporal resolution comparison between 33 and 100 saved points per cycle. Metrics are computed over the third cycle.}
\label{tab:temporal_comparison}
\centering
\small
\begin{tabular}{lcccc}
\toprule
 & $\overline{C_T}$ (33) & $\overline{C_T}$ (100) & \% diff \\
\midrule
Fin 1 & 0.48 & 0.49 & 2.04 \\
Fin 2 & 0.62 & 0.65 & 4.61  \\
Fin 3 & 0.32 & 0.32 & 0.00  \\
System Avg & 0.47 & 0.49 & 4.08  \\
\bottomrule
\end{tabular}
\end{table}

\section{Mesh comparison}
\label{sec:mesh_comparison}
To ensure that the spatial resolution was sufficient to capture the dominant viscous structures in the flow, baseline simulations were performed using 128 cells per chord. At the present Reynolds number ($Re = 5000$), the laminar boundary layer thickness scales as $\delta \sim 5c/\sqrt{Re}$, corresponding to approximately $0.07c$. With $c = 128$ cells per chord, this yields roughly nine grid cells spanning the boundary layer thickness, providing adequate resolution of near-wall velocity gradients and shear-layer development without reliance on wall models. The 128-cell configuration was therefore adopted as the reference grid for the spatial convergence study. 

\begin{figure}[h!]
\centering
\includegraphics[width=1.0\textwidth]{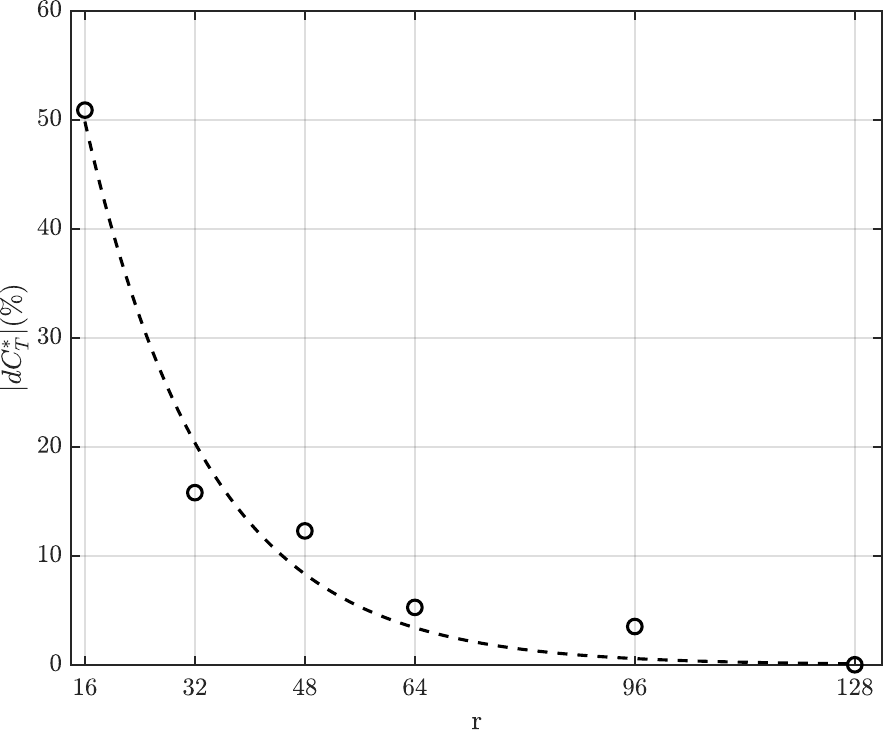}
\caption{Grid refinement study showing relative error of system-average $\overline{C_T}$ from 16 to 128 cells per chord, using 128 cells per chord as the reference case.}
\label{fig:ch1_grid}
\end{figure}

To evaluate the minimum resolution required to capture the primary force-producing mechanisms while reducing computational cost, additional simulations were conducted at progressively coarser resolutions. When the grid was reduced to 32 cells per chord, the predicted thrust coefficient $C_T$ was approximately $16\%$ lower than the 128 cell reference solution, indicating a much coarser model that would not be sufficient in final performance measurement but resolved enough for our coarse search of performance maps indicating zones of high or low performing interactions. Increasing the resolution to 64 cells per chord reduced the deviation to approximately $5.26\%$. In figure \ref{fig:ch1_grid} the results of this grid refinement study are presented as the relative change in system average $C_T$ versus cells per chord, $r$.

Based on this trade-off between accuracy and computational expense, 64 cells per chord was selected for high-fidelity single-case simulations, whereas 32 cells per chord was employed as a coarse-resolution model for large parameter sweeps focused on identifying global vortex interaction trends.

\begin{figure}[h!]
\centering
\includegraphics[width=0.8\textwidth]{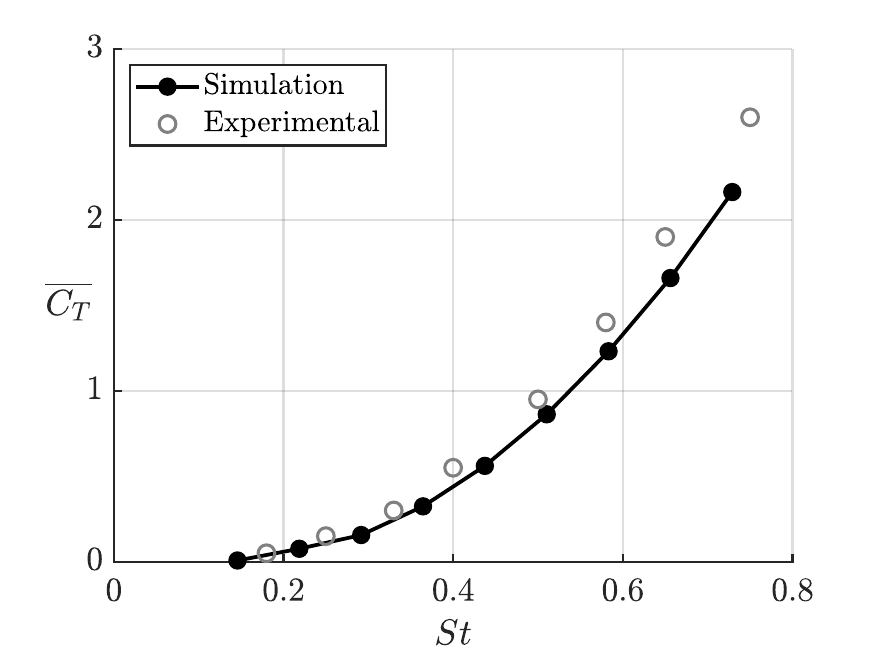}
\caption{Cycle-averaged coefficient of thrust, $\overline{C_T}$, for a single fin (NACA 0020) at our fine setting (c = 64 cells) as Strouhal number ($St$) varies.}
\label{fig:ch1_validation}
\end{figure}

To further assess whether 32 cells per chord provided sufficient resolution for capturing the dominant flow physics, an external validation was performed against the experimental results reported by Van Buren et al \cite{van2019foil}. In figure \ref{fig:ch1_validation}, the average coefficient of thrust $C_T$ is plotted against the Strouhal number of the system. The simulations reproduce the steady increase in $C_T$ with increasing Strouhal number, as well as the relative magnitude of thrust enhancement observed experimentally. This agreement indicates that, although the 32-cell grid underpredicts absolute thrust relative to the 128-cell reference solution, it preserves the correct physical trends associated with Strouhal dependent performance. Consequently, the 32-cell configuration was deemed suitable for large-scale parameter sweeps focused on comparative performance trends rather than precise force magnitudes. 

\begin{figure}[H]
    \centering
    
    \begin{subfigure}[b]{0.48\textwidth}
        \centering
        \includegraphics[width=\textwidth]{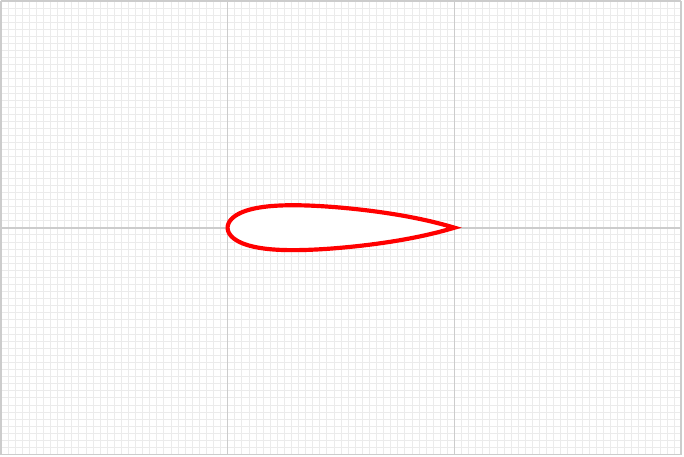}
        \caption{$L = 32$ cells per chord}
        \label{fig:mesh_L32}
    \end{subfigure}
    \hfill
    \begin{subfigure}[b]{0.48\textwidth}
        \centering
        \includegraphics[width=\textwidth]{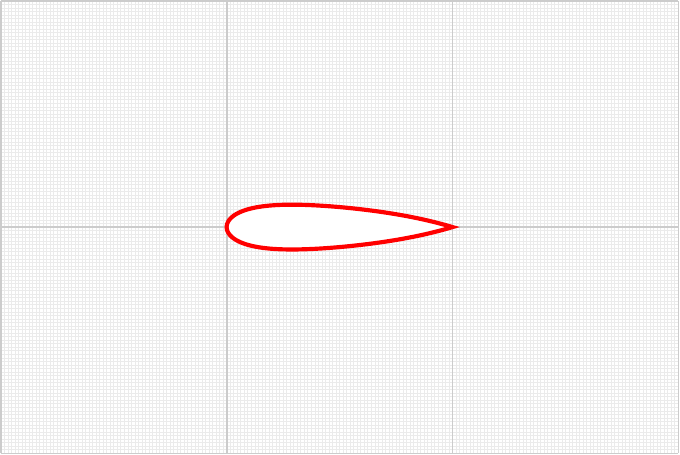}
        \caption{$L = 64$ cells per chord}
        \label{fig:mesh_L64}
    \end{subfigure}
    
    \caption{Computational mesh comparison showing refinement from $L=32$ (a) to $L=64$ (b) cells per chord. The refinement increases spatial resolution near the foil while preserving identical domain dimensions.}
    \label{fig:ch1_mesh_compare}
\end{figure}

To visually assess the influence of grid density on flow-field representation, figure \ref{fig:ch1_mesh_compare} compares the computational mesh for $L = 32$ and $L = 64$ cells per chord in the vicinity of the fin. The increased grid density at $L = 64$ provides finer resolution of the immersed boundary geometry. In contrast, the $L = 32$ configuration exhibits coarser discretization near the body, allowing faster large parameter sweeps.

\chapter{Single fin system}
\label{ch2:single_fin}
\section{Introduction}

In this chapter, we examine how a single fin generates forward thrust through oscillatory motion. We begin by exploring how different kinematic modes; pure heave, pure pitch, and combined pitch and heave motion, produce thrust through distinct physical mechanisms. By analyzing time varying forces over an oscillation cycle, we show how symmetric sinusoidal motion results in zero net lateral force while still producing net positive forward thrust.

We then quantify the thrust and efficiency performance of a rigid single fin, establishing a baseline for comparison using our NACA 0020 model introduced in chapter \ref{chap:validation}. Building upon this foundation, we introduce a numerical model incorporating a leading edge torsional spring to enable passive pitching. This formulation allows the pitch response to arise from fluid structure interaction rather than being fully prescribed.

We examine both constant and time varying stiffness models to emulate flexible foil behavior. Using scaling arguments, we derive an analytical expression to estimate force production over an oscillation cycle for the leading edge spring configuration. Finally, we compare the thrust performance of the passive system to that of a baseline rigid foil undergoing prescribed pitch and heave, highlighting how time-varying stiffness can enhance or diminish forward thrust generation.

\section{Methods}
\begin{figure}[H]
    \centering
    
    \begin{subfigure}[b]{0.44\textwidth}
        \centering
        \includegraphics[width=\textwidth]{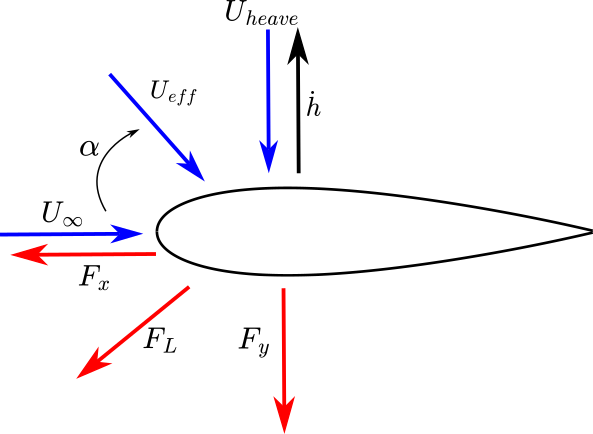}
        \caption{}
        \label{fig:pure_heave}
    \end{subfigure}
    \hfill
    \begin{subfigure}[b]{0.5\textwidth}
        \centering
        \includegraphics[width=\textwidth]{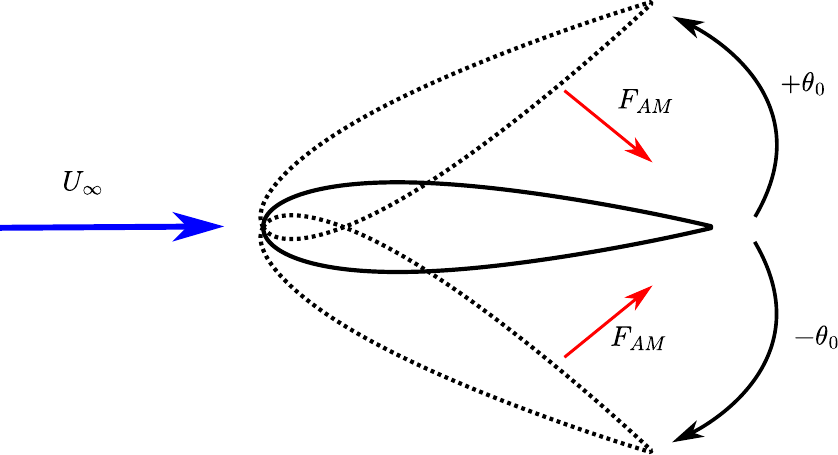}
        \caption{}
        \label{fig:pure_pitch}
    \end{subfigure}
    
    \caption{Comparison of two fundamental thrust-generation mechanisms for an oscillating fin. 
    (a) Pure heave generates lift based thrust through the interaction of the freestream velocity, $U_{\infty}$, and the heave velocity, $\dot{h}$, which combine to produce an effective inflow velocity, $U_{\mathrm{eff}}$. 
    (b) Pure pitching motion generates thrust primarily through unsteady added-mass forces, denoted $F_{\mathrm{AM}}$, arising from fluid acceleration around the rotating foil.}
    \label{fig:ch2_heave_vs_pitch}
\end{figure}

Beginning with a single fin, we discuss how different modes of thrust generation occur through pure heaving vs pure pitching, shown in figures \ref{fig:pure_heave} and \ref{fig:pure_pitch}, respectively. Although both motions can produce forward thrust, they do so through fundamentally different physical mechanisms.

For pure heaving motion, the fin undergoes a prescribed vertical displacement $h(t)$ in a steady freestream of velocity $U_\infty$. The instantaneous heave velocity $\dot{h}$ combines with the freestream velocity to produce an effective inflow velocity \cite{Floryan2017},
\[
U_{\mathrm{eff}} = \sqrt{U_\infty^2 + \dot{h}^2},
\]
oriented at an angle 
\[
\alpha = \tan^{-1}\left(\frac{\dot{h}}{U_\infty}\right),
\]
relative to the horizontal. This induced angle of attack generates a lift force proportional to the square of the effective velocity,
\[
L = \frac{1}{2} \rho U_{\mathrm{eff}}^2 s c C_L,
\]
where $c$ is the chord length and $s$ the span of the foil. And $C_L$ being defined as \cite{Theodorsen1935},
\[
C_L = 2 \pi sin(\alpha)+(3/2)\pi\dot{\alpha}c/U_{\mathrm{eff}}.
\]

The lift force acts perpendicular to the effective flow direction and may be decomposed into lateral and streamwise components. The streamwise projection of lift produces thrust,
\[
F_x = -L \sin(-\alpha),
\]
while the lateral component is given by
\[
F_y = L \cos(-\alpha).
\]
In pure heave, forward thrust is therefore generated through lift-based mechanisms resulting from the reorientation of the lift vector relative to the freestream. 

In contrast, for pure pitching motion about the leading edge, the foil rotates with angular displacement $\theta(t)$ while maintaining a fixed vertical position. In this case, quasi-steady lift does not generate net streamwise thrust when averaged over a cycle. Instead, thrust arises primarily from added mass forces associated with accelerating the surrounding fluid. When a body accelerates in a fluid, it must impart momentum to an effective mass of fluid, resulting in a reactive force proportional to the fluid density and the acceleration of the body. For pitching motion, the dominant added mass contribution scales with
\[
F_x \sim \rho c^2 \ddot{\theta},
\]
reflecting the acceleration of the fluid induced by rotational motion \cite{Floryan2017, sedov1965two}. These unsteady inertial forces produce a net streamwise component of thrust when integrated over a cycle.

Thus, pure heave generates thrust predominantly through lift based mechanisms governed by the instantaneous angle of attack, whereas pure pitch generates thrust predominantly through added mass effects governed by angular acceleration. The relative contribution of these mechanisms becomes especially important in combined pitch and heave motion, where both lift based and added mass forces act simultaneously. Pitch also directly influences the circulatory forces when combined with heave, because from the lift-based perspective the pitch is modulating the angle of attack.

\begin{figure}[H]
    \centering
    
    \begin{subfigure}[b]{0.44\textwidth}
        \centering
        \includegraphics[width=\textwidth]{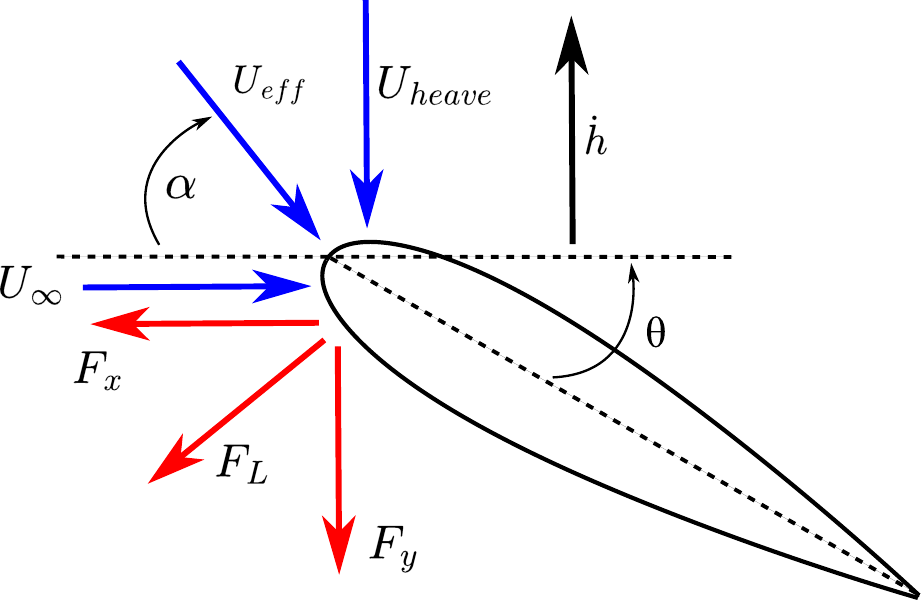}
        \caption{}
        \label{fig:pitch_and_heave}
    \end{subfigure}
    \hfill
    \begin{subfigure}[b]{0.5\textwidth}
        \centering
        \includegraphics[width=\textwidth]{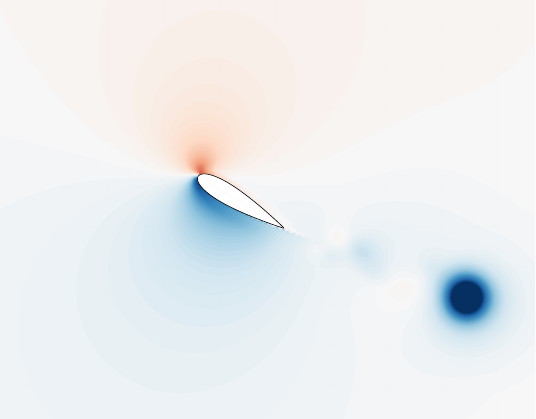}
        \caption{}
        \label{fig:pitch_and_heave_sim}
    \end{subfigure}
    
    \caption{A NACA 0020 undergoing combined pitch and heave motion. Kinematic schematic (a) illustrating the combination of pitching about the leading edge with a heave velocity. Corresponding non-dimensional pressure field (b) from the simulation, showing low pressure regions (blue) and high pressure regions (red) on the opposite surface  and vortex shed into the freestream.}
    \label{fig:ch2_pitch_and_heave}
\end{figure}

When heaving and pitching motions are combined, as illustrated in figure \ref{fig:pitch_and_heave}, the fin benefits from two distinct thrust-generating mechanisms: lift-based forces associated with the effective angle of attack, and added mass forces associated with unsteady acceleration. The superposition of these motions allows constructive interaction between circulatory and inertial contributions to thrust. In particular, introducing pitch in phase with the heave displacement effectively tilts the fin into the direction of motion, slightly decreasing the instantaneous angle of attack while mitigating large-scale flow separation. This reduces the likelihood of dynamic stall and permits larger heave velocities without a corresponding loss in performance. Additionally when oscillating the foil through this motion we shed vortices into the wake downstream form the fin as show in figure \ref{fig:pitch_and_heave_sim}.

Following established scaling arguments \cite{floryan2018efficient}, thrust may be approximated as

\[
T \sim \rho S_p V^2 - D_o,
\]
where $S_p$ is the propulsor area, $V \sim fA$ is the characteristic transverse velocity scale, and $D_o$ represents the offset drag associated with the projected frontal area of the oscillating fin. The quadratic dependence on $V$ reflects the scaling of both lift-based and added-mass contributions to thrust. Similarly, the power input scales as
\[
P \sim \rho S_p f L \left(V^2 - V_h V_\theta \right),
\]
where $L$ is the chord length and $V_h$, $V_\theta$ are characteristic velocity scales associated with heave and pitch, respectively. These relations highlight the balance between thrust production and drag penalties in determining efficiency. At low Strouhal numbers, thrust is insufficient to overcome offset drag, leading to poor performance. At high Strouhal numbers, increasing power requirements reduce efficiency. The resulting competition gives rise to an optimal Strouhal band for efficient propulsion.

Thus, combined pitch heave motion enables the foil to simultaneously exploit lift-based force reorientation and added-mass acceleration effects, while operating within a Strouhal regime that maximizes propulsive efficiency. With this basis we begin with our analysis of single fin systems at $St = 0.36$, approximately in the middle of this zone.

\section{Results and discussion}

Starting with the computational setup described in Chapter \ref{chap:validation}, we simulate a NACA 0020 foil undergoing prescribed pitching and heaving motion over a full oscillation cycle. The foil translates one full chord above and below its midline position while pitching between $-30^\circ$ and $+30^\circ$. The complete motion over one cycle is shown in figure \ref{fig:ch2_pitch_and_heave}. The instantaneous coefficient of thrust vs heave motion for this case is also shown.

\begin{figure}[H]
    \centering
    
    \begin{subfigure}[b]{0.35\textwidth}
        \centering
        \includegraphics[width=\textwidth]{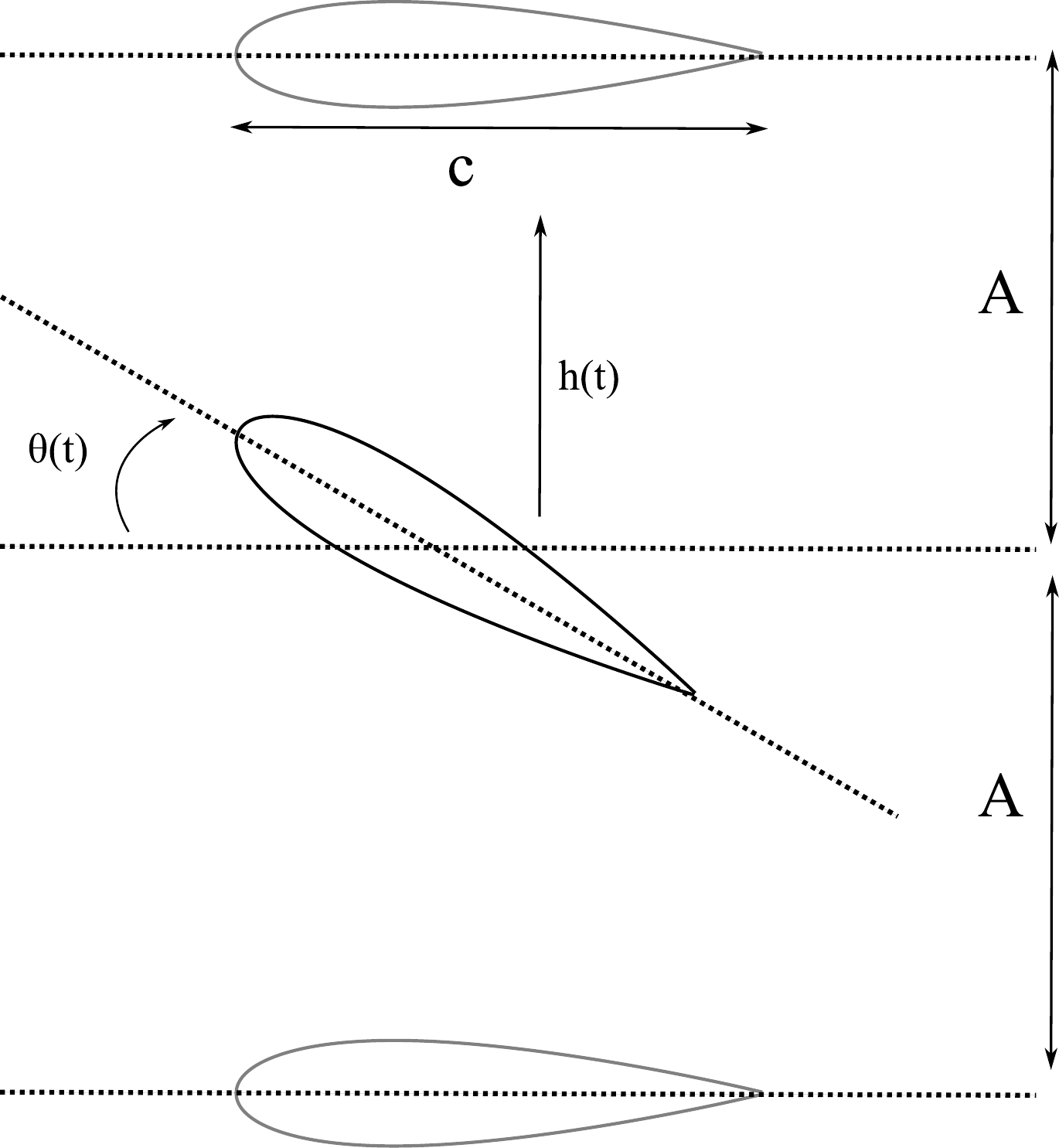}
        \caption{}
        \label{fig:amplitude}
    \end{subfigure}
    \hfill
    \begin{subfigure}[b]{0.6\textwidth}
        \centering
        \includegraphics[width=\textwidth]{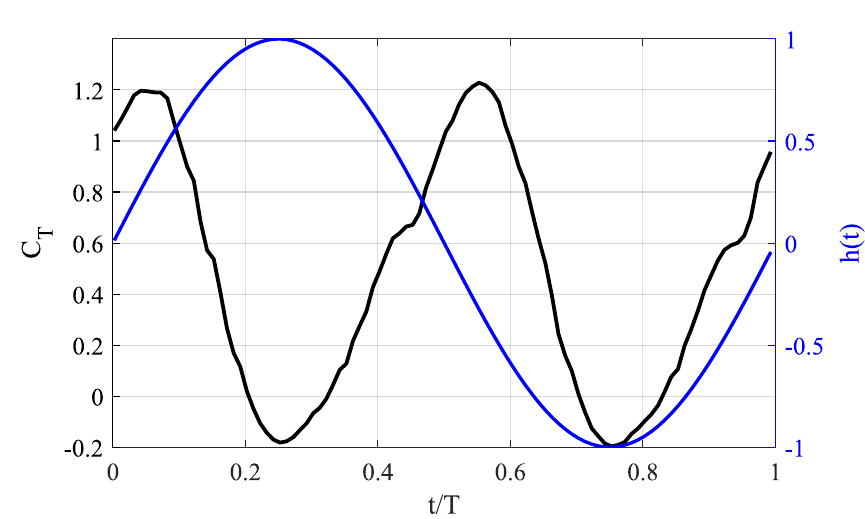}
        \caption{}
        \label{fig:single fin force}
    \end{subfigure}
    
    \caption{Kinematics and force generation for a single oscillating fin. In (a) prescribed heaving motion over one oscillation cycle, illustrating the heave amplitude $A$, chord length $c$, and instantaneous vertical displacement $h(t)$. In (b) the instantaneous coefficient of thrust $C_T$ is plotted (black) over one non-dimensional cycle, shown together with the corresponding heave position (blue) to illustrate the relationship between fin location and thrust production.}
    \label{fig: ch2_force_vs_cycle}
\end{figure}

Peak thrust occurs midway through the heave motion when the foil reaches its maximum pitch amplitude while translating through the center of the stroke. At this point relative inflow velocity from the combination of heave and freestream velocities produce the largest streamwise force contribution. The cycle averaged performance of the single fin yields a coefficient of thrust and Froude efficiency of $\overline{C_T} = 0.47$ and $\eta = 0.37$, respectively. 

\subsection{Leading edge spring model}
We now replace the prescribed pitch motion with a passive rotational degree of freedom governed by a leading edge torsional spring. In this configuration the heave motion remains prescribed and symmetric, while the pitch angle $\theta(t)$ is no longer imposed kinematically. Instead, the instantaneous pitch angle emerges from a balance between the hydrodynamic moment acting on the foil and the restoring torque of the spring.

The rotational dynamics are governed by a moment balance about the leading edge,
\begin{equation}
(I_0 + m_a I_0)\,\ddot{\theta}
=
M(t)
-
k\,\theta
-
\frac{1}{2} L m \ddot{h}(t),
\label{eq:LEspring_base}
\end{equation}
where $M(t)$ is the hydrodynamic moment computed directly from the pressure and viscous stress distribution along the foil surface, $I_0$ is the foil mass moment of inertia about the leading edge, $m_a$ is an added rotational inertia coefficient, and $m$ is the foil mass. The final term represents the inertial coupling between the prescribed heave acceleration and the rotational motion.

Unlike the rigid case, the pitch amplitude and phase are not specified. They are determined by the coupled interaction between the flow field and the structural stiffness. The spring stiffness $k$ therefore becomes the primary parameter controlling the rotational response of the foil and, in turn, the resulting force production.

\subsection{Time varying stiffness}
Finally, we extend the leading edge spring model by allowing the torsional stiffness to vary in time over the oscillation cycle. We do this by altering the spring stiffness term $k$ from equation \ref{eq:LEspring_base} to become,

\begin{equation}
k(t) = p\, k_{\mathrm{amp}} \cos\!\left( 2\pi k_f t \right) + k_{\mathrm{amp}}.
\label{eq:spring_stiffness}
\end{equation}

\noindent This introduces a controlled time dependence into the rotational restoring torque and provides a mechanism to further modify the passive pitch response and resulting force production. 

The motivation for both the constant and time varying stiffness cases builds upon the work of \cite{yudin2023propulsive}, in which numerical simulations identified optimal stiffness regimes for passively pitching plates. Their results demonstrated that maximum propulsive performance occurs when stiffness variation is tuned relative to the forcing frequency. Additionally, they note that the spring stiffness variation term $k_f$ must be double the oscillation heave frequency term $f$ in order to retain symmetric stiffness variation across both half cycles, allowing for net zero lateral force. In this regime, the parametric resonance is induced, leading to performance benefits.

\begin{figure}[h!]
\centering
\includegraphics[width=0.9\textwidth]{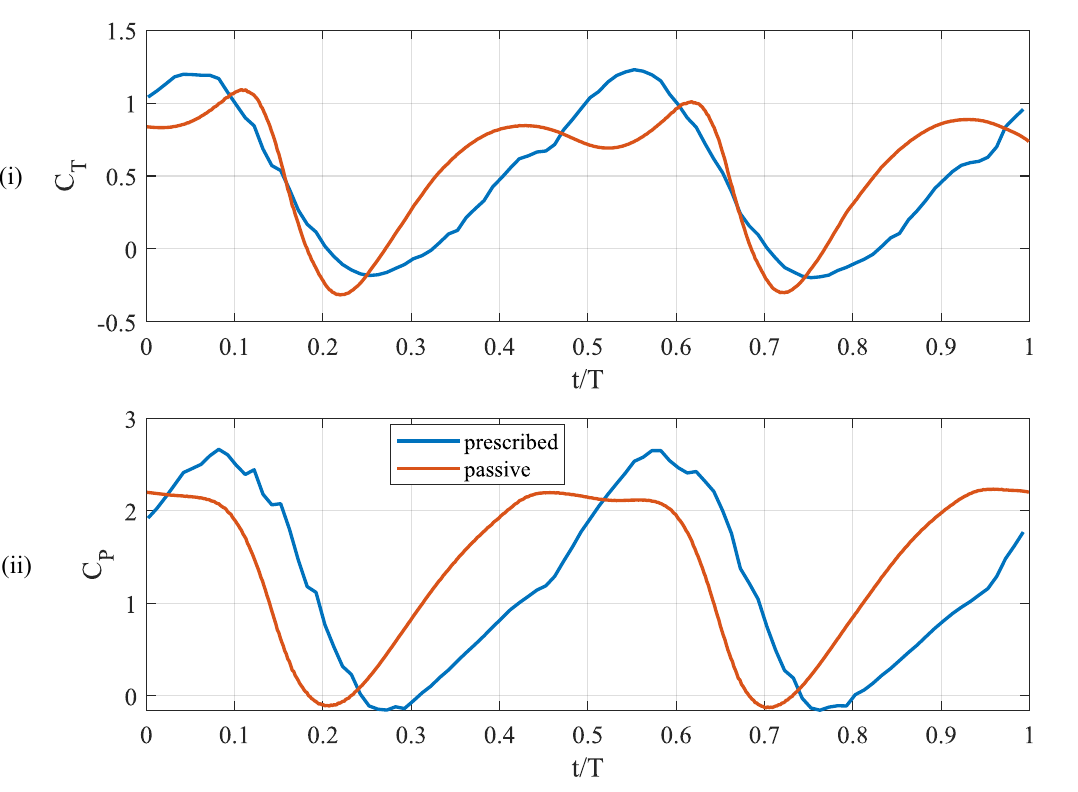}
\caption{Comparison between a rigid (prescribed pitching) fin and a leading edge spring (passively pitching) fin over one oscillation cycle. Plot (a) Instantaneous coefficient of thrust, $C_T$, plotted versus normalized time $t/T$. Plot (b) Instantaneous coefficient of power, $C_P$, for the same two cases.}
\label{fig:ch2_prescribed_vs_passive}
\end{figure}

From their findings we set our stiffness to be at its peak mid heave and most flexible at the oscillation turn around points. Additionally from their plots it showed that a $50\%$ alteration in spring stiffness yielded the highest performance increase. We set out model to follow these two heuristics. With all other kinematic settings remaining in our base settings, this resulted in a $\overline{C_T} = 0.51$ and an $\eta = 0.39$. By adding these heuristics into a simple leading edge spring system, we increase thrust by $8\%$ and efficiency by $6.4\%$. We plot the time instantaneous $C_T$ for both a rigid foil and our idealized flexible foil in figure \ref{fig:ch2_prescribed_vs_passive}.

\section{Conclusion}

In this chapter we examined the thrust generation mechanisms of a single oscillating fin and established a baseline performance for the NACA 0020 foil model. Pure heaving motion was shown to generate thrust primarily through lift-based mechanisms arising from the effective angle of attack created by the interaction between the freestream velocity and the heave velocity. In contrast, pure pitching motion generates thrust predominantly through unsteady added-mass effects associated with the acceleration of the surrounding fluid. When these motions are combined, the foil simultaneously exploits both circulatory lift and inertial added-mass contributions, enabling efficient thrust production within the optimal Strouhal regime.

Using prescribed pitch–heave kinematics at $St = 0.36$, the rigid foil produced a cycle-averaged thrust coefficient of $\overline{C_T} = 0.47$ and a Froude efficiency of $\eta = 0.37$. This configuration served as a reference case for evaluating the leading edge spring based pitching mechanisms. Introducing a leading-edge torsional spring allowed the pitch response to emerge from fluid–structure interaction rather than being kinematically imposed. Extending this model to include time-varying stiffness further modified the phase relationship between pitch and heave, redistributing hydrodynamic loading over the oscillation cycle.

Applying stiffness variation heuristics motivated by prior work resulted in measurable improvements in propulsion performance, the time-varying stiffness model simultaneously increased the thrust coefficient and efficiency. These results demonstrate that even simple passive compliance mechanisms can meaningfully influence oscillating fin performance. The insights gained from this single-fin analysis provide the foundation for the maneuvering and multi-fin interaction studies examined in the following chapters.    

 \chapter{Maneuvering via asymmetric motions} \label{ch3:manuevering} 
 
 \section{Introduction} 
 
 Building on the findings of the previous chapter, we now examine how a fin-based propulsion system can be manipulated to generate lateral forces and controlled maneuvering moments. So far our focus has been on symmetric oscillatory motions that maximize forward thrust and propulsive efficiency. However, propulsion alone does not define vehicle performance. True capability requires maneuverability. 
 
 Continuing through the lens of biological inspiration, nature provides numerous examples of swimmers that inherently integrate propulsion and maneuvering \cite{webb2015stability, bandyopadhyay2002maneuvering}. Fish and marine mammals are capable of executing turns within their own body length while maintaining stability and control \cite{weihs1972hydrodynamical}. In contrast, traditional human-designed marine vehicles rely on separate subsystems for thrust generation, steering, and stabilization. Propellers generate forward thrust, while rudders or other control surfaces (canards, bow thrusters, bow planes, etc) provide directional control. As a result, conventional vessels typically require several body lengths to alter direction and exhibit comparatively large turning radii. This distinction becomes particularly relevant in the design of uncrewed underwater vehicles, which are of increasing interest due to their low cost, deployability, and wide range of mission capabilities. 
 
 Oscillatory fin-based propulsion has already demonstrated efficient forward thrust generation through the combined action of lift-based forces and unsteady added-mass effects. The advantages of this approach may become even more pronounced when asymmetric motions are introduced. By intentionally breaking kinematic symmetry, oscillatory fins may generate not only forward thrust but also lateral forces and maneuvering moments. This opens the possibility of a unified actuation system capable of propulsion, steering, and stabilization simultaneously. Such integration could reduce the need for additional control surfaces, decrease drag-producing appendages, and simplify vehicle geometry. The result would be a potentially more efficient, lower-drag, and mechanically streamlined underwater vehicle architecture. In this chapter, we explore how asymmetric fin kinematics can be used to produce controlled lateral forces and maneuvering performance, advancing fin-based propulsion from efficient thrust generation toward fully integrated vehicle control.
 
 \section{Methods} 
 
 We explore three primary modes of breaking kinematic symmetry. The first for a rigid fin with prescribed motion, we may add a pitch bias, $\theta_{pb}$ generating a non-net zero lift vector in the lateral direction of that bias. 
 
 \begin{align} 
 \theta(t) = \theta_{pb} + \theta_0 \sin(2\pi f t + \phi) \end{align} 
 
 Secondly, we can alter our prescribed heave motion by increasing the speed, $\dot{h}$, that way we heave our fin faster in one direction and slower in the other, increasing our effective velocity, $U_{eff}$ in said direction. We start with our base heave motion: 
 
 \begin{align}
 h(t) &= h_0 \sin(2\pi f t), 
 \end{align} 
 
 \noindent we replace the pure sinusoid by a tilted waveform by applying a Fourier series with tilt parameter $\lambda$ and $N$ modes: 
 
 \begin{equation} 
 s(\tau;\lambda)=\sum_{k=1}^{N}\frac{\lambda^{k-1}}{k}\sin(k\tau) \qquad \tau\equiv 2\pi f t 
 \label{fourier}
 \end{equation}

 \noindent We normalize by the maximum magnitude over one period,
 
 \begin{equation} 
 \mathcal{N}(\lambda)=\max_{\tau\in[0,2\pi]}\left|s(\tau;\lambda)\right|, \qquad \hat{s}(\tau;\lambda)=\frac{s(\tau;\lambda)}{\mathcal{N}(\lambda)}, \qquad |\hat{s}|\le 1
 \end{equation} 
 
 \noindent so that the heave remains bounded by $\pm h_0$ even as the waveform is skewed: 
 
 \begin{equation} 
 h(t)=h_0\,\hat{s}(2\pi f t;\lambda) 
 \end{equation} 
 
 To similarly skew the pitch rate while keeping the pitch amplitude bounded by $\pm\theta_0$, we define a tilted pitch equation 
 \begin{equation} 
 \Psi(t)=(2\pi f t+\phi)+\lambda\ \hat{s}(2\pi f t+\phi;\lambda)
 \end{equation} 
 
 \noindent and prescribe pitch as 
 
 \begin{equation} 
 \theta(t)=\theta_0\sin\!\big(\Psi(t)\big) \qquad 
 \end{equation} 
 
 This construction preserves the peak amplitudes $|h|\le h_0$ and $|\theta|\le \theta_0$ while increasing the slope (speed) over one half-cycle and decreasing it over the other, producing a fast upstroke and a slow downstroke (or vise versa depending on $\lambda$ and phase). 
 
\begin{figure}[h!]
\centering
\includegraphics[width=1.0\textwidth]{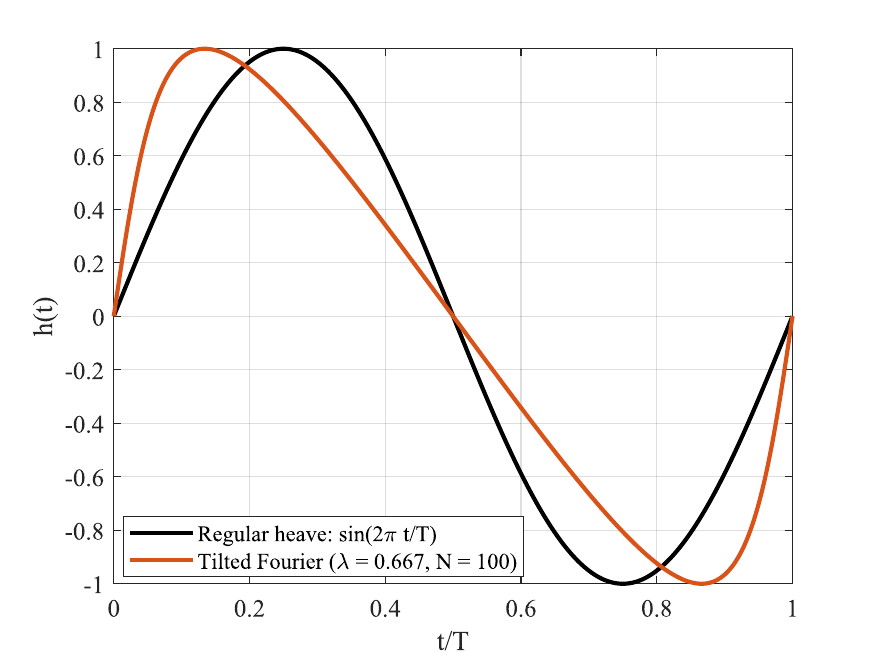}
\caption{Waveform comparison between a regular sinusoidally symmetric heave (black) and our asymmetric heave form tilted via a Fourier series and re-normalized to maintain bounded extrema (orange).}
\label{fig:ch3_waves}
\end{figure}
 
\noindent In figure \ref{fig:ch3_waves} we show the effects of this Fourier series transformation, setting the tilt factor $\lambda = .667$, we have created a 5 to 1 ratio in which the maximum speed in the heaved up direction is 5 times higher than the maximum in the downward direction. 
 
 Lastly, looking at our passive pitch leading edge spring model, we can generate asymmetric lateral forces by means of asymmetric stiffness. Looking to our time varying stiffness equation \ref{eq:spring_stiffness}: 
 
 \[ k(t) = k_{\mathrm{amp}} 
 \left[1 + p \cos(2\pi k_f t)\right] 
 \] 
 
\begin{figure}[h!]
\centering
\includegraphics[width=1.0\textwidth]{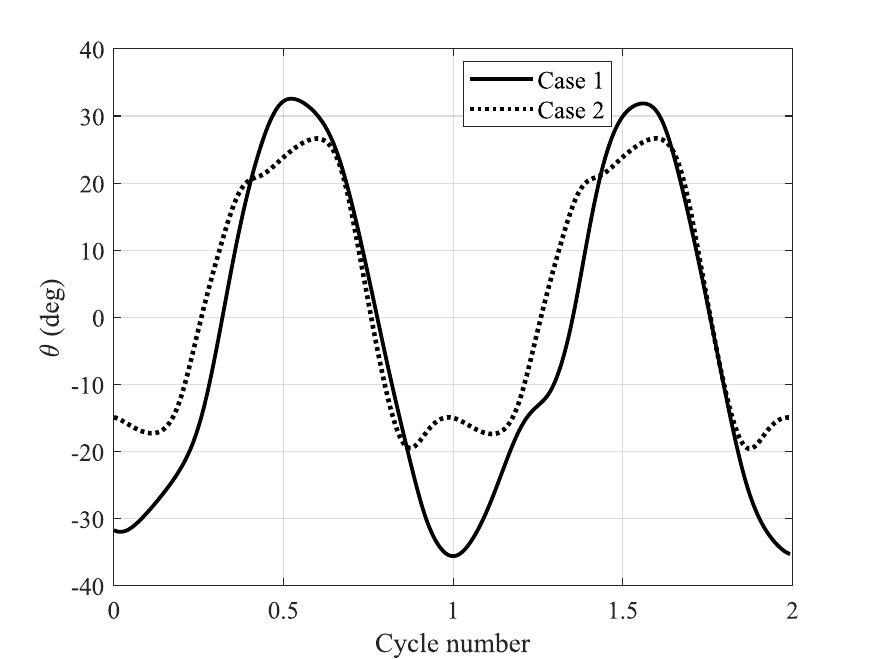}
\caption{Instantaneous $\theta$ comparison between symmetrically varying stiffness leading edge spring fin (case 1) and an asymmetric varying stiffness leading edge spring (case 2).}
\label{fig:ch3_theta_comparison}
\end{figure}

 If we no longer set $k_f = 2f$, we no longer alter our flexibility symmetrically throughout each half oscillation cycle. We then are creating a half cycle variance, where one half experience's a stiffer fin and the other half heave experience's the fin as flexible. Looking to figure \ref{fig:ch3_theta_comparison}, we show the instantaneous recording $\theta(t)$ of a fin where one half cycle is stiff the other is flexible. We see that the symmetric case (case 1) reaches a $\theta_0 = \pm32^\circ$, where the asymmetric case reaches $\theta_0 = 28^\circ$ and $\theta_0 = -19^\circ$ for its max and min respectively. Illustrating how we can effectively alter angle of attack by changing flexibility.

 \section{Discussion} 
 
 We begin with the scaling laws governing force generation for a fin undergoing simultaneous pitching and heaving motion \cite{van2019scaling}. The lateral force is decomposed into lift-based and added-mass contributions, allowing us to examine how each mechanism contributes independently before considering their combined effect. The total lateral force $F_y$ is written as 
 
 \begin{equation} 
 F_y \;=\; F_{y,L} + F_{y,AM}. 
 \label{eq:Fy_split} 
 \end{equation} 
 
 This decomposition provides a structured framework for understanding how modifications to kinematics influence force production. 
 
 \subsection{Lift-based contribution} 
 
 The lift force scales with the square of the effective velocity and the instantaneous lift coefficient. Following the scaling arguments, 
 \begin{equation} 
 F_L \sim \rho s c U_{\mathrm{eff}}^2 C_L, \qquad C_L \sim \alpha + \frac{c\dot{\alpha}}{U_{\mathrm{eff}}} 
 \end{equation} 
 
\noindent where $s$ is the span, $c$ is the chord, and $U_{\mathrm{eff}}$ is the effective inflow velocity experienced by the fin. The first term in $C_L$ represents the quasi-steady contribution proportional to angle of attack, while the second term accounts for unsteady rotational effects associated with pitch rate. Projecting the lift force into the lateral direction by multiplying by $U_\infty/U_{\mathrm{eff}}$ gives 
 
 \begin{equation}
 F_{y,L} \sim \rho s c\left( \underbrace{\alpha U_\infty U_{\mathrm{eff}}}_{\text{(1) steady lift}} + \underbrace{c\dot{\alpha}U_\infty}_{\text{(2) unsteady lift}} \right). 
 \end{equation} 
 
 The grouped terms represent the steady and unsteady lift components, respectively. The steady term depends directly on the instantaneous angle of attack and effective velocity, while the unsteady term depends on the rate of change of angle of attack. The instantaneous angle of attack is modeled as 
 
 \begin{equation}
 \alpha(t) = -\theta(t) - \tan^{-1}\!\left(\frac{\dot{h}(t)}{U_\infty}\right) 
 \end{equation} 
 
\noindent which accounts for both geometric pitch and induced angle from vertical velocity. For heave velocities that remain small relative to the freestream velocity, the small-angle approximation yields 
 
 \begin{equation} 
 \alpha(t) \approx -\theta(t) - \frac{\dot{h}(t)}{U_\infty}. \end{equation} 
 
 This approximation simplifies the scaling while preserving the dominant physical contributions. 
 
 \subsubsection{Added-mass contribution} 
 
 The added-mass force originates from fluid inertia associated with translational and rotational accelerations of the foil. Following previous studies \cite{sedov1965two, Floryan2017, van2019scaling}, the lateral component is scaled as 
 
 \begin{equation} 
 F_{y,AM} \sim \rho s c^{2} \left( \underbrace{c\ddot{\theta} + \ddot{h}}_{\text{(1) acceleration}} + \underbrace{\dot{h}\dot{\theta}\theta + \dot{\theta}(1+\theta^2)U_\infty}_{\text{(2) Coriolis}} + \underbrace{c\dot{\theta}^{2}\theta}_{\text{(3) centrifugal}} \right). 
 \end{equation}
 
 Group (1) represents translational and rotational accelerations. Group (2) captures Coriolis-type coupling between translation and rotation, and group (3) represents centrifugal effects associated with finite pitch amplitude. For small pitch amplitudes, $\theta = O(\theta_0) \ll 1$, we adopt the approximation $1 + \theta^2 \approx 1$, simplifying the Coriolis contribution without altering its leading-order behavior. Combining the lift-based and added-mass contributions yields the full scaling for the lateral force, 
 
 \begin{equation} 
 F_y \sim \rho s c \left( \alpha U_\infty U_{\mathrm{eff}} + c\dot{\alpha}U_\infty \right) + \rho s c^{2} \left( c\ddot{\theta} + \ddot{h} + \dot{h}\dot{\theta}\theta + \dot{\theta}U_\infty + c\dot{\theta}^{2}\theta \right). \label{eq:Fy_full_vanburen} 
 \end{equation} 
 
 Considering the baseline case of temporally symmetric sinusoidal heave and pitch. This represents the reference condition from which all asymmetric cases will be evaluated. Next, the prescribed kinematics as introduced previously are 
 
 \begin{equation}
 h(t) = h_0 \sin(\omega t), \qquad \theta(t) = \theta_0 \sin(\omega t + \phi),
 \end{equation} 
 where $h_0$ and $\theta_0$ denote the heave and pitch amplitudes, $\phi$ is the phase offset between heave and pitch, and $\omega = 2\pi f$ is the oscillation frequency. The corresponding time derivatives are 
 
 \begin{equation} 
 \dot{h}(t) = h_0 \omega \cos(\omega t), \qquad \ddot{h}(t) = -h_0 \omega^2 \sin(\omega t), 
 \end{equation}
 
 \begin{equation} \dot{\theta}(t) = \theta_0 \omega \cos(\omega t + \phi), \qquad \ddot{\theta}(t) = -\theta_0 \omega^2 \sin(\omega t + \phi)
 \end{equation} 
 
 \noindent Using the small-angle approximation for the instantaneous angle of attack, 
 
 \begin{equation} 
 \alpha(t) \approx -\theta(t) - \frac{\dot{h}(t)}{U_\infty}, \end{equation} 
 
 \noindent we obtain 
 
 \begin{equation}
 \alpha(t) \approx -\theta_0 \sin(\omega t + \phi) - \frac{h_0 \omega}{U_\infty}\cos(\omega t)
 \end{equation}
 
\noindent All kinematic quantities remain purely sinusoidal and therefore symmetric about zero over a full oscillation cycle. Under these symmetric kinematics, all first-derivative terms are sinusoidal with zero mean over one period, 
 
 \begin{equation} 
 \langle \dot{h} \rangle = \langle \dot{\theta} \rangle = \langle \alpha \rangle = 0 
 \end{equation} 
 
\noindent and all acceleration terms are likewise sinusoidal with zero mean. As a result, the cycle-averaged lateral force vanishes. Any non-zero mean lateral force must therefore arise from an intentional breaking of temporal symmetry. This expression provides the foundation for examining how intentional asymmetry modifies the cycle-averaged lateral force.

From this point, we modify the lateral force formulation for our three asymmetric cases: 1) pitch bias, 2) asymmetric heave and pitch speed, and 3) asymmetric stiffness variation. Each mechanism alters specific terms within equation \ref{eq:Fy_full_vanburen}, allowing us to isolate how symmetry breaking produces a non-zero mean lateral force. 
 
\subsection{Static pitch bias}

We first introduce asymmetry through a constant pitch bias $\theta_{pb}$. The baseline kinematics are modified to 
 
\begin{equation}
h(t)=h_0\sin(2\pi f t), \qquad \theta(t)=\theta_0\sin(2\pi f t+\phi)+\theta_{pb}.
\end{equation}
 
\noindent Using the small angle approximation, 
 
 \begin{equation} 
 \alpha(t)\approx -\theta(t)-\frac{\dot h(t)}{U_\infty}, \end{equation} 
 
\noindent the angle of attack may be decomposed into a constant bias component and a zero-mean oscillatory component,
 \begin{equation} 
 \alpha(t)=\alpha_{pb}+\alpha_{\mathrm{osc}}(t), \qquad \alpha_{pb} \equiv -\theta_{pb}, \qquad \langle \alpha_{\mathrm{osc}}\rangle=0. 
 \end{equation} 
 
\noindent The presence of $\alpha_{pb}$ introduces a non-zero mean angle of attack over the oscillation cycle. Beginning from equation \ref{eq:Fy_full_vanburen} and taking a cycle average, all total time derivative terms vanish due to periodicity. The remaining mean force scaling reduces to 
 
\begin{equation} \frac{\overline{F_y}}{\rho s c} \sim \langle \alpha\,U_\infty U_{\mathrm{eff}} \rangle + c\,\langle \dot h\,\dot\theta\,\theta \rangle + c^2\,\langle \dot\theta^2\,\theta \rangle 
\label{eq:s1_mean_reduced} 
\end{equation} 
 
\noindent where $U_{\mathrm{eff}}=\sqrt{U_\infty^2+\dot h^2}$. For the biased motion, the dominant mean contribution arises from the steady lift term involving the bias multiplied by the mean effective velocity, 
 
 \begin{equation} 
 \langle \alpha\,U_\infty U_{\mathrm{eff}} \rangle \approx U_\infty\,\alpha_{pb}\,\overline{U}_{\mathrm{eff}} = -U_\infty\,\theta_{pb}\,\overline{U}_{\mathrm{eff}}. \end{equation} 
 
\noindent The remaining correlation terms retain contributions proportional to $\theta_{pb}$, yielding the compact scaling
 
 \begin{equation} 
 \frac{\overline{F_y}}{\rho s c} \sim -U_\infty\,\theta_{pb}\,\overline{U}_{\mathrm{eff}} +\frac12\,c\,(2\pi f)^2 h_0\theta_0\,\theta_{pb}\cos\phi +\frac12\,c^2\,(2\pi f)^2\theta_0^2\,\theta_{pb}. 
 \label{eq:s1_mean_force_final}
 \end{equation} 
 
\noindent Thus, the mean lateral force scales linearly with the imposed pitch bias. In quasisteady aerodynamics, the linear relationship between the lift coefficient and angle of attack is a major finding in Thin Airfoil Theory. In order to non-dimensionalize, we introduced Strouhal numbers based on heave and pitch, and defined the mean lift coefficient as 
 \begin{equation}
 \overline{C_L}\equiv \frac{\overline{F_y}}{\tfrac12\rho U_\infty^2 s c}, \qquad U^*\equiv \frac{\overline{U}_{\mathrm{eff}}}{U_\infty}, 
 \end{equation} 
 
 \begin{equation} 
 St_h=\frac{2fh_0}{U_\infty}, \qquad St_\theta=\frac{2fc\theta_0}{U_\infty}. 
 \end{equation}
 
\noindent Substituting equation \ref{eq:s1_mean_force_final} into the non-dimensional form yields 
 
 \begin{equation} 
 \overline{C_L} \sim \theta_{pb}\left[ -2U^* + \pi^2\left(St_hSt_\theta\cos\phi + St_\theta^2\right) \right], \label{eq:s1_CL_final} 
 \end{equation} 
 
 \noindent demonstrating theoretically that the cycle-averaged lateral force should scale linearly with the static pitch bias,
 
 \begin{equation}
 \overline{C_L}\propto \theta_{pb}. 
 \end{equation} 

\begin{figure}[h!]
\centering
\includegraphics[width=1.0\textwidth]{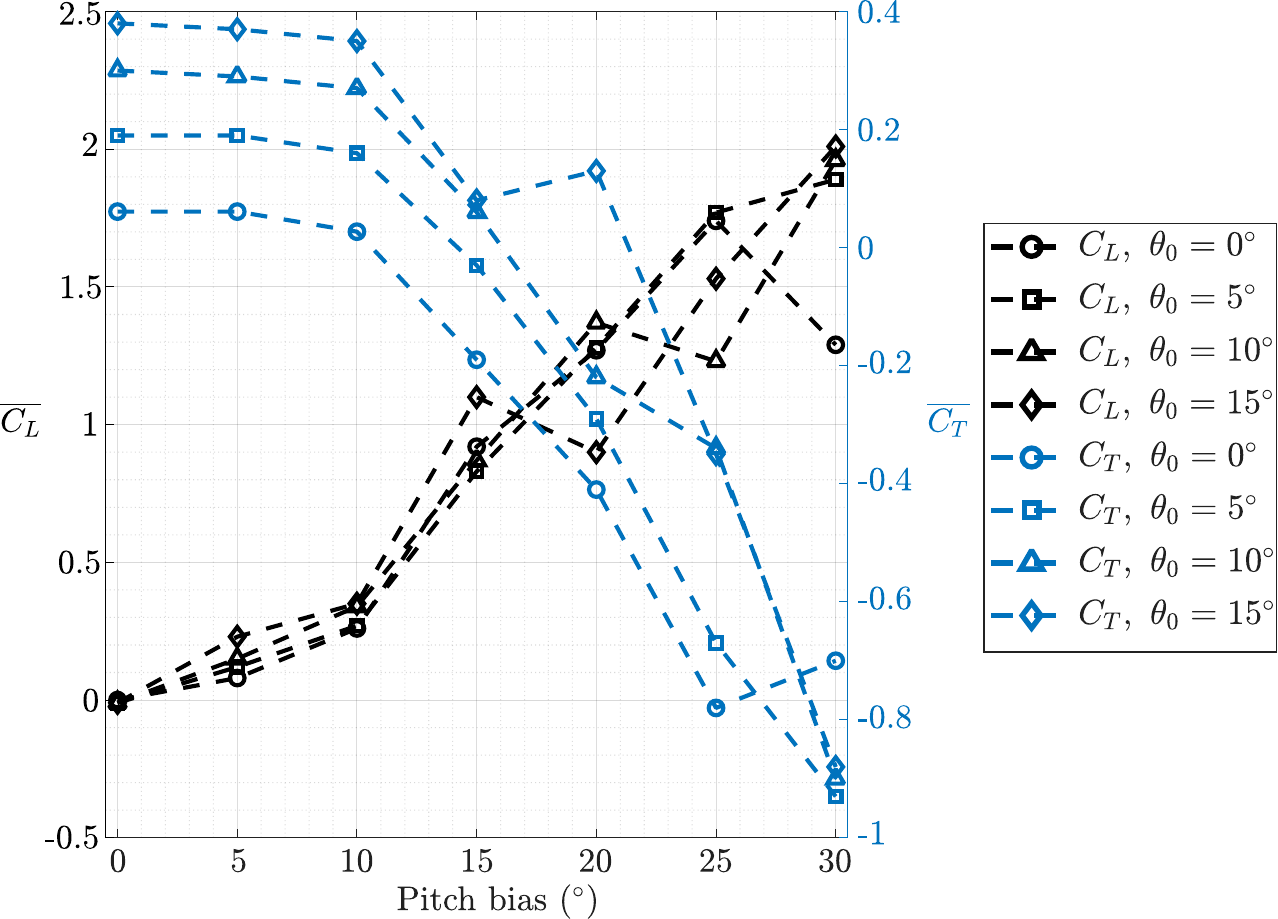}
\caption{Mean side-force coefficient $\overline{C_L}$ (black) and mean thrust coefficient $\overline{C_T}$ (blue) as functions of imposed pitch bias for multiple pitch amplitudes $\theta_0 = 0^\circ, 5^\circ, 10^\circ,$ and $15^\circ$. Increasing pitch bias produces a linear rise in $\overline{C_L}$ across all $\theta_0$ cases, consistent with the predicted linear scaling with $\theta_{pb}$. Simultaneously, $C_T$ decreases as bias increases}
\label{fig:ch3_pitcbias}
\end{figure}

\noindent We compare the scaling derived above with the results in figure \ref{fig:ch3_pitcbias}. We ran four separate cases of our oscillating fin system varying $\theta_0$ from $0^\circ \sim  15^\circ$ every $5^\circ$, with $\theta_0 = 0^\circ$ representing a heaving only fin case. 

Figure \ref{fig:ch3_pitcbias} illustrates how introducing a constant pitch bias to an oscillating fin undergoing combined heave and pitch modifies the cycle averaged lateral force. Across all tested pitch amplitudes, the mean side-force coefficient increases approximately linearly as the bias angle grows. The slope of this increase remains consistent for each $\theta_0$ case, indicating that the oscillatory pitch amplitude and the imposed static bias act largely independently in setting the mean lateral force. Additionally we plot the each system's $\overline{C_T}$ to show as lateral force production increase forward thrust production is perpetually diminished. With pitch bias greater than $10^\circ$ we see an increase penalty in forward thrust as lateral force generation becomes larger.

This behavior directly matches the scaling derived earlier in this chapter, which predicted that the average lateral force across the cycle should scale linearly with the imposed pitch bias, $\overline{C_L} \propto \theta_{pb}$. The numerical results therefore validate the theoretical decomposition, confirming that the steady lift term associated with the biased angle of attack dominates the mean lateral force production.

While the nearly linear trend resembles the classical lift response predicted by Thin Airfoil Theory for steady, inviscid flow, the proportionality observed here differs significantly from the traditional $C_L \sim 2\pi \alpha$ scaling. This difference is expected, as the present case involves unsteady oscillatory motion and operates at a moderate Reynolds number where viscous effects, phase interactions, and added-mass contributions meaningfully alter the force response.

\subsection{Asymmetric heave and pitch} 
\label{sec: asymmetric heave}
 
Next we looked at altering our heave and pitch motion by form of a tilted sine wave introduced in section \ref{sec: asymmetric heave} of this chapter. Beginning with our baseline lateral force equation,
 
 \begin{equation}
 F_y(t)=F_{y,L}(t)+F_{y,AM}(t). 
 \end{equation}
 
 \noindent Using the same scaling framework introduced previously, the lift and added mass contributions were written as 
 
 \begin{equation} \frac{F_{y,L}}{\rho s c} \sim \alpha\,U_\infty\,U_{\mathrm{eff}} + c\,U_\infty\,\dot{\alpha}, \end{equation}
 
 \begin{equation} 
 \frac{F_{y,AM}}{\rho s c} \sim c\left( c\,\ddot{\theta} +\ddot{h} +\dot{h}\dot{\theta}\theta +U_\infty\dot{\theta} +c\dot{\theta}^2\theta \right). 
 \end{equation} 
 
\noindent With effective terms of velocity and angle of attack defined as 
 
 \begin{equation}
 U_{\mathrm{eff}}(t)=\sqrt{U_\infty^2+\dot{h}(t)^2}, \qquad \alpha(t)\approx-\theta(t)-\frac{\dot{h}(t)}{U_\infty}. \end{equation} 
 
Temporal asymmetry enters through differences in heave speed, heave acceleration, and pitch rate between half cycles. Rather than evaluating the full time history directly, we instead compare the ratio of predicted forces based on these asymmetries. To fully capture the asymmetric heave and pitch motions, it would be necessary to incorporate the tilted Fourier representations introduced in equation \ref{fourier}. While this could be evaluated numerically, it introduces additional complexity beyond what is required for the present analysis. Instead, a piecewise approach is adopted, where the performance of sinusoidal motions corresponding to the distinct heave and pitch velocity regimes is evaluated separately and their resulting force ratios are compared. This approach provides a simplified yet effective estimate of how asymmetry influences the overall force generation and correlates with the trends observed in the full simulations. To simplify this into a ratio of heave speed, acceleration, and pitch motions we take our full oscillation cycle and split it into two half cycles. We define the upstroke and downstroke intervals, $\mathcal{I}_u$ and $\mathcal{I}_d$, as the two half cycles separated by maximum and minimum heave displacements. The half cycle mean forces are 
 
\begin{equation} \langle F_y\rangle_u = \frac{1}{\Delta t_u}\int_{\mathcal{I}_u} F_y(t)\,dt, \qquad \langle F_y\rangle_d = \frac{1}{\Delta t_d}\int_{\mathcal{I}_d} F_y(t)\,dt. 
 \end{equation} 
Defining the half cycle force ratios as 
\begin{equation} \frac{\langle F_y\rangle_u}{\langle F_y\rangle_d} = \frac{\langle F_{y,L}\rangle_u+\langle F_{y,AM}\rangle_u} {\langle F_{y,L}\rangle_d+\langle F_{y,AM}\rangle_d}. 
\end{equation}
 
\noindent For symmetric sinusoidal kinematics, the half cycles are symmetric in time duration and waveform shape. Consequently, 
 
 \begin{equation} 
 \langle F_y\rangle_u = -\langle F_y\rangle_d, 
 \end{equation} 
 
\noindent and the magnitude of there forces are the same, resulting in a cycle averaged lateral force that cancels one another out. When temporal symmetry is broken, the two half-strokes no longer generate equal and opposite forces. The asymmetry enters primarily through differences in peak heave speed and peak heave acceleration. To introduce temporal asymmetry while keeping the heave amplitude bounded to remain at $\pm h_0$, we replace the pure sinusoid with the normalized tilted Fourier waveform introduced earlier, 
 
 \begin{equation} 
 h(t)=h_0\,\hat{s}(2\pi f t;\lambda) 
 \end{equation} 
 
\noindent where $\lambda$ controls the degree of tilt and $|\hat{s}|\le 1$. This construction preserves the displacement bounds $|h|\le h_0$ while redistributing slope within the cycle, producing a fast half-stroke and a slow half-stroke. The corresponding derivatives are 
 \begin{equation} 
 \dot{h}(t)=h_0(2\pi f)\hat{s}'(2\pi f t;\lambda), \qquad \ddot{h}(t)=h_0(2\pi f)^2\hat{s}''(2\pi f t;\lambda), \end{equation}
 
\noindent so all tilt dependence enters through changes in peak speed and peak acceleration. Therefore, the magnitude of the lift and added mass contributions to force ar scaled as 
 
 \begin{equation} |F_{y,L}|\sim \rho s c \, U_{\mathrm{eff}}\,|\dot{h}|, \qquad |F_{y,AM}|\sim \rho s c^2|\ddot{h}|. 
 \end{equation}
 
\noindent Averaging over each half-stroke, the dominant dependence reduces to peak values of speed and acceleration, \begin{equation} \frac{|\langle F_y\rangle_u|}{|\langle F_y\rangle_d|} \;\approx\; \frac{ V_u + c\,A_u }{ V_d + c\,A_d }, \end{equation} 
 where 
 
 \begin{equation}
 V_{u,d}=\max_{\mathcal{I}_{u,d}}|\dot{h}|, \qquad A_{u,d}=\max_{\mathcal{I}_{u,d}}|\ddot{h}|. 
 \end{equation} 
 
This expression shows explicitly that temporal asymmetry enters through two independent kinematic ratios: speed and acceleration. Regardless of how the waveform is generated, the resulting lateral force asymmetry collapses when expressed in terms of these peak kinematic measures.  

\begin{figure}[h!]
\centering
\includegraphics[width=0.8\textwidth]{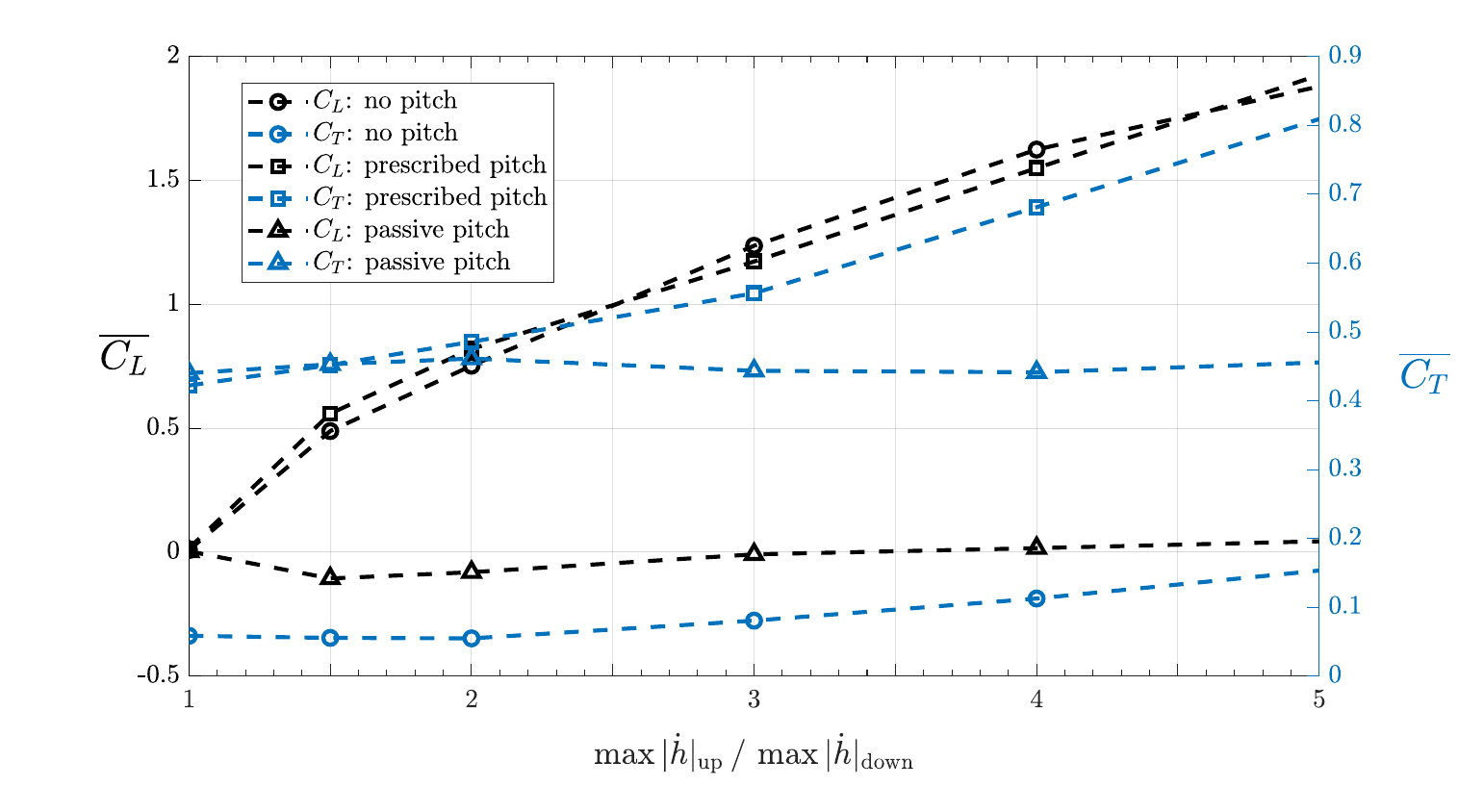}
\caption{Cycle averaged lateral force $\overline{C_L}$ plotted against tilt factor $\lambda$ for pure heave ($\theta_0 = 0^\circ$), a standard pitching and heaving fin ($\theta_0 = 30^\circ$), and  a leading edge spring fin ($\theta_0 = passive$).}
\label{fig:ch3_heavetilt}
\end{figure}

To compare our scaling analysis of the affects of the asymmetric heaving and pitching of our fin system we applied the Fourier series to tilt the waveform, presented earlier to our fin simulation model. We vary the tilt parameter $\lambda$ to result in a max to min speed ratio of $1.5, 2, 3, 4, 5$ across the two half cycles and plot the resulting $C_L$ against $\lambda$ in figure \ref{fig:ch3_heavetilt}.

Figure \ref{fig:ch3_heavetilt} shows how introducing asymmetry into the heave motion, quantified through the velocity ratio $\lambda$ between the fast and slow half strokes, generates a non-zero mean lift coefficient. We considered three configurations: 1) heave only with $\theta_0 = 0^\circ$, 2) prescribed heave and pitch with $\theta_0 = 30^\circ$, and 3) heave with a leading-edge spring allowing passive pitch. For the prescribed pitching case, the pitch waveform was also made asymmetric so that it followed the tilted heave motion.

For the rigid configurations, the trend is clear. As the heave becomes more asymmetric, the mean lift coefficient increases. The strongest response occurs in the heave-only case, which produces the largest average lift coefficient of all cases tested here. This matches the scaling developed earlier in this chapter, where we showed that breaking temporal symmetry through velocity and acceleration differences between half strokes leads directly to a non-zero mean lateral force. Increasing $\lambda$ increases the peak heave velocity and acceleration on one half cycle, and the resulting lift scales accordingly.

When prescribed pitching is added, the fin still produces a substantial mean lift force, with its magnitude is surpassing the heave only in the final case. Physically, this makes sense. Adding pitch to heave helps align the fin more favorably with the flow, increasing thrust production efficiency. In doing so, it reduces the peak effective angle of attack that would otherwise occur in pure heave. Since lift scales with angle of attack, this reduction limits the maximum lateral force that can be generated. Additionally, pitching motion contributes to force production through added mass effects tied to acceleration. It is not immediately obvious how a velocity bias modifies the added mass contribution in this coupled scenario, but it is clear that the added pitching motion redistributes how force is generated between lift-based and inertial mechanisms.

\begin{figure}[h!]
\centering
\includegraphics[width=1.0\textwidth]{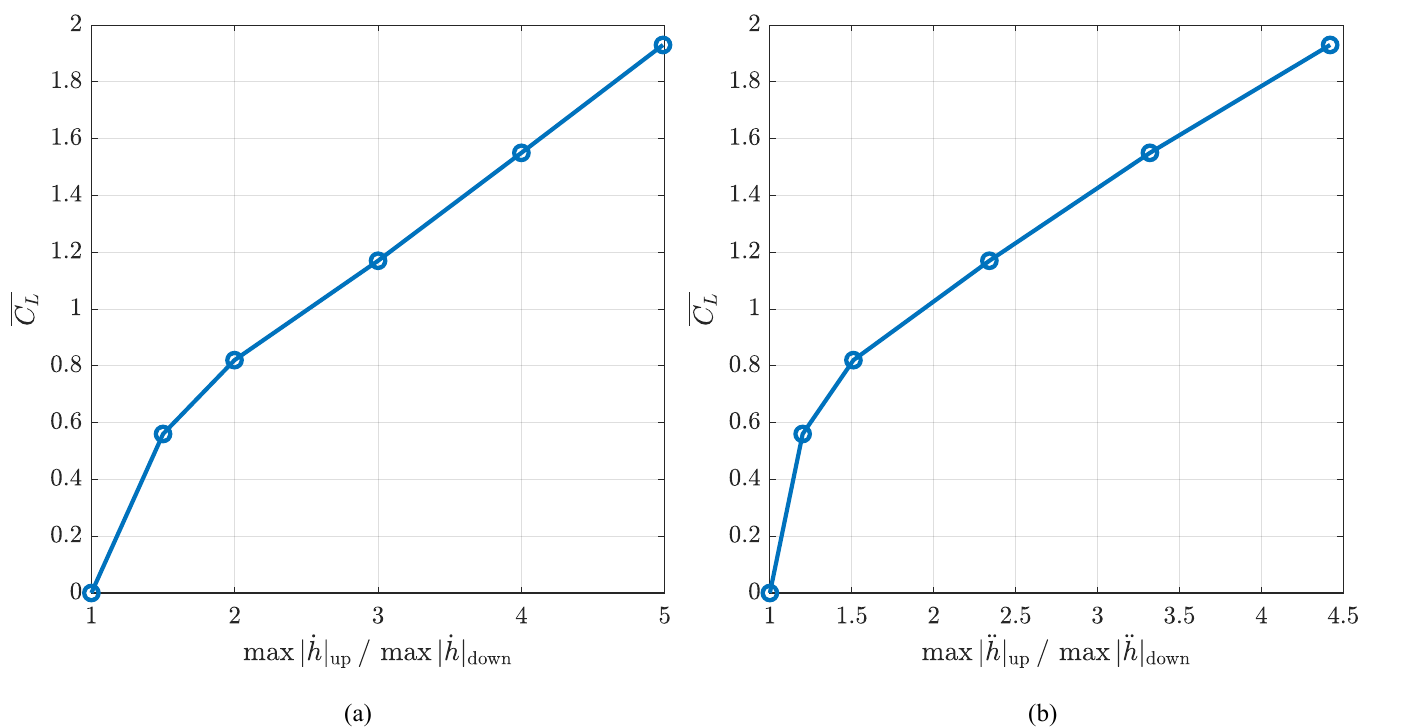}
\caption{Coefficient of lateral force $\overline{C_L}$ as a function of waveform asymmetry generated by the tilted heave motion. (a) $\overline{C_L}$ plotted against the ratio of maximum heave speed during the upstroke and downstroke. (b) $\overline{C_L}$ plotted against the ratio of peak heave acceleration between the two half strokes.}
\label{fig:ch3_scaling_heavetilt}
\end{figure}

To compare the proposed scaling analysis with the results of the simulations, the average coefficient of lateral force, $\overline{C_L}$, is plotted against two measures of asymmetry generated by the tilted heave waveform: the ratio of the maximum heave speeds between the upstroke and downstroke, and the ratio of the peak accelerations over the same portions of the cycle. We show the results in figure \ref{fig:ch3_scaling_heavetilt}. These ratios provide a compact measure of how strongly the motion is biased toward one half of the oscillation cycle.

As the asymmetry of the waveform increases, the generated lateral force also increases. When the motion is symmetric, corresponding to a ratio of one, the cycle produces essentially zero mean lateral force because the forces generated during each half stroke cancel over a period. As the waveform becomes increasingly skewed, the faster half stroke produces larger instantaneous forces due to the higher effective velocity and acceleration of the fin relative to the surrounding fluid. The opposing half stroke occurs more slowly and therefore generates weaker forces that do not fully cancel those produced during the faster portion of the motion. The result is a non-zero cycle averaged lateral force.

The nearly linear trends observed in both relationships indicate that the generated lateral force scales directly with the magnitude of the kinematic imbalance. The agreement between the velocity ratio and acceleration ratio trends also supports the physical interpretation that the lateral force arises from both lift based forces associated with relative velocity and added mass forces associated with fluid acceleration.

The flexible leading-edge spring case behaves differently. Across most values of $\lambda$, the mean lift remains near zero, with only the $\lambda = 3$ case showing a meaningful deviation. In this configuration, the pitch response is not prescribed but instead determined by the instantaneous hydrodynamic torque. When the heave velocity increases on one half stroke, the resulting increase in lateral force produces a larger passive pitch deflection. That rotation reduces the effective angle of attack, which in turn limits additional lift growth. Because the angle of attack is coupled directly to the spring stiffness and hydrodynamic loading, there appears to be an intrinsic feedback mechanism that suppresses the net mean lift. This coupling likely explains why asymmetric heave does not translate into sustained lateral force in the flexible case.
 
\subsection{Leading-edge spring stiffness asymmetry}  
 
 For the leading edge spring configuration, the heave motion is prescribed and the pitch response of the fin is governed by a torsional spring at the leading edge. Using the quarter-chord as an approximation for location of the hydrodynamic moment about the leading edge, 
 
 \begin{equation} 
 M_z(t)\approx \frac{c}{4}\,F_y(t), 
 \end{equation} 
 \noindent the spring torque balance becomes 
 
 \begin{equation} k(t)\,\theta(t)\approx -\,\frac{c}{4}\,F_y(t). \label{eq:LEspring_balance_final}
 \end{equation} 
 
\noindent Rearranging gives a direct relationship between pitch and lateral force, 
 
 \begin{equation}
 \theta(t)\approx -\,\frac{c}{4k(t)}\,F_y(t). \label{eq:theta_from_k_final} 
 \end{equation} 
 
\noindent For small angles, the instantaneous angle of attack is approximated as 
 
 \begin{equation} \alpha(t)\approx -\theta(t)-\frac{\dot{h}(t)}{U_\infty}. 
 \label{eq:alpha_smallangle_final} 
 \end{equation} 
 
\noindent Using the same lift based scaling introduced earlier in this chapter, the lateral force scales as 
 \begin{equation} 
 F_y(t)\sim \rho s c \,\alpha(t)\,U_\infty\,U_{\mathrm{eff}}(t), \qquad U_{\mathrm{eff}}(t)=\sqrt{U_\infty^2+\dot{h}(t)^2}. \label{eq:Fy_lift_scaling_final} 
 \end{equation} 
 
\noindent Substituting equation \ref{eq:theta_from_k_final} into equation \ref{eq:alpha_smallangle_final} and then into equation \ref{eq:Fy_lift_scaling_final} yields an approximation for the the instantaneous lateral force, 
 
 \begin{equation} F_y(t) \sim -\frac{\rho s c\,\dot{h}(t)\,U_{\mathrm{eff}}(t)} {1+\dfrac{\rho s c^2}{4k(t)}\,U_\infty\,U_{\mathrm{eff}}(t)}. \label{eq:Fy_k_closed_final} 
 \end{equation} 
 
\noindent Equation \ref{eq:Fy_k_closed_final} is an example of how stiffness controls lateral force generation. As stiffness increases, the fin rotates less into the flow, the effective angle of attack increases, and the lateral force consequently grows. In the limit of very large stiffness, the motion approaches pure heave and the lateral force approaches a finite bound. For low stiffness, the fin rotates into the flow, reducing effective angle of attack, subsequently decreasing lateral force.
 
\subsubsection{Half cycle stiffness asymmetry} To examine net lateral-force generation, we split each oscillation cycle into two half cycles like we did when looking at asymmetric heave and pitch speed in section \ref{sec: asymmetric heave}. Using the same method here, we take $\mathcal{I}_u$ and $\mathcal{I}_d$, separated by consecutive maximum and minimum heave displacements. Over each half cycle the stiffness is approximated as piecewise constant, 
 
 \begin{equation}
 k(t)\approx k_u \quad \text{for } t\in\mathcal{I}_u, \qquad k(t)\approx k_d \quad \text{for } t\in\mathcal{I}_d. \end{equation} 
 
\noindent Since equation \ref{eq:Fy_k_closed_final} depends on $k(t)$ in the denominator, different stiffness values on the two half cycles produce different lateral force magnitudes. Collapsing the cycle average by the stiffness gives the net scaling 
 
 \begin{equation} 
 \overline{F_y} \;\propto\; \left( \frac{k_u}{k_u+B} - \frac{k_d}{k_d+B} \right), 
 \label{eq:Fy_mean_k_scaling_final} 
 \end{equation} 
 
\noindent where 
 
 \begin{equation} 
 B \equiv \frac{\rho s c^2}{4}\,U_\infty\,\overline{U}_{\mathrm{eff}}, \qquad \overline{U}_{\mathrm{eff}}=\frac{1}{T}\int_0^T U_{\mathrm{eff}}(t)\,dt, 
 \end{equation} 
 
\noindent is a hydrodynamic torque scale that is fixed for a given operating condition. For the stiffness amplitude used in this study, 
 
 \begin{equation} 
 k_u = k_0(1+p), \qquad k_d = k_0(1-p), 
 \end{equation} 
 so the mean lateral force scaling becomes 
 
 \begin{equation} 
 \overline{F_y} \;\propto\; \left( \frac{k_0(1+p)}{k_0(1+p)+B} - \frac{k_0(1-p)}{k_0(1-p)+B} \right). \label{eq:Fy_mean_p_scaling_final}
 \end{equation} 
 
\noindent When the imposed stiffness is small relative to the hydrodynamic scale, $k_0\ll B$, the expression simplifies to
 
 \begin{equation}
 \overline{F_y} \;\propto\; \frac{k_u-k_d}{B} = \frac{2k_0}{B}\,p, 
 \label{eq:Fy_linear_regime_final}
 \end{equation} 
 
\noindent predicting a linear increase in net lateral force production with stiffness asymmetry. This scaling captures the central mechanism of this configuration. A stiffer half cycle produces a larger effective angle of attack and greater lateral force. A more flexible half cycle rotates into the flow, reducing lateral force production. The difference between these two responses produces a non-zero cycle averaged lateral force that increases linearly with stiffness asymmetry. 

\begin{figure}[h!]
\centering
\includegraphics[width=1.0\textwidth]{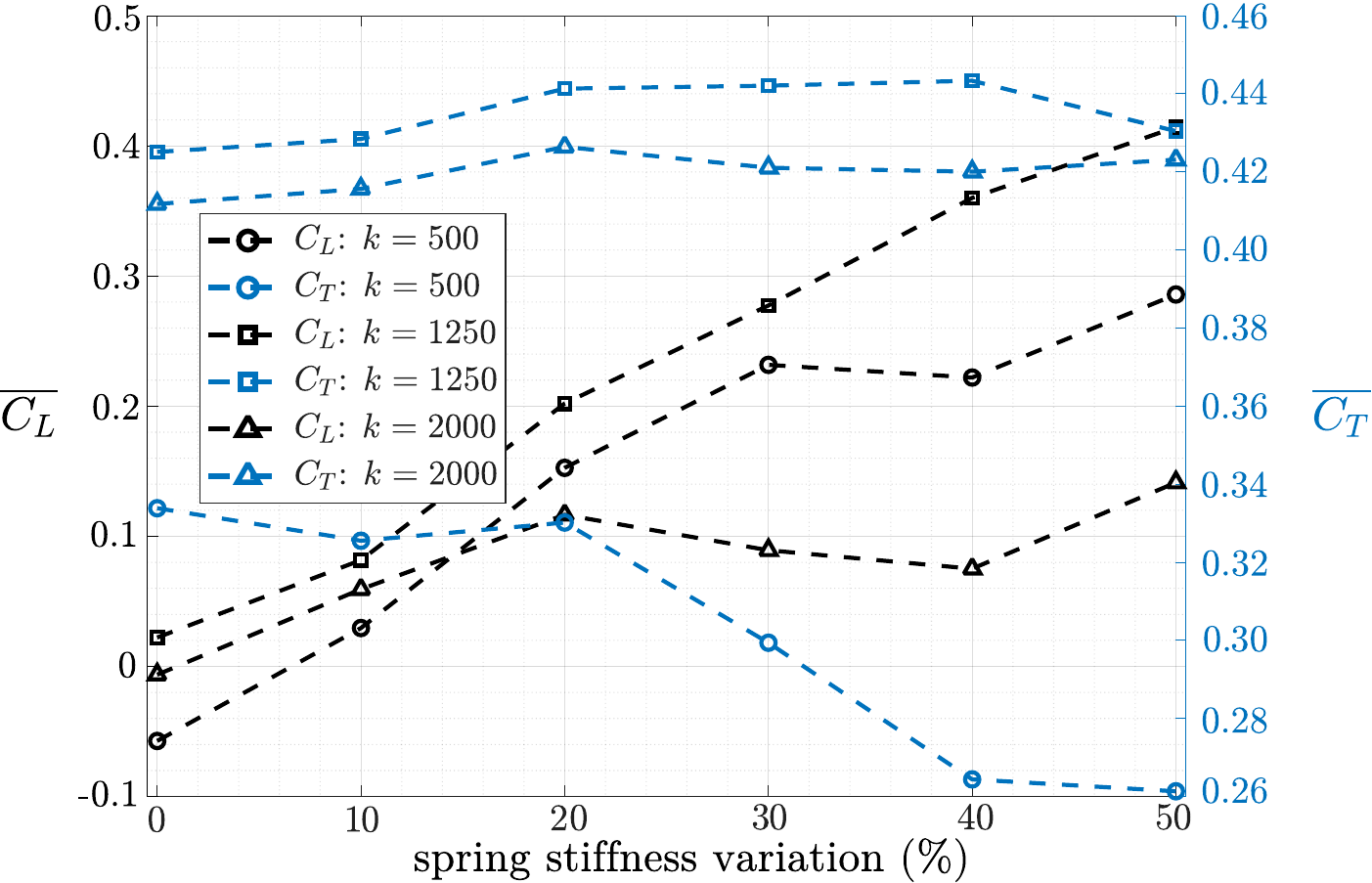}
\caption{Cycle average coefficient of lateral force $\overline{C_L}$, and streamwise coefficient of thrust $\overline{C_T}$ plotted against percent of stiffness variation amplitude $p$ for three varying stiffness cases, $k = 500, 1250, 2000$.}
\label{fig:ch3_stiffness_variation}
\end{figure}

Running our final parameter sweep where we varied $p = 0 \sim 0.5$ every $0.1$ ($0.1 \approx 10\%$ in spring stiffness variation), across three different $k$ values, $500, 1250, 2000$, representing a light, medium, and heavy stiffness case. We plot the resulting cycle averaged $C_L$ and $C_T$ for each case against $p$ in figure \ref{fig:ch3_stiffness_variation}.

Figure \ref{fig:ch3_stiffness_variation} shows how stiffness asymmetry, varied by a percent, $p$, of baseline stiffness $k$, alters the cycle averaged lateral force for three baseline stiffness values, $k = 500$, $1250$, and $2000$. In the light and medium cases, increasing $p$ produces a linear increase in the mean lateral force coefficient. The heavy case stops and plateaus after $p = 0.2$. This trend is consistent with the scaling derived earlier, where we showed that a difference in stiffness between the two half cycles produces a net lateral force proportional to the stiffness imbalance. As $p$ increases, one half stroke becomes effectively stiffer while the other becomes more compliant, breaking temporal symmetry in the pitch response and generating a non-zero mean lateral force.

The magnitude of the response depends strongly on the baseline stiffness. The intermediate stiffness case, $k = 1250$, produces the largest lift coefficient across the tested range. The lowest stiffness case, $k = 500$, shows a weaker response at small $p$, but grows steadily as modulation increases. The highest stiffness case, $k = 2000$, generates smaller overall lift coefficients compared to the intermediate case, indicating that there exists an optimal stiffness range where asymmetric modulation most effectively translates into lateral force. This behavior follows directly from the closed form scaling presented earlier, where the mean force depends on the ratio of stiffness to the hydrodynamic torque scale. When the system is too compliant, the fin rotates excessively and reduces effective angle of attack. When it is too stiff, the modulation has diminished influence because the motion approaches rigid behavior. 

The thrust coefficient, $\overline{C_T}$, shown on the secondary axis, exhibits a more subtle response to stiffness modulation. For the lowest stiffness case, thrust decreases as $p$ increases, reflecting the growing asymmetry and increased lateral loading. In contrast, the medium and high stiffness cases show relatively small variation in $\overline{C_T}$ across the modulation range. This indicates that stiffness asymmetry can be used to generate meaningful lateral forces without severely compromising forward thrust, particularly in the moderate to high stiffness regimes. To compare the results of our simulation to our theoretical prediction we plot the medium case ($k = 1250$) against the lateral scaling parameter presented in equation \ref{eq:Fy_linear_regime_final}. 

\begin{figure}[h!]
\centering
\includegraphics[width=1.0\textwidth]{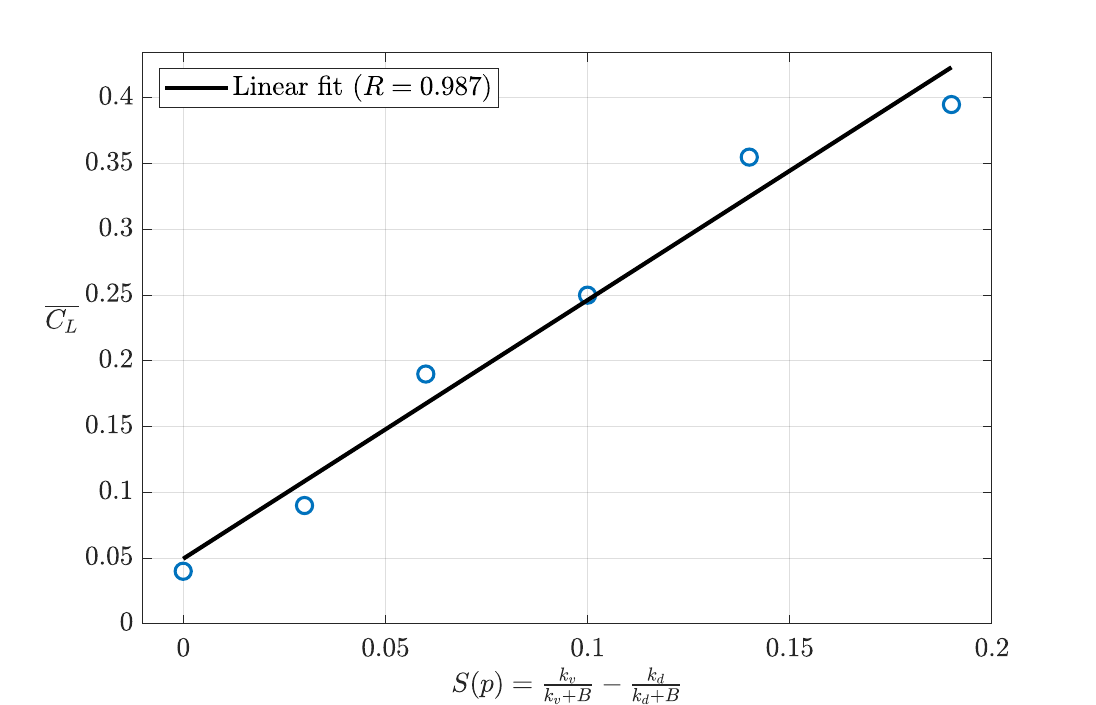}
\caption{Collapse of the medium stiffness case ($k = 1250$) against the stiffness-asymmetry scaling variable $S(p) = \frac{k_u}{k_u + B} - \frac{k_d}{k_d + B}$. The mean lift coefficient $\overline{C_L}$ from simulation is plotted against the theoretical scaling parameter, with a linear fit yielding a correlation coefficient of $R = 0.987$, indicating a strong linear relationship.}
\label{fig:ch3_stiffness_scaling}
\end{figure}

Figure \ref{fig:ch3_stiffness_scaling} plots the medium stiffness case, $k = 1250$, directly against the stiffness-asymmetry scaling variable  
\[
S(p) = \frac{k_u}{k_u + B} - \frac{k_d}{k_d + B}.
\]
We chose to plot the $k = 1250$ case as it had the most consistent trend in response to change in $p$. Rather than viewing the mean lift coefficient as a function of $p$ alone, this comparison tests the simulation results against the structure of the derived scaling law that scales the net lateral force production as linear relative to the increase of stiffness $k$. The trend shown in the plot is approximately linear. The simulation data follow the predicted relationship closely, confirming that the mean lateral force scales with the stiffness imbalance relative to the hydrodynamic torque scale. This matches the behavior predicted in equation \ref{eq:Fy_linear_regime_final}.

This figure is important because it directly connects the analytical model to the CFD results. It shows that stiffness modulation produces a predictable and controllable lateral force, and that the dominant physics are captured by the simplified scaling.
 
\section{Conclusion}

This chapter examined how an oscillating fin can be driven away from purely propulsive behavior and used to generate controlled lateral forces for maneuvering. Starting from the same lift-based and added-mass scaling framework used for thrust production, we showed that a temporally symmetric heave and pitch cycle produces zero cycle-averaged lateral force. Any non-zero mean lateral force therefore requires an intentional breaking of kinematic symmetry.

We considered three symmetry-breaking mechanisms. First, a static pitch bias introduces a non-zero mean angle of attack and produces a mean lateral force that scales linearly with the imposed bias, consistent with the reduced cycle averaged scaling derived from the full force decomposition. Second, asymmetry introduced through a tilted heave waveform that generates lateral forces by increase speed in one half cycle. When expressed in terms of peak heave speed and peak heave acceleration ratios between half cycles, the results collapse and appear near linear following the analytical derivation. Third, for a leading-edge spring fin, asymmetric stiffness variation produces a predictable lateral force response. The cycle averaged lateral force increases approximately linearly with the stiffness asymmetry parameter, and the simulation data collapsed well against the derived stiffness scaling variable, demonstrating that the dominant physics are captured by the closed form torque balance and subsequent assumptions.

A key outcome is that the same fin can, in principle, provide both thrust and steering authority through simple modifications to kinematics or stiffness scheduling. Pitch bias and rigid waveform tilt provide strong lateral force authority but can reduce thrust, while stiffness modulation offers a way to generate lateral forces with smaller penalties in forward performance, particularly near an intermediate stiffness regime where modulation is most effective. Taken together, these results provide a compact set of scaling relationships and validated control inputs for producing maneuvering forces in fin-driven vehicles, and they establish a foundation for extending asymmetric actuation to multi-fin configurations.    

\chapter{Two fin systems}
\label{ch4:2_fins}

\section{Introduction}

Transitioning from a single fin system to a two in-line fin system introduces additional complexity but also presents significant opportunities for performance gains. Previous studies have shown that the addition of a second fin can dramatically alter propulsive output. In some cases the thrust produced by a two-fin system can approach twice that of a single fin, while in other configurations thrust can be nearly eliminated entirely \cite{Muscutt2017}. These outcomes depend strongly on the timing and interaction between the two fins.

\begin{figure}[h!]
\centering
\includegraphics[width=0.8\textwidth]{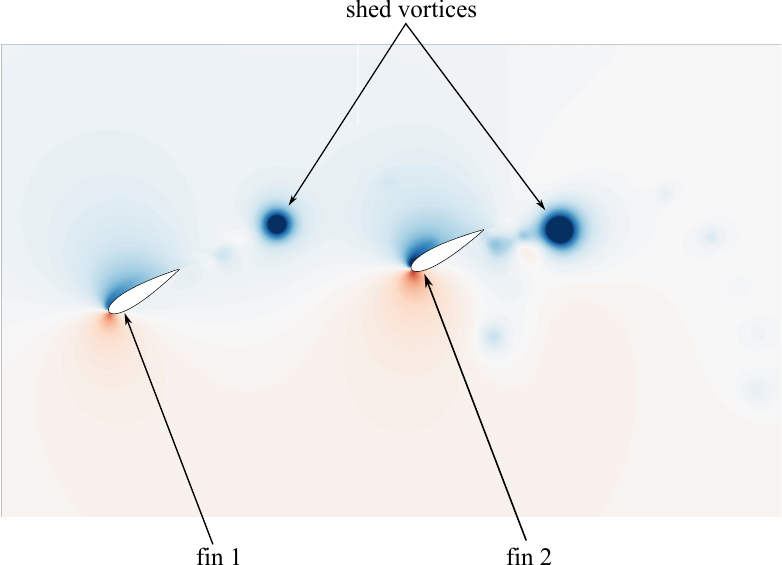}
\caption{Example of a pressure field for a generic oscillating 2-fin system depicting 2 in-line NACA 0020 fins, with both fins shedding vortices into the freestream.}
\label{fig:ch4_2fin_system}
\end{figure}

The performance of a two-fin system is therefore governed not only by the motion of each individual fin, but also by the hydrodynamic environment created by the upstream fin. As the lead fin oscillates, it sheds vortex structures into the wake. These vortices are advected downstream and interact with the second fin as they pass through the flow field. This process is illustrated in the pressure field shown in figure \ref{fig:ch4_2fin_system}, where regions of low pressure indicate the presence of shed vortices that convect downstream with the freestream.

For optimal performance, the downstream fin must encounter these vortices at the correct phase of its motion. If properly timed, the second fin can exploit the vortex generated by the upstream fin, using the associated low-pressure region to enhance the lift forces generated at its leading edge. This interaction can substantially increase the thrust and efficiency of the overall system. In this sense, the timing between fins plays a role analogous to timing a swing in baseball or cricket, where performance depends on striking the incoming ball at precisely the right moment.

Computational studies have demonstrated that the most favorable interactions occur when the downstream fin passes close to, but does not directly intersect, the vortex shed by the upstream fin \cite{Muscutt2017, Guo2023}. Maintaining this proximity allows the downstream fin to benefit from the vortex-induced pressure field while avoiding destructive interference that can degrade performance.

In addition to the pressure field effects, the rotational velocity field associated with the vortex also modifies the effective inflow experienced by the downstream fin. The induced velocities alter the local angle of attack, which can further amplify or diminish thrust production depending on the phase relationship between the fins. This mechanism has been described in detail by \cite{SeoMittal2022}. Evidence of these interactions can be observed in figure \ref{fig:ch4_2fin_system}, where the leading edge of the second fin aligns with and intensifies the vortex shed by the upstream fin.

\begin{figure}[h!]
\centering
\includegraphics[width=1.0\textwidth]{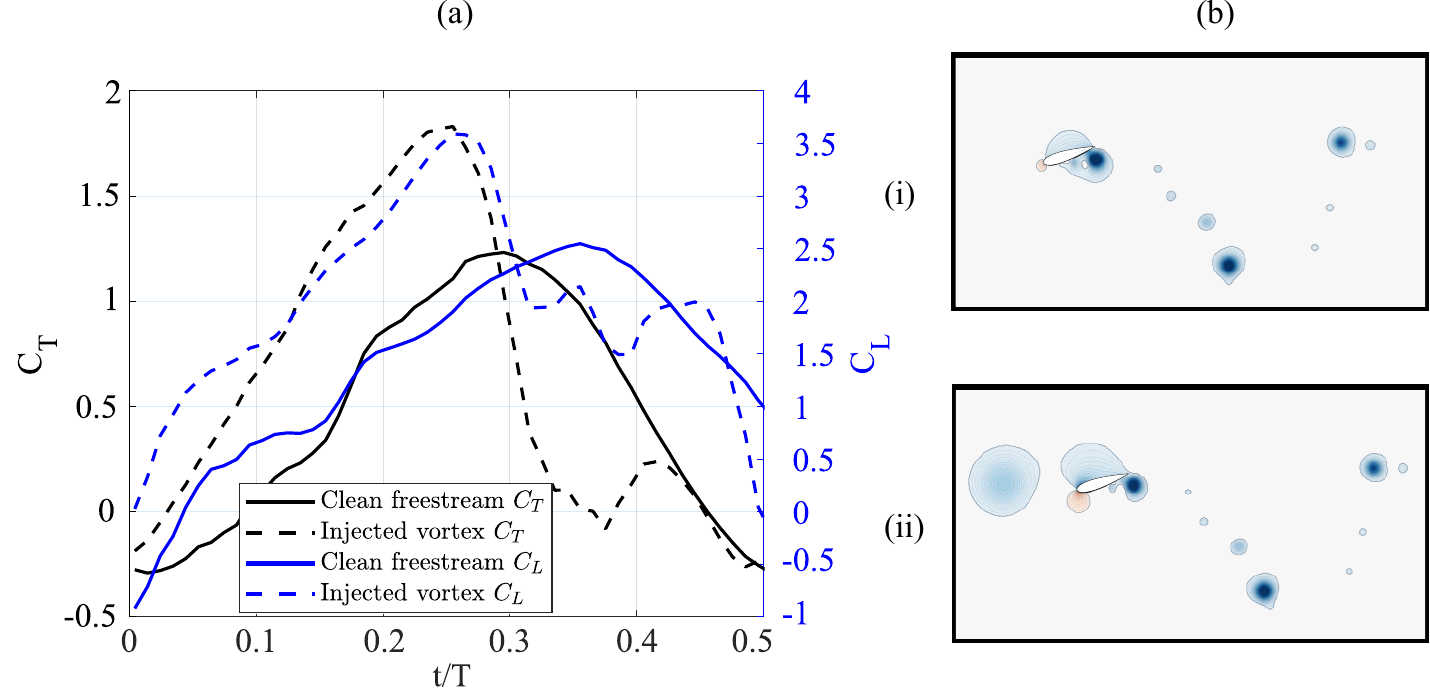}
\caption{Example of two flow fields one without (b.i) and one with (b.ii) a vortex being injected into the freestream. The instantaneous coefficients of thrust and lateral force, $C_T$ and $C_P$, of each case is plotted in (a).}
\label{fig:ch4_injected_vortex}
\end{figure}

To illustrate the effect of a fin encountering a generic vortex, we present two cases in figure \ref{fig:ch4_injected_vortex}. The instantaneous coefficient of thrust $C_T$ of a single fin in a clean freestream with no flow structures is compared with a case in which a vortex is injected into the freestream for the fin to encounter. Flow field images are shown in panels b.i (clean freestream) and b.ii (injected freestream) to the right of the time-varying force plot in (a). The fin encountering the injected structure produces a cycle-averaged thrust coefficient of $\overline{C_T} = 0.51$. In comparison, the fin operating in a clean freestream produces $\overline{C_T} = 0.47$, corresponding to an increase in thrust of approximately $9\%$.

Given the sensitivity of these interactions, the performance of two-fin systems depends strongly on the relative phase offset between the fins and the spacing between them. Small changes in either parameter can significantly alter how the downstream fin encounters the vortex shed by the upstream fin, leading to large differences in thrust production and efficiency. In the following sections we systematically explore this parameter space by varying fin spacing and phase offset to map the resulting performance of two-fin propulsion systems.

\section{Methods and results}
The hydrodynamic interactions of in-line oscillating fins have been studied extensively in the literature, particularly in the work of \cite{Muscutt2017}, which established that downstream performance is highly sensitive to the relative timing between fins. To better understand the timing relationship between shed vortical structures and a downstream fin, we first construct a performance map for a two-fin system. 

\begin{figure}[h!]
\centering
\includegraphics[width=0.8\textwidth]{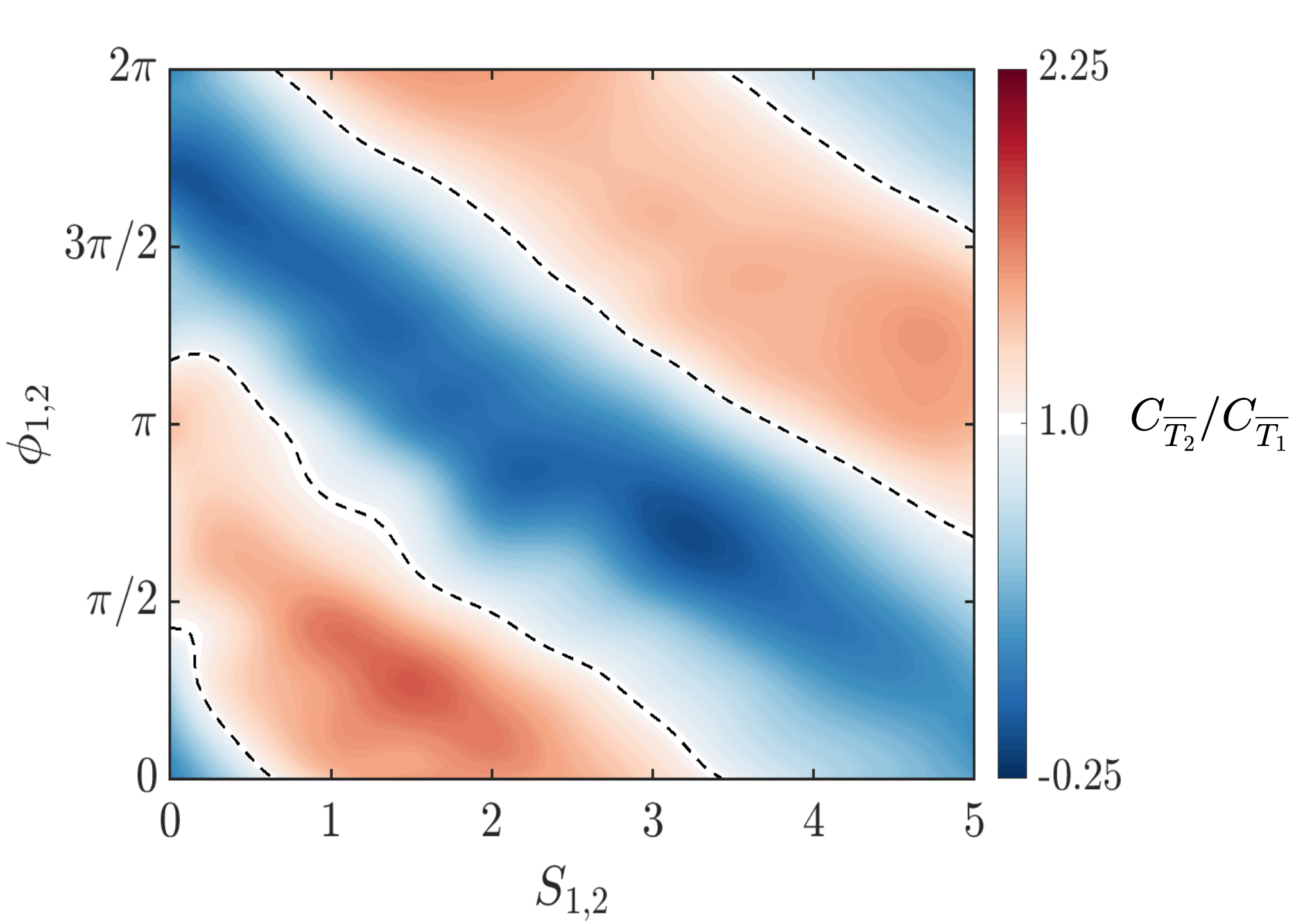}
\caption{Color contour performance map of the thrust of the downstream fin relative to the upstream fin in a 2-fin system as it varies with phase offset $\phi_{1,2}$ and spacing relative to the upstream fin $S_{1,2}$}
\label{fig:ch4_2fin_map}
\end{figure}

Using the coarse search method described in chapter \ref{chap:validation}, we systematically explore the parameter space of a two-fin configuration. Two in-line NACA 0020 fins are simulated, and two primary parameters are varied: the spacing $S_{1,2}$ and phase offset $\phi_{1,2}$ of the downstream fin relative to the upstream fin. We run the parameters across $0 \sim 2\pi$ for $\phi_{1,2}$ every $24^\circ$ and $0 \sim 5c$ for $S_{1,2}$ every $0.5c$, with $c$ denoting chord lengths. The results are presented as a two-dimensional contour map in figure \ref{fig:ch4_2fin_map}.

For each configuration we compute the cycle-averaged thrust produced by both fins and evaluate the performance of the downstream fin relative to the upstream fin. Isolating the effect of vortex interaction on the downstream fin, deeming this the normalized thrust ratio
\begin{equation}
\frac{C_{\overline{T}_2}}{C_{\overline{T}_1}},
\end{equation}
where $C_{\overline{T}_1}$ and $C_{\overline{T}_2}$ represent the cycle-averaged coefficients of thrust generated by fins 1 and 2 respectively. This normalization highlights the relative performance gain or loss experienced by the downstream fin as its phase and spacing change. Regions of high thrust enhancement are shown in red, while regions of degraded thrust performance appear in blue. We zero the color contour around ${C_{\overline{T}_2}}/C_{\overline{T}_1} = 1.0$. The purpose of this is to show us the relative benefit or detriment to the system by the addition the downstream fin relative to the performance of a single fin. The boundary or transition across the value of $1.0$ is denoted by a dotted black line. 

The highest and lowest performing cases were re-run at our fine search settings. The highest performance case was at $S_{1,2} = 1.5c$ and $\phi_{1,2} = 51^\circ$ and has a $C_{T_2}/C_{T_1} = 1.37$. The lowest performance at $S_{1,2} = 3.3c$ and $\phi_{1,2} = 124^\circ$ with a $C_{T_2}/C_{T_1} = 0.01 $. 

The resulting performance map reveals distinct diagonal bands where the thrust of the downstream fin rapidly transitions between high and low values. These bands appear at approximately $45^\circ$ in the $(S,\phi_{1,2})$ parameter space. This pattern indicates that maintaining optimal performance requires coordinated changes in both spacing and phase. In other words, if the spacing between fins increases, the phase offset must also increase proportionally to preserve the timing between the downstream fin and the vortex shed by the upstream fin.

This behavior reflects the convective transport of vortices through the wake. The vortices shed by the upstream fin travels downstream with the freestream velocity, and the downstream fin must encounter this structure at a specific point in its oscillation cycle to benefit from the associated low pressure region. The diagonal structure of the performance bands therefore reflects the relationship between vortex convection time, fin spacing, and phase offset.

\subsection{Non-dimensional phase offset}

The diagonal performance bands observed in figure \ref{fig:ch4_2fin_map} suggest that the optimal phase relationship between fins is governed by the time required for vortices shed by the upstream fin to convect downstream. Because these vortices travel approximately with the freestream velocity, the delay between vortex shedding at the first fin and its interaction with the second fin scales with the convective travel time between fins. This convective delay can be approximated as
\begin{equation}
t_{conv} \sim \frac{S}{U_\infty},
\end{equation}
where $S$ is the spacing between fins and $U_\infty$ is the freestream velocity. In contrast, the motion of the fins occurs over the oscillation period
\begin{equation}
T = \frac{1}{f},
\end{equation}
where $f$ is the oscillation frequency. The relevant non-dimensional quantity therefore becomes the ratio of the vortex convection time to the oscillation period. Expressed in phase units, this gives an ideal phase shift between fins of

\begin{equation}
\phi_0 \sim 2\pi \frac{S f}{U_\infty}.
\label{eqn:phaseMod}
\end{equation}

Equation \ref{eqn:phaseMod} represents a convective timing correction that accounts for the travel time of the vortex between fins. Physically, this quantity represents the phase lag required for the downstream fin to encounter the vortex shed by the upstream fin at a favorable point in its oscillation cycle.

\begin{figure}[h!]
\centering
\includegraphics[width=1.0\textwidth]{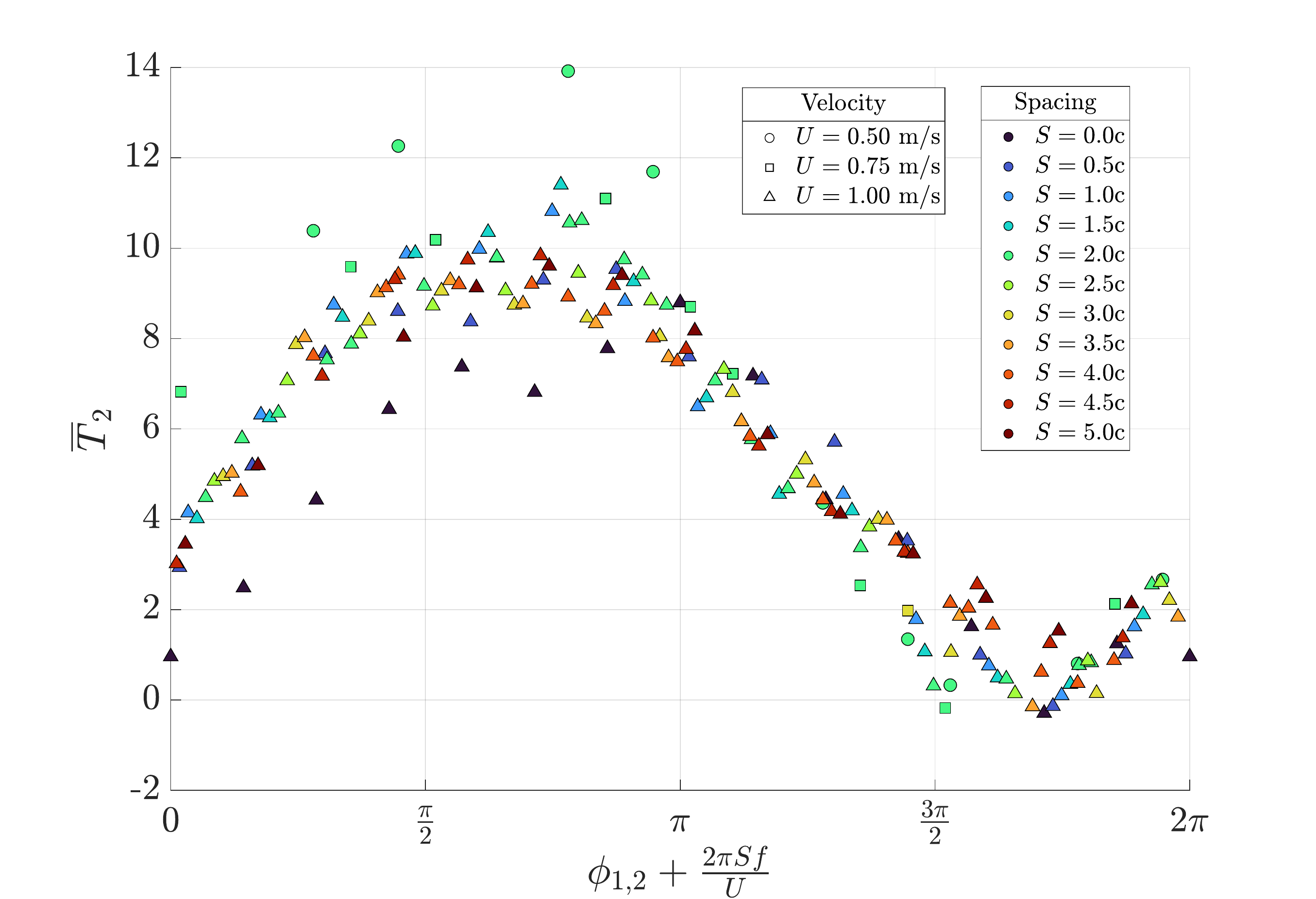}
\caption{The thrust of the downstream fin relative to the upstream fin in a 2-fin system collapse onto a phase modifier to account for structure wake timing.}
\label{fig:ch4_nondimensional offset}
\end{figure}

To evaluate this relationship, we re-plot the thrust coefficient of the downstream fin while incorporating the phase modifier $\phi_0$. Specifically, we examine the thrust produced by the second fin as a function of the adjusted phase $\phi_{1,2} + \phi_0$. The results are shown in figure \ref{fig:ch4_nondimensional offset}. When this convective phase correction is applied, the data collapse across a wide range of fin spacings, phase offsets, and freestream velocities.

This collapse indicates that the dominant timing mechanism governing two-fin interactions is the convection of vortices between fins. Once this effect is accounted for, the optimal relative phase between fins becomes approximately constant. Across the range of simulations considered here, the maximum thrust of the downstream fin occurs near

\begin{equation}
\phi_{1,2} \approx \frac{2\pi}{3}.
\end{equation}

These results hold across Strouhal numbers in the range $St = 0.20$--$0.50$, suggesting that the convective scaling captured in equation \ref{eqn:phaseMod} provides a useful framework for predicting optimal timing in two-fin propulsion systems.

\subsection{Thrust and power correlation}
After examining the relationship between timing and fin performance, we explored closer as to why. As previously mentioned the downstream fin ideally passes near to but not into the vortice. The shed vortices from the lead fins and the leading side of the fin are both low pressure regions. The low pressure regions combine, increasing the lift based thrust generated by the foil. Also, the velocity field of the vortex itself will act to increase the downstream foil effective angle of attack.

\begin{figure}[h!]
\centering
\includegraphics[width=1.0\textwidth]{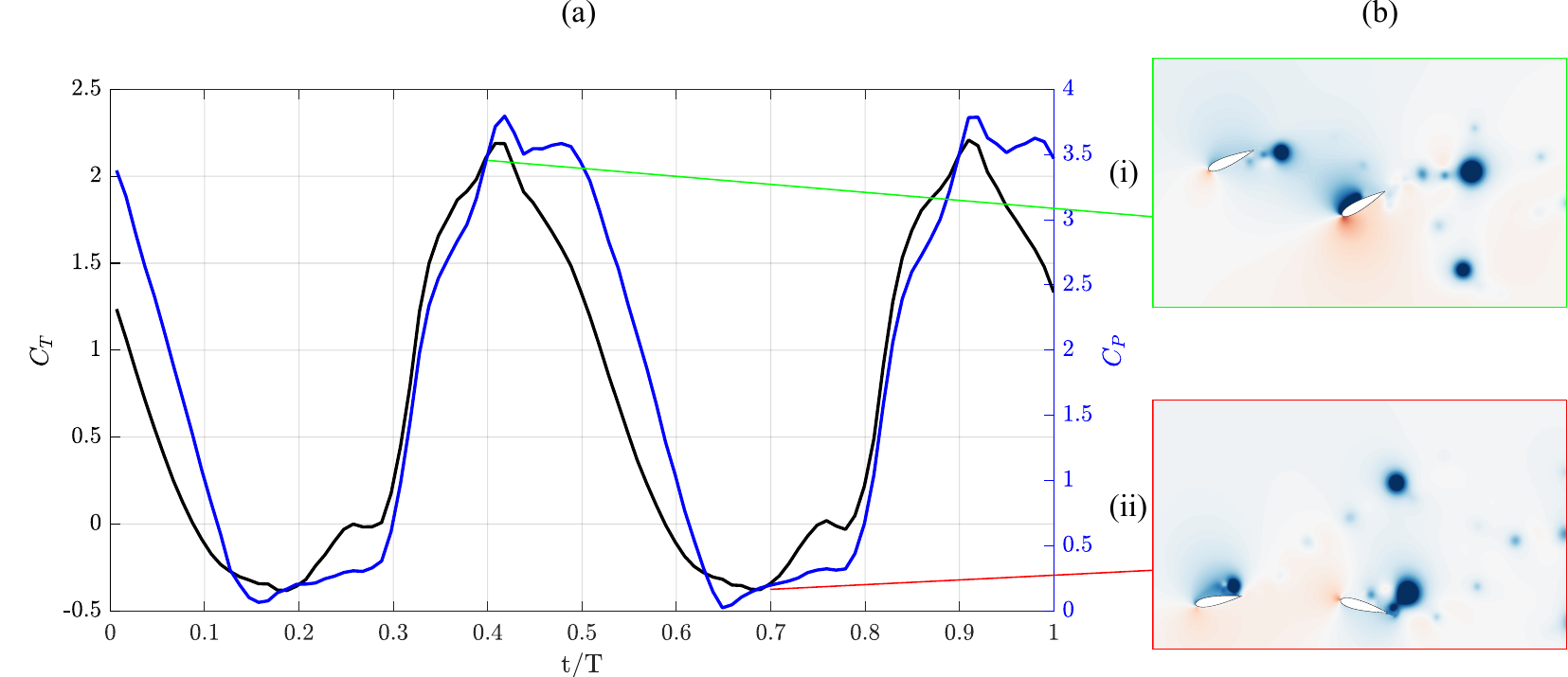}
\caption{Instantaneous coefficient of thrust $C_T$ (black) and coefficient of power $C_P$ (blue) for the downstream fin over the final oscillation cycle for a high-performance case. The upper panel (green) shows the moment of maximum thrust, the lower panel (red) corresponds to the minimum thrust condition.}
\label{fig:ch4_time_varying_good}
\end{figure}

We illustrate this interaction in figure \ref{fig:ch4_time_varying_good} (a), where the time-varying coefficients of thrust and power are shown across the final oscillation cycle for the highest performing case from the performance map in figure \ref{fig:ch4_2fin_map} (at spacing $S = 1.5c$ and phase offset $\phi_{1,2} = 51^\circ$). In the left panel we plot the instantaneous coefficient of thrust $C_T$ and the instantaneous coefficient of power $C_P$ generated by the downstream fin as a function of normalized time $t/T$ across the final oscillation cycle.

Two corresponding flow field snapshots are shown to the right of the force history (a) to illustrate the hydrodynamic mechanisms responsible for these variations in thrust. The upper image (b.i), highlighted in green, corresponds to the moment of peak thrust production for the downstream fin. At this instant the leading edge of the fin is mid drive of its oscillation and interacts favorably with a vortex shed by the upstream fin, resulting in a strong low-pressure region near the fin surface that enhances lift-based thrust production. 

In contrast, the lower image (b.ii) highlighted in red corresponds to the minimum thrust produced by the downstream fin during the cycle. In this configuration the fin is altering its direction and pitching about the leading edge shedding its vortice into the freestream.

The instantaneous power coefficient follows a similar trend. When the downstream fin generates high thrust, the forces acting on the fin are larger and the power required to drive the motion correspondingly increases. Conversely, during periods of reduced thrust production the power consumption decreases. 

For comparison, we examine a low-performing configuration from the performance map in Figure~\ref{fig:ch4_2fin_map}, corresponding to a spacing of $S = 3.3c$ and a phase offset of $\phi_{1,2} = 124^\circ$. The results are presented using the same structure as in Figure~\ref{fig:ch4_time_varying_good}, with the instantaneous coefficients of thrust and power plotted over the final oscillation cycle and accompanied by representative flow field snapshots.

As expected, the downstream fin in this configuration produces significantly lower thrust throughout the cycle compared to the high-performing case. Similar to the previous example, the peaks in thrust and power occur at approximately the same time during the oscillation cycle, reflecting the fact that periods of high thrust generation correspond to increased hydrodynamic loading and therefore increased power input.

\begin{figure}[h!]
\centering
\includegraphics[width=1.0\textwidth]{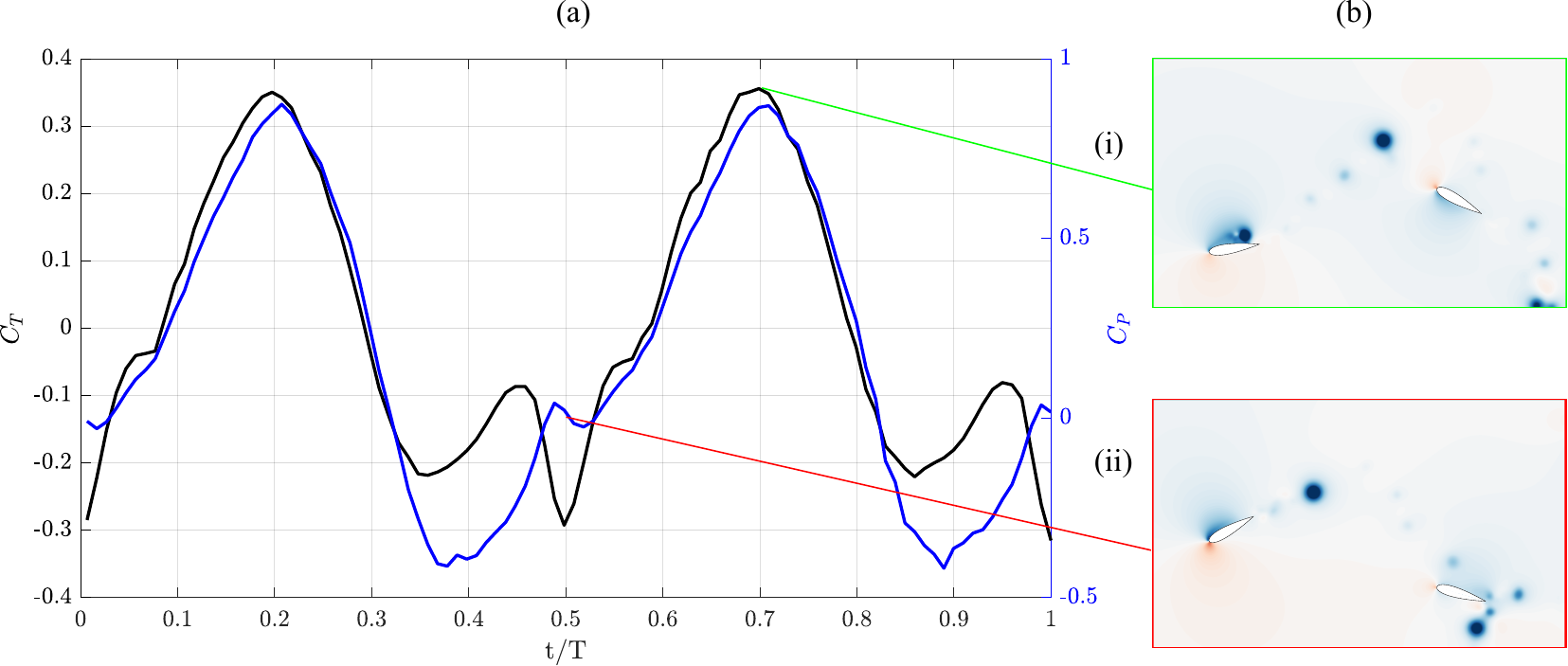}
\caption{Instantaneous coefficient of thrust $C_T$ (black) and coefficient of power $C_P$ (blue) for the downstream fin over the final oscillation cycle for a low-performance case. The upper panel (green) shows the moment of maximum thrust, the lower panel (red) corresponds to the minimum thrust condition.}
\label{fig:ch4_time_varying_bad}
\end{figure}

In figure \ref{fig:ch4_time_varying_bad}(b.i), the downstream fin reaches its maximum thrust production near mid-heave, where the fin experiences relatively undisturbed inflow and the pressure distribution around the fin supports lift-based thrust generation.

A more complex interaction occurs later in the cycle. In figure \ref{fig:ch4_time_varying_bad}(b.ii), a sharp spike in power occurs simultaneously with a sudden drop in thrust. Prior to this event, the thrust signal follows an upward trend that would normally continue toward a peak. However, at approximately $t/T = 0.45$, the thrust abruptly decreases before slowly recovering. Notably, this drop does not coincide with the stroke reversal where the fin transitions into the next heave motion, which is where a change in thrust might typically be expected.

Instead, examination of the flow field reveals that a vortex shed from the upstream fin passes very close to the downstream fin at this instant. The proximity of this vortex disrupts the pressure field surrounding the fin, weakening the pressure differential responsible for lift-based thrust generation. In this case the vortex interaction is detrimental rather than beneficial, effectively suppressing thrust production despite the continued motion of the fin. Not only does this damage thrust production it also causes a sharper increase in power usage required to drive the fin through its prescribed motion path. This affect and implications on performance is explored further in the next chapter.

\section{Conclusion}

In this chapter we examined the hydrodynamic interactions present in two-fin systems and demonstrated how the timing between fins governs overall propulsion performance. By constructing a performance map across spacing $S_{1,2}$ and phase offset $\phi_{1,2}$, we showed that the thrust produced by the downstream fin can vary dramatically depending on how it encounters the vortex structures shed by the upstream fin. Small changes in spacing or phase can therefore shift the system between constructive and destructive vortex interactions.

The performance map revealed distinct diagonal bands in the $(S_{1,2},\phi_{1,2})$ parameter space where the thrust of the downstream fin rapidly transitions between high and low values. These bands reflect the convective transport of vortices through the wake and the requirement that the downstream fin encounter these structures at the correct time in its oscillation cycle. By introducing a non-dimensional phase modifier based on the ratio of vortex convection time to oscillation period, we showed that these interactions collapse onto a common scaling. When this convective timing correction is applied, the optimal phase relationship between fins becomes approximately constant, with peak downstream thrust occurring near $\phi_{1,2} \approx 2\pi/3$ across a range of spacings and freestream conditions.

Time resolved analysis of thrust, power, and flow fields further clarified the physical mechanisms responsible for these performance differences. In high-performing configurations, the downstream fin passes near a vortex shed by the upstream fin, allowing the low pressure region and induced velocity field to enhance lift-based thrust generation. In contrast, poorly timed interactions can cause the vortex to pass too close to the downstream fin, disrupting the pressure field around the fin and reducing lift based thrust production. These detrimental interactions can also increase the power required to drive the fin through its prescribed motion.

Taken together, these results demonstrate that the performance of two-fin propulsion systems is fundamentally governed by the convective timing of wake structures between fins. Properly phased interactions allow the downstream fin to exploit the wake of the upstream fin for enhanced thrust, while poorly timed interactions can significantly degrade performance. Understanding and predicting this timing relationship provides a foundation for designing multi $N$-fin propulsion systems that harness beneficial wake interactions rather than suffer from destructive interference. The implications of these interactions will guide us in the next chapter when looking at efficiency and overall propulsion system performance.    
\chapter{Three fin systems}
\label{ch5:3_fins}

\section{Introduction}

In the previous chapter we examined the effects of adding a second fin in-line with a leading fin. The performance of the system was shown to become strongly governed by the interaction between the downstream fin and the vortex structures shed by the upstream fin. The timing of the second fin relative to these wake structures was therefore a dominant factor in determining overall system performance. For the parameter space examined in that study, an optimal combination of phase offset and spacing produced a system with an approximately $\sim20\%$ increase in average thrust compared to the single-fin case. Previous experimental and numerical studies \cite{Muscutt2017} have shown that at higher Strouhal numbers the gains from multi-fin interactions can be even larger, with thrust increases approaching $\sim50\%$ under favorable conditions.

In this chapter we extended this investigation by introducing a third fin into the system. The addition of a third fin significantly increased the complexity of the hydrodynamic interactions, as the wake structures generated by multiple upstream fins could interact with downstream fins in several different ways. The goal of this study was therefore to better understand how the interaction mechanisms identified in the two-fin system evolved when a third fin was added.

Three primary questions guided this investigation. First, we examined whether the heuristics identified for the optimal timing in a two-fin system continued to hold for a three-fin configuration. In the two-fin case, optimal thrust generation occurred when the downstream fin encountered the vortex structures shed by the upstream fin at a favorable phase in its oscillation cycle. Here we investigated whether the ideal timing for the third fin was governed by the same mechanisms that dictated the optimal phase relationship for the second fin, or whether new interaction effects emerged as additional wake structures were introduced into the system.

Second, we sought to better understand the directional nature of the interactions between upstream and downstream fins. From the two-fin study it was clear that upstream fins influenced downstream fins through the vortices they shed into the wake. However, it remained unclear whether downstream fins could also influence upstream fins through pressure field interactions. In particular, we investigated whether the pressure field generated by a downstream fin could propagate upstream and alter the forces experienced by the leading fins. Additionally, we examined how this interaction changed with increasing spacing between fins. At sufficiently large spacing distances, it was expected that the influence of upstream fins would diminish. In such cases, it was important to determine whether the downstream fin was primarily influenced by the nearest upstream fin or whether the cumulative wake from multiple upstream fins continued to play a role.

Finally, we considered an additional performance metric beyond thrust production, namely the Froude efficiency of the system. In the previous chapter (chapter \ref{ch4:2_fins}) we analyzed the instantaneous coefficient of power and showed how vortex interactions could significantly alter the power requirements of individual fins. In many propulsion systems, thrust and efficiency tend to exhibit a trade-off, where configurations that maximize thrust often do so at the expense of efficiency. Therefore in this chapter we investigated whether this relationship persisted for multi-fin ($N =3$) oscillatory propulsion systems. Specifically, we examined whether certain interaction mechanisms between fins could simultaneously produce favorable thrust and efficiency characteristics, and sought to identify the interaction regimes that led to either high or low system efficiency.

\section{Methods and analysis}
\subsection{Three fin performance map}

\begin{figure}[h!]
\centering
\includegraphics[width=0.9\textwidth]{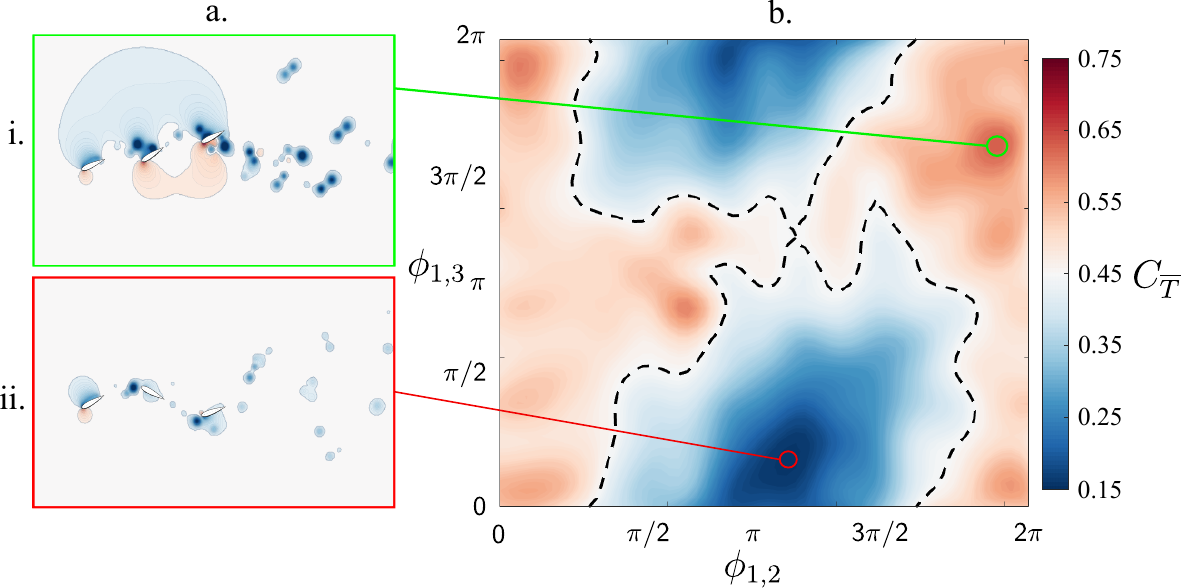}
\caption{Color contour plot showing the system average coefficient of thrust $\overline{C_T}$ for an $N = 3$ fin system (b) as the phase offsets of fins 2 and 3 relative to fin 1 are varied. The axes correspond to $\phi_{1,2}$ and $\phi_{1,3}$. Example flow fields highlighting the highest performing case (a.i) and lowest performing case (a.ii) within the parameter space are also shown.}
\label{fig:ch5_3fin_map}
\end{figure}

Beginning as we did in chapter \ref{ch4:2_fins}, we constructed a performance map around our base-case $N = 3$ fin system shown in figure \ref{fig:ch5_3fin_map}. In order to simplify the parameter space and allow the results to be visualized on a two-axis plot, the spacing between fins was held constant at $S = 2c$. We then varied the phase offsets of the two downstream fins relative to the leading fin, denoted as $\phi_{1,2}$ and $\phi_{1,3}$ for fins 2 and 3 respectively. System performance was evaluated using the mean coefficient of thrust of the system, $\overline{C_T}$.

As with the two-fin analysis, performance was visualized using a color contour map where red indicates regions of high thrust corresponding to strong propulsive performance, while blue denotes regions of low thrust and poor performance. To provide a reference baseline, the color contour was centered around $\overline{C_T} = 0.45$, which represents the approximate mean thrust coefficient produced by a single fin operating under the same coarse conditions. This normalization allowed the map to clearly illustrate whether the addition of downstream fins provided a performance benefit or detriment relative to a single-fin system. As in the two-fin study presented in chapter \ref{ch4:2_fins}, the boundary corresponding to $\overline{C_T} = 0.45$ is indicated by a dotted black contour line, marking the transition between regions where the multi-fin configuration outperforms or underperforms the single-fin baseline (see Figure~\ref{fig:ch4_2fin_map}).

Alongside this performance map in figure \ref{fig:ch5_3fin_map}, we also present the global pressure fields corresponding to the best and worst performing configurations within the parameter space. Across the explored phase combinations, the system thrust varied substantially, with values ranging approximately from $\overline{C_T} \approx 0.15$ to $\overline{C_T} \approx 0.60$. The distribution of thrust across the phase space exhibited several notable structural features. In particular, the map displayed vertical banding of alternating high and low thrust regions, indicating a strong periodic dependence on the phase relationship between fin 1 and fin 2.

In addition to this vertical structure, a diagonal band of elevated thrust appeared across the phase space at approximately a $45^\circ$ orientation. This feature suggested that while fin 3 possessed an optimal phase relationship with the upstream wake, this optimal timing was not independent of the motion of fin 2. Instead, changes in the phase of fin 2 required corresponding adjustments in the phase of fin 3 to maintain favorable wake interactions. This observation implied that the downstream fins were not interacting with the upstream wake independently, but rather that the combined wake structures produced by the first two fins governed the optimal timing for the third fin.

Although instantaneous snapshots of an inherently unsteady flow field have limitations, the pressure field visualizations still provided qualitative insight into the mechanisms responsible for these performance differences. In the highest thrust cases, the low-pressure regions associated with vortex structures shed by upstream fins aligned favorably with the low-pressure region on the upper surface of the downstream fins during their lift-producing portion of the cycle. This alignment reinforced the lift forces responsible for thrust generation. In contrast, in the lowest thrust cases, the incoming vortex structures disrupted this pressure distribution. In these configurations, the low-pressure regions from upstream vortices interfered with the high-pressure region that typically forms on the lower surface of a lifting fin, effectively weakening the lift force and reducing thrust production.

\begin{figure}[h!]
\centering
\includegraphics[width=1.0\textwidth]{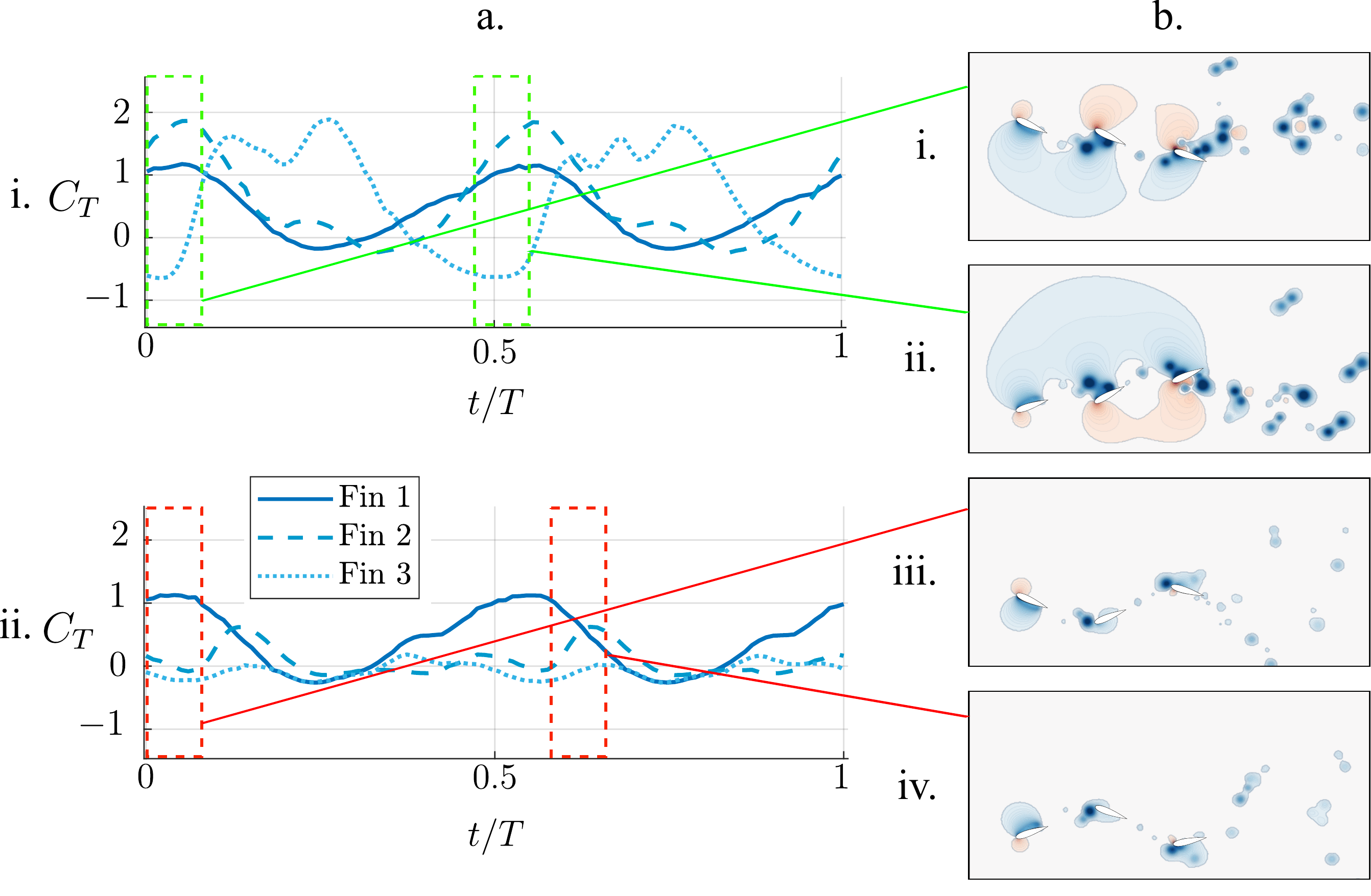}
\caption{Instantaneous coefficient of thrust over the final oscillation cycle for two three-fin systems corresponding to the high and low thrust configurations identified in figure \ref{fig:ch5_3fin_map}. Panel (a) shows the temporal variation in thrust for each fin in both high performing (a.i) and low performing (a.ii) configurations. Selected flow-field snapshots highlighting key moments in the force cycle are shown in (b).}
\label{fig:ch5_thrustforces}
\end{figure}

Taking the two highlighted cases shown in figure \ref{fig:ch5_3fin_map} (a.i) and (a.ii), we analyzed their time-varying coefficient of thrust $C_T$ across the final oscillation cycle $t/T$, shown in figure \ref{fig:ch5_thrustforces}. Examining the instantaneous force histories alongside the flow fields allows us to identify the physical mechanisms responsible for high and low thrust production in the three-fin system.

Starting with the high performing configuration ($\phi_{1,2}= 335^\circ$ and $\phi_{1,3}=283^\circ$), shown in panel (a.i), we observe that the thrust produced by the downstream fins is periodically amplified by favorable interactions with vortices shed from the upstream fins. In panel (b.i), fin 2 experiences such an interaction as it moves downward toward mid-heave. The low pressure region associated with the fin aligns with a nearby shed vortex of similar sign, strengthening the local pressure differential across the fin and increasing thrust generation.

Later in the cycle, panel (b.ii) highlights the interaction experienced by fin 3 at approximately $t/T \approx 0.55$. At this instant the fin is moving upward through the midpoint of its heave cycle and the thrust produced by the fin decreases rapidly. Examination of the flow field reveals that the high-pressure side of the fin is weakened by a nearby upstream vortex. This reduces the instantaneous pressure differential across the fin surface and consequently diminishes thrust production during this portion of the cycle.

We now examine the lowest performing configuration ($\phi_{1,2}=180^\circ$ and $\phi_{1,3}=26^\circ$), shown in panel (a.ii). In contrast to the previous case, the vortex interactions here are consistently unfavorable. In panel (b.iii), fin 2 encounters a vortex shed from the upstream fin just before $t/T \approx 0.10$. Rather than reinforcing the pressure field around the fin, this interaction disrupts the local flow and produces a noticeable drop in thrust.

A similar effect is observed for fin 3 in panel (b.iv). In this case there are no nearby vortices that constructively interact with the fin to augment its effective inflow velocity. Instead, the flow surrounding the fin is dominated by a broad region of disturbed wake. This weakened pressure differential across the fin significantly reduces its ability to generate thrust. Overall, the downstream fins operate in a disorganized wake environment, which severely limits their propulsive contribution compared with the favorable case discussed previously.

\subsection{Variable fin spacing}

Up to this point the spacing between fins has been held constant. We now investigate how altering the distance between fins changes the interaction dynamics within the system. Physically, reducing the spacing between fins should strengthen the hydrodynamic coupling between them, while increasing the spacing should allow the wake of each fin to evolve before interacting with downstream fins. To examine this effect we consider three configurations with uniform spacing between fins of $S = 0.5c$, $S = 2c$, and $S = 10c$.

\begin{figure}[h!]
\centering
\includegraphics[width=1.0\textwidth]{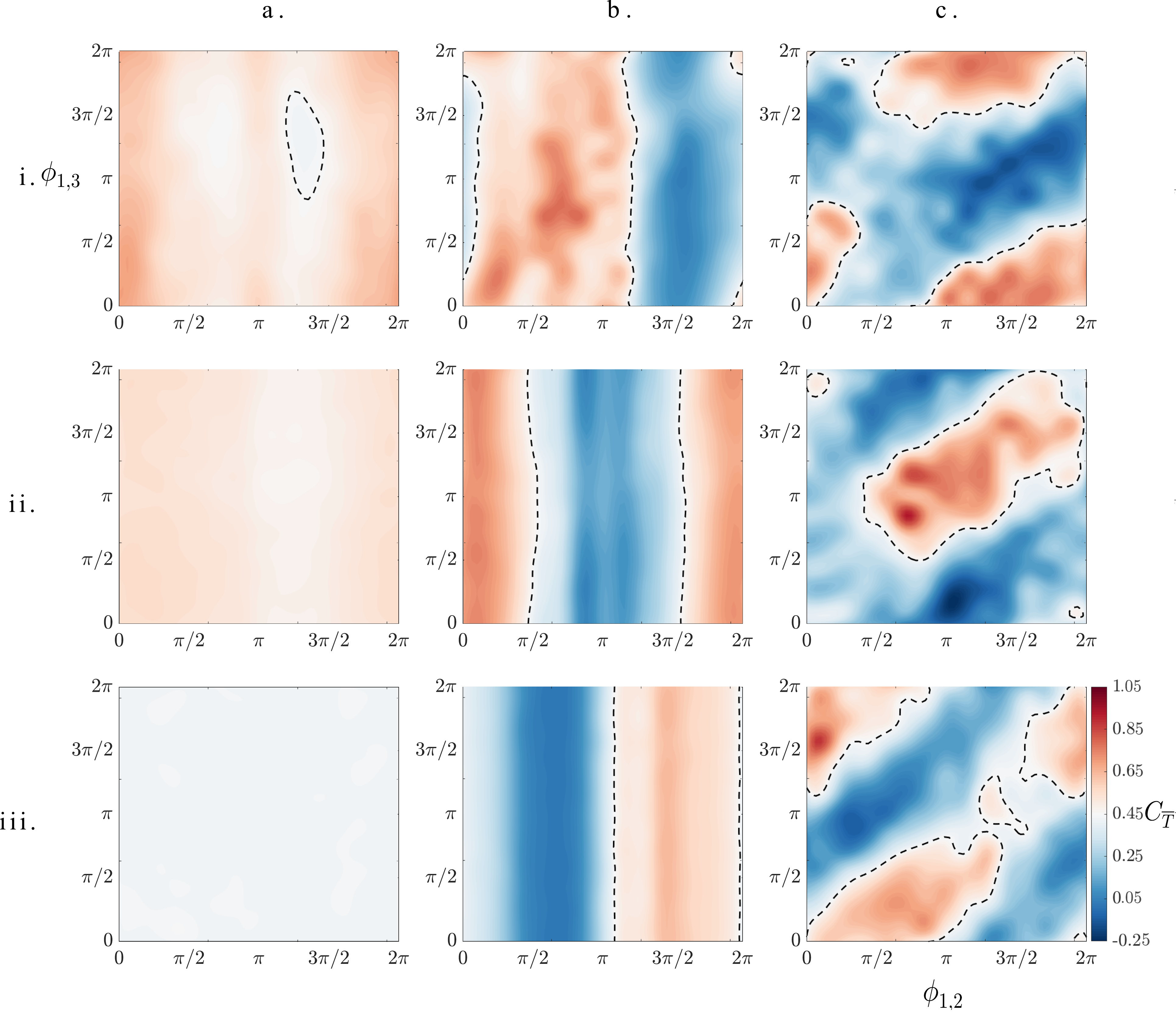}
\caption{Mean coefficient of thrust $\overline{C_T}$ for each fin in a three-fin system as the phase offsets $\phi_{1,2}$ and $\phi_{1,3}$ are varied for three different fin spacings. Columns correspond to the thrust produced by fin 1 (a), fin 2 (b), and fin 3 (c), while rows correspond to spacings of $S=0.5c$ (i), $S=2c$ (ii), and $S=10c$ (iii). Red regions indicate higher thrust and blue regions indicate lower thrust. The maps illustrate that the upstream fin is largely unaffected by the motion of downstream fins, while downstream fins are primarily influenced by the wake generated by their nearest upstream neighbor.}
\label{fig:ch5_spacing_map}
\end{figure}

Figure \ref{fig:ch5_spacing_map} shows the mean coefficient of thrust for each fin across the phase space defined by $\phi_{1,2}$ and $\phi_{1,3}$. Each column corresponds to a different fin within the system, while each row corresponds to a different spacing between fins. Red regions indicate high thrust production and blue regions indicate low thrust production.

Looking first at the upstream fin (column a), we observe that its thrust production is largely independent of the motion of the downstream fins. Even when the fins are closely spaced ($S = 0.5c$), only a small variation in thrust is observed across the parameter space. As the spacing increases to $S = 2c$, the impact diminishes, and finally this variation essentially disappears by $S = 10c$. This indicates that the influence of downstream fins on upstream ones is weak, even when the fins are placed relatively close together.

The second fin (column b) shows a different behavior. Here the thrust varies primarily as a function of $\phi_{1,2}$, producing vertically banded regions of high and low thrust. This indicates that the dominant interaction controlling the performance of fin 2 is its phase relationship with the upstream fin, fin 1. Changes in the phase of the third fin ($\phi_{1,3}$) have little noticeable effect on the thrust of fin 2, particularly as spacing increases. In other words, the second fin is governed almost entirely by the wake generated by the first fin.

Finally, the third fin (column c) exhibits diagonal bands of performance across the parameter space. These bands indicate that the thrust produced by fin 3 depends on both $\phi_{1,2}$ and $\phi_{1,3}$. Physically this occurs because the third fin is influenced by the vortex structures shed by the second fin, and its performance depends on the relative timing between these two motions. As the phase of fin 2 changes, the phase of fin 3 must shift correspondingly to maintain a favorable interaction with the incoming wake. This produces the approximately $45^\circ$ diagonal structures visible in the maps.

Taken together these results reveal a clear hierarchy of interactions within the system. Upstream fins are largely unaffected by downstream fins, while downstream fins are primarily influenced by the motions of upstream structures. Increasing the spacing between fins further reinforces this behavior by allowing the wake to evolve before interacting with the next fin. 

This observation has important implications for the design and optimization of multi-fin propulsion systems. As downstream fins are dominated by the wake of the upstream neighbors, but not vice versa, the system can be optimized sequentially. In practice this means that once the phase relationship between two fins is optimized, additional fins can be added downstream, needing only to be optimized via their own phase. This sequential optimization, rather than re-optimizing the entire system, will be further discussed in the next chapter.

\subsection{Phase modifier}
Using the data from the spacing study in figure \ref{fig:ch5_spacing_map}, we now test the validity of the phase modifier $\phi_0$ introduced earlier in chapter \ref{ch4:2_fins} for predicting optimal timing between fins. Recall that the phase modifier accounts for the convection time required for wake structures generated by an upstream fin to reach a downstream fin. If this scaling holds for multi-fin systems, thrust production should collapse onto a single periodic curve when plotted using the modified phase $\phi_{n-1,n} + \phi_0$.

\begin{figure}[h!]
\centering
\includegraphics[width=0.8\textwidth]{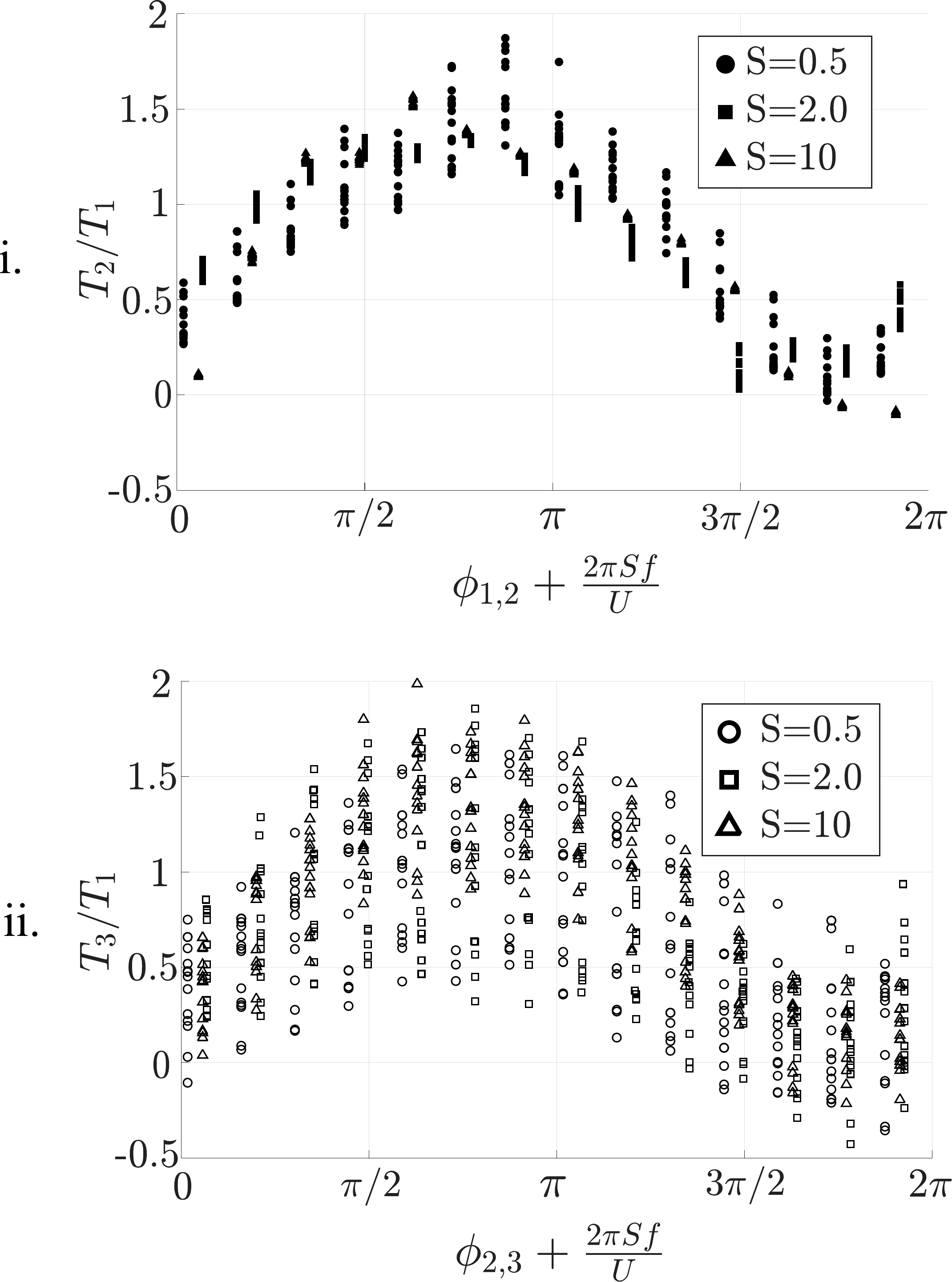}
\caption{Relative thrust produced by downstream fins after applying the phase modifier $\phi_0$ for three-fin systems with spacings of $S = 0.5c$, $S = 2c$, and $S = 10c$. (i) Thrust of the second fin normalized by the thrust of the first fin as a function of the modified phase $\phi_{1,2} + \phi_0$. (ii) Thrust of the third fin normalized by the thrust of the first fin as a function of the modified phase $\phi_{2,3} + \phi_0$. The data for the second fin collapsed well onto a single periodic curve, peaking near $\phi_{n-1,n} \approx 2\pi/3$, consistent with the two-fin system. The third fin exhibits a broader performance band due to the more complex wake environment produced by interactions between the wakes of the first and second fins.}
\label{fig:phase_modifier_3fin}
\end{figure}

Figure \ref{fig:phase_modifier_3fin} shows the mean thrust produced by the second and third fins normalized by the thrust of the first fin across the full range of phase offsets. The data include all three spacings considered previously ($S = 0.5c$, $S = 2c$, and $S = 10c$). The horizontal axis therefore represents the modified phase relationship between neighboring fins, while the vertical axis shows the relative thrust produced by each downstream fin.

For the second fin (figure \ref{fig:phase_modifier_3fin}i), the phase modifier successfully collapses the thrust data from all three spacings onto a single periodic curve. Despite the different wake development distances, the thrust peaks consistently around $\phi_{1,2} \approx 135^\circ$. This result mirrors the behavior observed in the two-fin system and suggests that the dominant interaction controlling the performance of the second fin is the vortex wake shed by the first fin. Since the second fin primarily interacts with this single upstream wake, the phase modifier captures the relevant timing relationship well.

The behavior of the third fin (figure \ref{fig:phase_modifier_3fin}b) is more complex. While a periodic trend is still visible and the peak thrust again occurs near $\phi_{2,3} \approx 270^\circ$, the data collapse is noticeably weaker and the spread in thrust values is significantly larger. This broader performance band arises from the increasingly complex wake environment encountered by the third fin. Unlike the second fin, which primarily interacts with the wake of a single upstream fin, the third fin experiences the combined influence of vortex structures shed by both the first and second fins. These interacting wakes introduce additional variability in the local flow field that is not captured by the simple phase modifier.

The influence of this wake complexity is particularly evident at smaller spacings, where vortices from upstream fins interact more strongly before reaching the third fin. As the spacing increases, the wakes have more distance to evolve and the spread in performance is somewhat reduced. Nevertheless, even in the larger spacing cases the third fin remains subject to multiple interacting vortex structures, which prevents the same level of collapse observed for the second fin.

Despite these limitations, the phase modifier still captures the dominant periodic trend in the thrust data. The fact that the thrust peaks consistently near $\phi_{n-1,n} \approx270^\circ$ across both fins and all spacings suggests that the convective timing of vortex interactions remains the primary mechanism governing thrust production. As a result, the phase modifier provides a useful first-order tool for predicting favorable phase relationships in multi-fin systems, even when the wake interactions become increasingly complex downstream.

Taken together, these results show that the heuristics for ideal vortex interactions in a three-fin system closely mirror those observed in the two-fin system. In both cases, optimal thrust production occurs when the downstream fin encounters the upstream vortex at a favorable phase, which consistently corresponds to a phase offset near $\phi_{n-1,n} \approx 270^\circ$. The phase modifier successfully collapses the performance of the second fin across different spacings, indicating that the dominant interaction mechanism remains the convective arrival of the upstream vortex. Although the third fin experiences a more complex wake environment due to the combined influence of two upstream fins, the same fundamental timing relationship still governs its performance. As a result, the primary design heuristic developed from the two-fin system---aligning downstream fin motion with the arrival of upstream vortices---remains applicable for multi-fin configurations.

Thus far, we have considered only the thrust produced by the system, which is directly related to swimming speed. In addition to thrust, however, it is also important to examine the energetic performance of the system. We therefore consider the propulsive efficiency $\eta$ across the same parameter space to determine how multi-fin interactions influence energetic cost. Previous studies \cite{floryan2018efficient} have shown that for individual propulsors thrust and efficiency are often inversely correlated. Whether this relationship persists in multi-fin systems, where vortex interactions between fins can alter both force production and power consumption, is not immediately clear.

\subsection{Efficiency}

\begin{figure}[h!]
\centering
\includegraphics[width=0.9\textwidth]{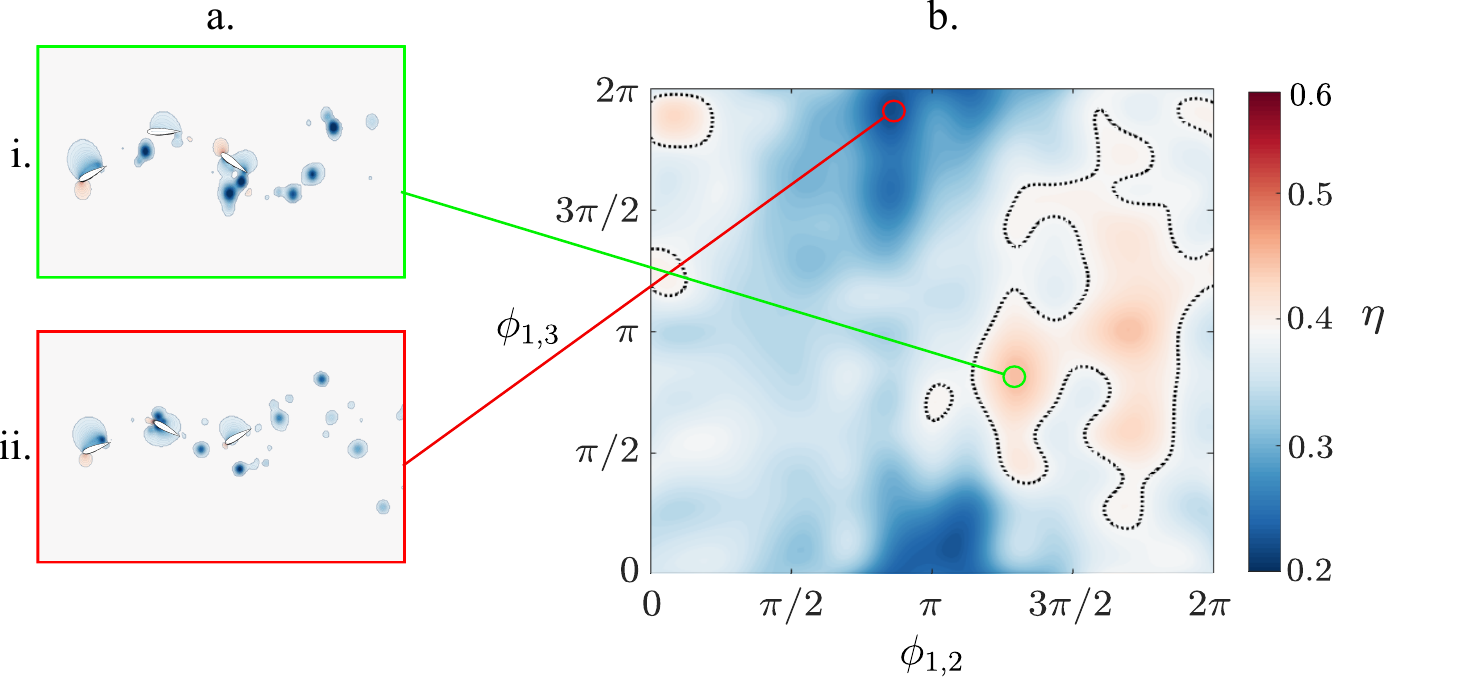}
\caption{Color contour plot showing the system average Froude efficiency $\eta$ for the $N = 3$ fin system as the phase offsets of fins 2 and 3 relative to fin 1 ($\phi_{1,2}$ and $\phi_{1,3}$) are varied. Example flow fields highlighting the highest efficiency case (a.i) and lowest efficiency case (a.ii) within the parameter space are also shown.}
\label{fig:ch5_3fin_eff_map}
\end{figure}

Figure \ref{fig:ch5_3fin_eff_map} shows the Froude efficiency as it varies with the phase offsets of the second and third fins, $\phi_{1,2}$ and $\phi_{1,3}$, respectively. Across the explored phase space the efficiency ranges from approximately $\eta = 0.22$ to $\eta = 0.45$. The overall distribution closely mirrors the thrust performance map shown previously in figure \ref{fig:ch5_3fin_map}. Regions of lower efficiency occur near $\phi_{1,2} \approx 180^\circ$ and for $-90^\circ < \phi_{1,3} < 90^\circ$, which correspond directly to regions of reduced thrust production. In contrast, the regions of higher efficiency are somewhat more broadly distributed and less sharply defined than the corresponding high-thrust regions.

This result suggests that, within this multi-fin configuration, operating at high thrust does not necessarily incur a strong efficiency penalty. Conversely, parameter combinations that produce low thrust also tend to result in low efficiency.

\begin{figure}[h!]
\centering
\includegraphics[width=1.0\textwidth]{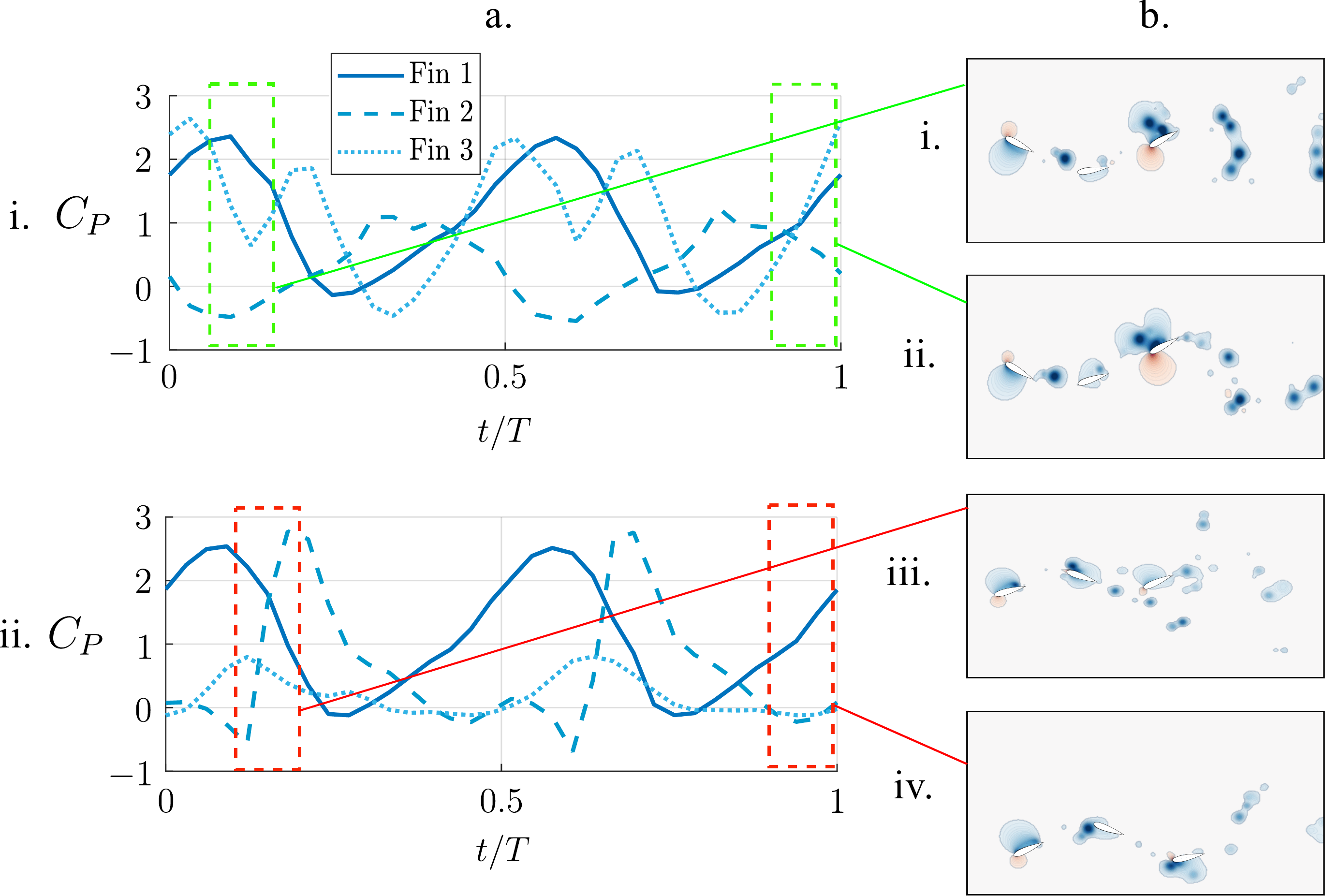}
\caption{Instantaneous coefficient of power $C_P$ over the final oscillation cycle for two three-fin systems corresponding to the high  and low efficiency configurations identified in figure \ref{fig:ch5_3fin_eff_map}. Column (a) shows the temporal variation in power for each fin for the high (a.i) and low (a.ii) cases, while panel (b) highlights selected flow-field snapshots corresponding to key points in the power cycle.}
\label{fig:ch5_efficiency_timeseries}
\end{figure}

To better understand how flow field timing influences efficiency, we examine the time histories of the power coefficient $C_P$ for the three fins over the final oscillation cycle, shown in figure \ref{fig:ch5_efficiency_timeseries}. Selected flow field snapshots are included to highlight key moments within the cycle. Since Froude efficiency depends on both thrust and power, periods of reduced power consumption tend to correspond to higher efficiency, while spikes in power can significantly reduce efficiency even if thrust remains relatively unchanged. The thrust histories shown previously in figure \ref{fig:ch5_thrustforces} can therefore be compared alongside these power histories to provide a more complete understanding of the system energetics.

The first fin in the system is least impacted by the change in phase between the high- and low-efficiency regions, which is consistent with the thrust behavior observed previously. The second fin, however, shows a much stronger response to the phase variations. In the high efficiency configuration, the power consumption of fin 2 remains relatively moderate compared to the other fins. In contrast, within the low efficiency region, fin 2 exhibits dramatic spikes in power consumption.

Notably, these spikes in power are not accompanied by corresponding increases in thrust in figure \ref{fig:ch5_thrustforces}. This indicates that the additional power input does not contribute to useful propulsive work, but instead represents wasted energetic expenditure that directly reduces efficiency.

Interestingly, the third fin exhibits the opposite trend. In the low efficiency configuration the power usage of fin 3 collapses relative to the high efficiency case. However, this reduction in power coincides with a significant reduction in thrust production. In other words, the third fin is not necessarily operating inefficiently; rather, it contributes little useful thrust to the system in this configuration.

Overall, the primary energetic penalty occurs in regions where thrust production is already poor. Operating in low thrust regions is therefore both hydrodynamically and energetically unfavorable. Conversely, the phase relationships that enhance thrust generation also tend to maintain favorable energetic performance. This indicates that the vortex interactions that constructively augment thrust production also help regulate the energetic cost of propulsion in the multi-fin system.

\section{Conclusion}

In this chapter we investigated three-fin in-line oscillatory propulsion using an immersed-boundary flow solver, extending the two-fin analysis of  chapter \ref{ch4:2_fins}. The primary focus was on how thrust production and Froude efficiency varied with the phase relationships between fins and with the spacing between fins. By constructing a performance map for a base-case $N=3$ system at fixed spacing ($S=2c$), we showed that system averaged thrust was highly sensitive to the relative timing of the downstream fins. Across the explored phase space, the mean thrust varied substantially, spanning approximately $\overline{C_T}\approx0.15$ to $\overline{C_T}\approx0.60$ (figure \ref{fig:ch5_3fin_map}), demonstrating that the addition of a third fin could either strongly enhance or severely degrade propulsive output depending on wake timing.

Time resolved force histories and accompanying flow fields clarified the mechanism responsible for these changes. High thrust configurations occurred when vortices shed by upstream fins passed near the suction side of downstream fins during their lift-producing portion of the cycle, reinforcing the local pressure differential and producing thrust amplification (figure \ref{fig:ch5_thrustforces}). Conversely, low thrust configurations occurred when incoming wake structures disrupted the pressure field around the downstream fin, either by weakening the high pressure region on the pressure side or by otherwise misaligning with the downstream fin motion. While the flow was inherently unsteady and snapshots were limited, the same qualitative picture emerged consistently: performance depended primarily on whether the downstream fin encountered upstream wake structures at a favorable phase in its oscillation.

Varying the spacing between fins further revealed a clear hierarchy of interactions within the three-fin system. The leading fin remained largely insensitive to downstream phase across all spacings considered, with only minor variation at the smallest separation ($S=0.5c$) and essentially no dependence by $S=10c$ (figure \ref{fig:ch5_spacing_map}). In contrast, the performance of fin 2 was governed almost entirely by its phase relationship with fin 1, while the performance of fin 3 depended most strongly on its timing relative to fin 2. These results indicated that downstream fins were primarily controlled by the wake of their nearest upstream neighbor, and that meaningful upstream influence from downstream fins was limited to very tight spacings.

Consistent with this interpretation, the phase modifier $\phi_0$ developed for two-fin systems continued to provide a useful first-order timing heuristic in the three-fin configuration. After applying the modified phase, thrust data for fin 2 collapsed well across spacings and peaked near a consistent phase offset of approximately $\phi_{n-1,n}\approx 270^\circ$ (figure \ref{fig:phase_modifier_3fin}), mirroring the two-fin result. The third fin exhibited a broader band of performance due to the more complex wake environment created by the interaction of wakes from fins 1 and 2, but still retained the same dominant periodic trend and a similar favorable phase region.

Finally, we examined energetic performance through the system averaged Froude efficiency. Efficiency varied over the phase space from approximately $\eta\approx0.22$ to $\eta\approx0.45$ and closely mirrored the thrust map (figure \ref{fig:ch5_3fin_eff_map}). Importantly, high thrust configurations did not necessarily incur a strong efficiency penalty. Instead, the lowest efficiency occurred primarily in regions where thrust was already poor. Time histories of the power coefficient showed that the dominant efficiency losses were associated with large spikes in power, particularly for fin 2, that were not accompanied by corresponding increases in thrust (figure \ref{fig:ch5_efficiency_timeseries}). In contrast, favorable vortex interactions that enhanced thrust also tended to maintain moderate power requirements and thus preserved efficiency.

A key implication of these results was that adding an additional fin downstream did not fundamentally alter the optimal configuration of the upstream fins. Because upstream fins were only weakly influenced by downstream fins except at extremely close spacings, multi-fin systems could be optimized sequentially: once an upstream pair was tuned, additional fins could be appended downstream and optimized primarily through their timing relative to the nearest upstream neighbor. This provides a practical design and optimization pathway for larger $N$-fin systems, which is pursued further in the next chapter.
\chapter{Bayesian optimization of multi-fin systems}
\label{ch6:bayesian}

\section{Introduction}

As the number of fins increased and additional control variables were introduced, the parameter space expanded rapidly and became impractical to explore visual via contour plots. While the detailed maps in chapters \ref{ch4:2_fins}--\ref{ch5:3_fins} provided clear physical understanding, they also highlighted a key limitation: even for a three-fin system, fully sampling the space of phase and spacing required a large number of expensive simulations. For larger fin systems, or when simultaneously varying spacing, phase, and kinematic parameters, brute-force mapping was no longer an efficient strategy.

To address this, we paired our immersed boundary flow solver with a Bayesian optimization framework. Various optimization regimes have been employed for fin optimization in prior studies \cite{kaya2007nonsinusoidal}. Bayesian optimization is well suited for computationally expensive objective functions because it can locate high performing regions of a complex parameter space using relatively few evaluations. In addition to identifying optimal configurations, the optimizer also provides a live plotting capability mid search, allowing us to interpret trends and limitations of the solver.

\section{Bayesian optimization framework}

Bayesian optimization begins by defining the parameter space, i.e. specifying the optimization variables and their allowable bounds. The procedure may evaluate an initial set of simulations to establish a starting dataset, for example you may allow 20 total runs of the searcher but to allow it a ``warm-up'' it may be given x number of runs prior to that to give a preliminary understanding of the space. After these preliminary samples, the optimizer selects new candidate points iteratively. At each iteration it constructs a surrogate model of the objective function using a Gaussian process. This surrogate model provides an estimate of the mean performance between observed data points and unknown areas. 

The Gaussian-process surrogate enables a balance between exploitation and exploration. Meaning when the optimizer identified a promising region (high predicted performance with relatively low uncertainty), it preferentially sampled more densely within that region to refine the estimate of the optimum. Simultaneously, it continued to test under-sampled regions where uncertainty remained high and where unrealized optima could exist. As additional simulations were performed, the surrogate model was updated, and uncertainty decreased in regions with denser observations.

In practice, our objective was typically a maximization of either mean thrust or efficiency. Since the optimization routine was formulated as a minimization problem, thrust maximization was implemented by minimizing the negative mean thrust (and similarly for efficiency where applicable).

\section{Validation against a known three-fin parameter space}

\begin{figure}[h!]
  \centering
  \includegraphics[width=1\linewidth]{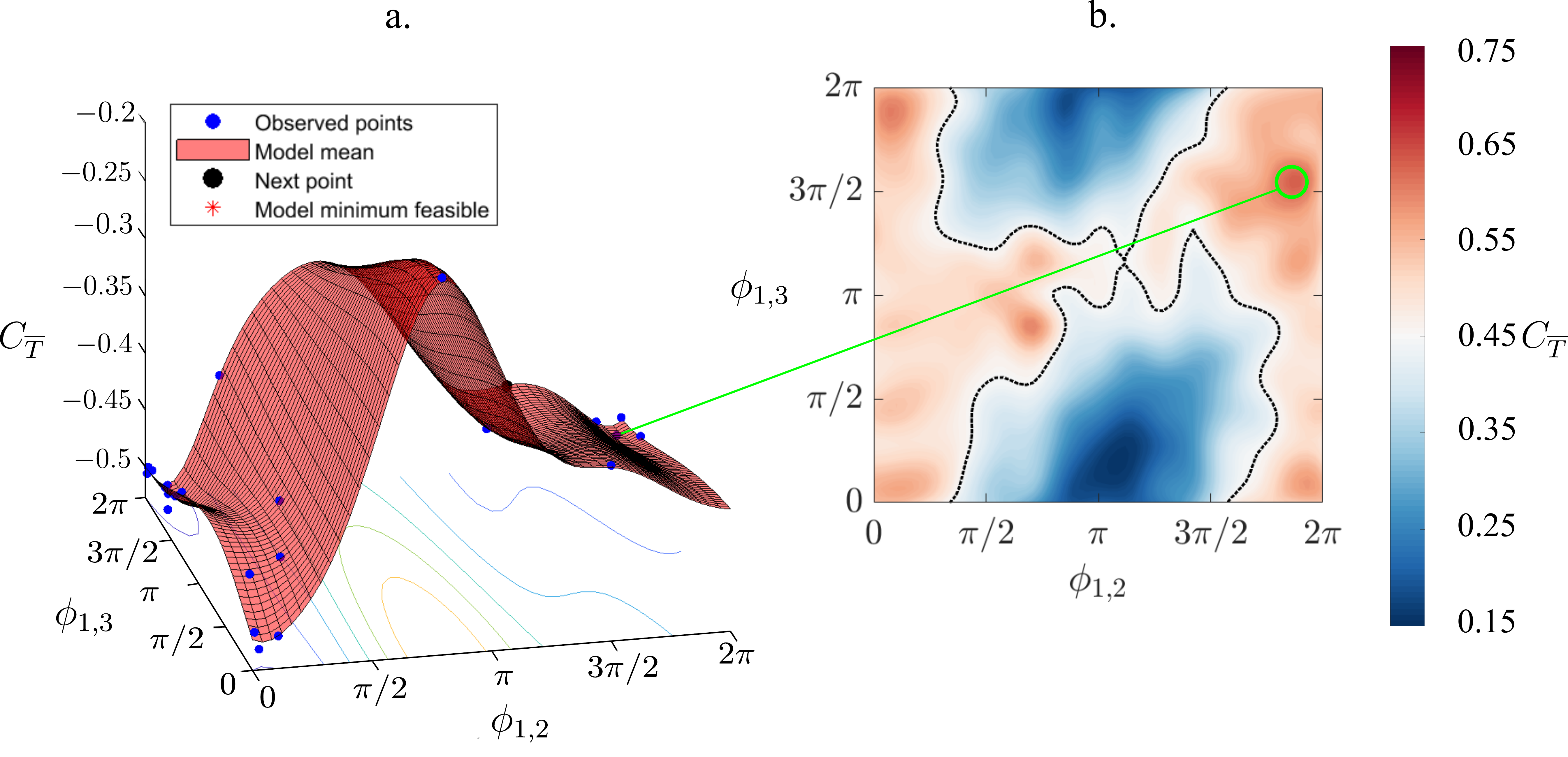}
  \caption{Bayesian optimization surrogate model for the three-fin system thrust objective as a function of fin phases (a), shown alongside the fully sampled parameter space (b). Observed samples are marked within the bounds as blue dots, and the surface represents the evolving model prediction of the objective. Note here, the Bayesian algorithm used is a minimizer, therefore a negative sign is placed in front of the output $C_{\overline{T}}$}.
  \label{fig:bayesian_explanation}
\end{figure}

Before applying the optimizer to larger and less interpretable parameter spaces, we first validated its behavior against the known three-fin phase map presented in chapter \ref{ch5:3_fins}. Figure \ref{fig:bayesian_explanation} shows the Bayesian optimization process for maximizing thrust in the three-fin system using $\phi_{1,2}$ and $\phi_{1,3}$ as the input parameters. Individual observed simulations are shown within the bounds, while the surface represents the surrogate model estimate of performance. We gave the solver one initial run at $\phi_{1,2} = \phi_{1,3} = 0$, and then allowed it 20 runs at our coarse search settings ($c = 32$ cells, 33 data points per cycle, 3 oscillation cycles) to attempt to find the optimal phases. 

The surrogate model captured the dominant features of the performance landscape, including broad low-performance regions (e.g., near $\phi_{1,2} \approx \pi$) as well as the high-performance region near $\phi_{1,2} \approx 0$. The global optimum found by the optimizer also corresponded to the optimum identified in the complete dataset. Most importantly, the optimizer identified this optimum substantially faster than an exhaustive grid search. Even in this two-parameter system, Bayesian optimization located the high-performing region within 20 simulation runs, where as the full performance map was made from 225 runs. This validation provided confidence that the optimizer could be applied to larger fin systems where exhaustive mapping was not feasible.

\section{Sequential optimization for larger $N$-fin systems}

The detailed spacing study in chapter \ref{ch5:3_fins} showed that for moderate and large spacings ($S > 2c$), downstream fins exerted little influence on the performance of upstream fins. Instead, each downstream fin was governed primarily by the wake of its nearest upstream neighbor. This directional hierarchy suggested a practical strategy for searching larger systems: rather than optimizing the full $N$-dimensional phase space simultaneously, the system could be optimized sequentially by adding one fin at a time and optimizing only the new fin’s phase relative to its upstream neighbor.

\begin{figure}[h!]
  \centering
  \includegraphics[width=.95\linewidth]{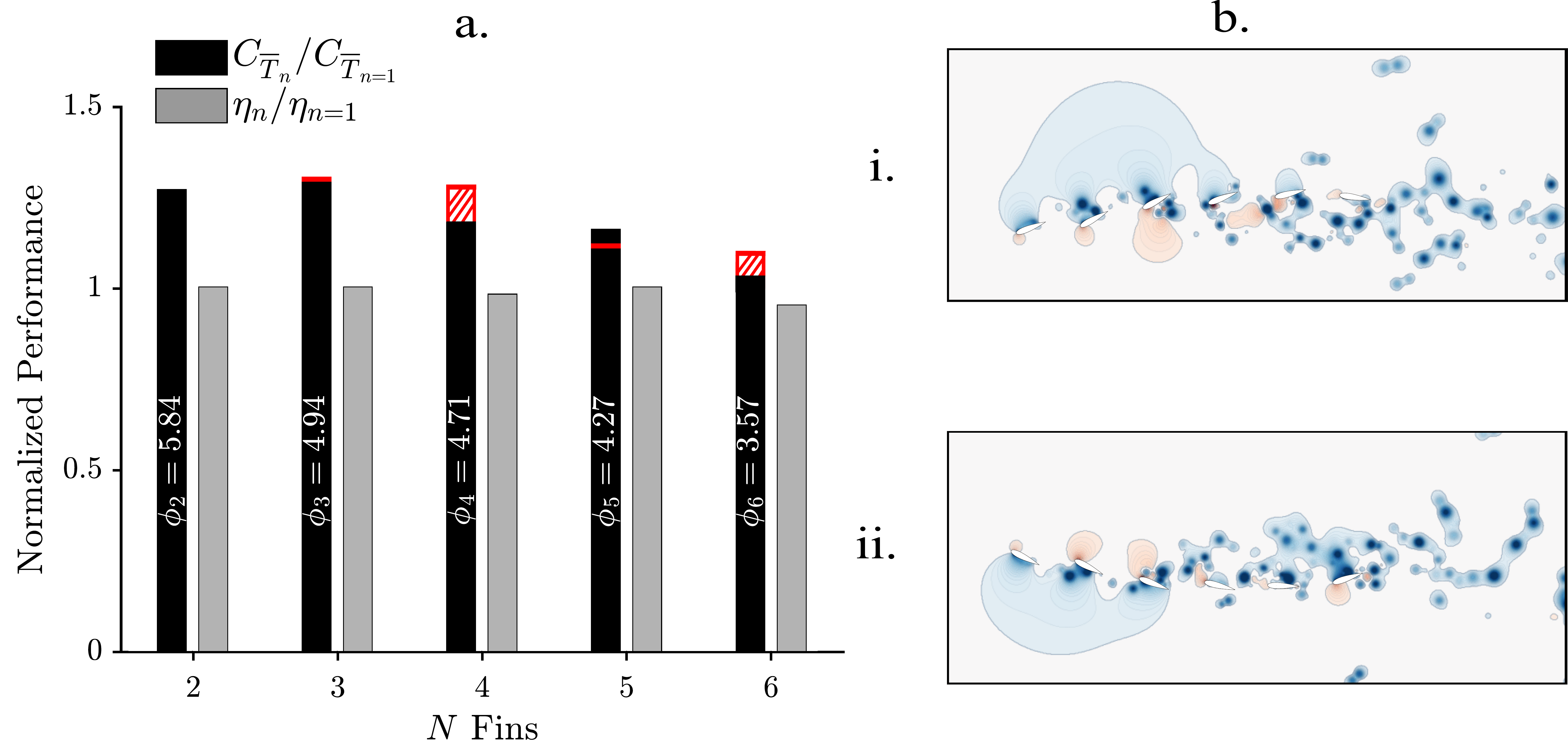}
  \caption{Thrust and efficiency of $N$-fin systems optimized for thrust using sequential Bayesian optimization. Performance is normalized by the single-fin baseline. Red markers show the performance of a reference configuration where each additional fin was assigned the same phase offset (equal phase case). Representative flow-field snapshots for the optimized $N=6$ case are shown at (b.i) $t/T=0.5$ and (b.ii) $t/T=1.0$.}
  \label{fig:bayesian_truncated_thrust}
\end{figure}

Using this sequential approach, we optimized fin systems with $N=2$ through $N=6$ fins under uniform spacing, with the objective of maximizing mean thrust. The resulting thrust and corresponding efficiency are shown in figure \ref{fig:bayesian_truncated_thrust}, normalized by the single-fin baseline ($C_T = 0.47$).

The thrust optimized results exhibited a modest increase in performance as fins were added up to approximately $N=3$, followed by diminishing returns for larger $N$. This behavior was consistent with the increasing wake complexity as additional fins were appended downstream. As the wake became more chaotic, it became progressively harder for downstream fins to consistently extract useful thrust from coherent upstream structures.

We also sought to compare the outcome of the Bayesian optimizer to the optimal phase offset discussed in chapters \ref{ch4:2_fins} and \ref{ch5:3_fins}. A simple first guess would be to apply the same optimal 2-fin phase to every fin pair (i.e., $\phi_{1,2} = \phi_{2,3} = \phi_{N,N+1}$). To test this, figure \ref{fig:bayesian_truncated_thrust} includes the equal phase reference case shown by red markers. This served to act as a test of the optimal phase modifier introduced in chapter \ref{ch4:2_fins} and \ref{ch5:3_fins}. Allowing us to see if this out performed the Bayesian search under coarse conditions and if or when the Bayesian search out performed the optimal phase. 

Noticeably, following the optimal phase modifier convention would set $\phi_{1,3} = 309^\circ$ rather than $283^\circ$. With the phase modifier system average thrust was slightly higher ($1.61\%$ higher) than $\phi_{1,3} = 283^\circ$, which was identified as the highest performing phase for fin-3 in our parameter sweep. Given we segmented over 15 equal sections from $ 0 \sim 2\pi$, $309^\circ$ was not present in our sweep. Interestingly though as shown previously, the Bayesian optimizer still selected $283^\circ$ as the ideal offset. This outcome may be due to a few factors. First, it could be 20 runs was not enough for the optimizer to come to a resolved enough selection. Second, it may be that the optimizer should search at fine grid settings as opposed to coarse grid settings, as the noise due to coarse settings may find false peaks. Lastly, some optimizer studies will have ''re-starts" in which you run the same set-up but alter the initial conditions. This acts to shake up the optimizer in a sense, preventing it from falling into one area and forcing it to explore more than it exploits. Regardless, we continue with the truncated optimization search and compare the results of the optimizer to the phase-modifier settings.

The optimized configurations was outperformed by the phase-modifier in $N = 3,4,6$ (within $1.61\%$ for $N = 3$ and $8\%$ in the $N = 4$ case). However, the Bayesian optimizer then outperformed the phase-modifier cases for $N = 5$. Indicating that optimal timing evolved downstream as the wake became increasingly structured by the upstream fin set. Overall the predicted phase modifier outperformed the Bayesian search for a majority of the cases in optimizing thrust. There are no peaks that strongly outperform the logical guess, and Bayesian might not be ideal for this chaotic system. Looking at the phase modifier timing, it appears the best configuration resembled a tuned downstream progression of phase offsets rather than a uniform phase shift repeated across the array.

\section{Efficiency-optimized systems}

\begin{figure}[h!]
  \centering
  \includegraphics[width=.95\linewidth]{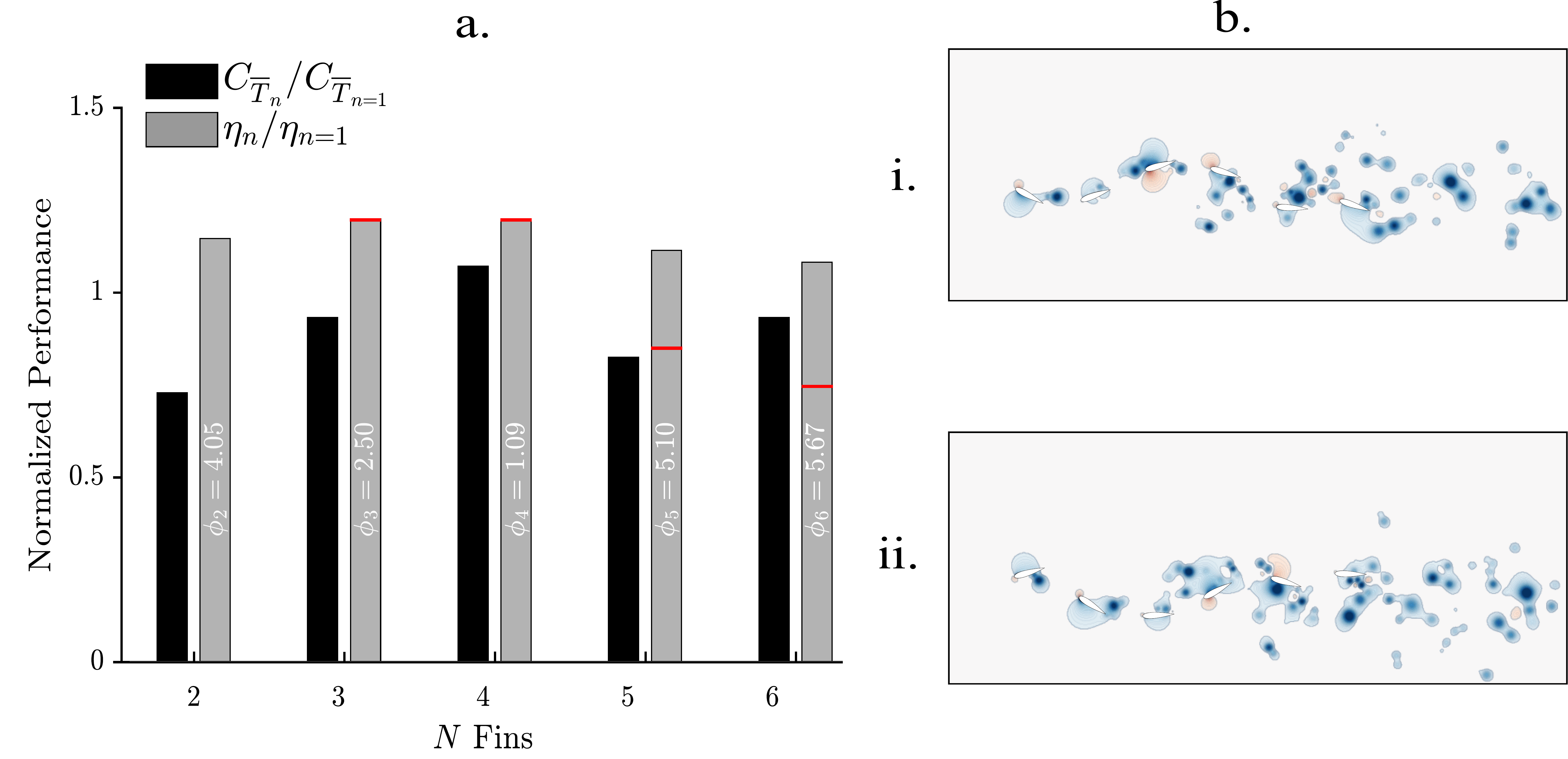}
  \caption{Thrust and efficiency of $N$-fin systems optimized for efficiency using sequential Bayesian optimization. Performance is normalized by the single-fin baseline. Red markers show the equal phase reference configuration. Representative flow field snapshots for the optimized $N=6$ case are shown at (b.i) $t/T=0.5$ and (b.ii) $t/T=1.0$.}
  \label{fig:bayesian_truncated_eta}
\end{figure}

We next applied the same sequential optimization strategy with the objective of maximizing efficiency. Figure \ref{fig:bayesian_truncated_eta} shows the resulting efficiency values (normalized by the single-fin baseline) along with the corresponding thrust levels.

The efficiency optimized cases displayed a similar trend to thrust optimization: efficiency increased as fins were added and then decayed for larger $N$. Peak efficiency occurred around $N=3$--$4$, reaching roughly $\sim12\%$ higher efficiency than the single fin baseline. However, operating at maximum efficiency produced a non-negligible thrust penalty. Even so, the $N=4$ case achieved relatively high efficiency with comparatively the same thrust relative to a single fin, suggesting that moderate fin counts can provide a favorable compromise between thrust and energetic cost.

Comparing the optimizer to the phase modifier performance, the phase modifier does not seem to apply to efficiency as the optimizer strongly outperforms the modifier. Looking at efficiency (what we're modifying for), the optimizer matches the phase modifier for $N = 3,4$ and then  beats it for $N = 5,6$. Additionally, the optimizer not only found cases that match or beat in efficiency but strongly outperforms in thrust generated at those phases.

Flow field snapshots from the optimized $N=6$ cases in figures \ref{fig:bayesian_truncated_thrust} and \ref{fig:bayesian_truncated_eta} further indicated that the fin kinematics maintained coherence across the array. In both objectives the fin motions resembled a traveling wave propagating downstream. The thrust optimized case exhibited a longer apparent wavelength, while the efficiency optimized case produced a shorter wavelength. This qualitative difference was most evident when viewed in the full flow-field animations corresponding to these cases.

\section{Non-uniform spacing optimization in three-fin systems}

Up to this point we considered fin arrays with uniform spacing. However, the spacing study in chapter \ref{ch5:3_fins} suggested that interaction strength depended strongly on fin separation and where small fin spacing ($S < 2$) upstream fins see some affect from downstream fins, motivating the question of whether non-uniform spacing could further improve system performance. To explore this, we performed Bayesian optimization on the 3-fin system while allowing both phase and spacing to vary. Spacings were bounded as $1 \leq S \leq 5$ (in chord lengths), and the system was optimized for either thrust or efficiency.

\begin{figure}[h!]
  \centering
  \includegraphics[width=.85\linewidth]{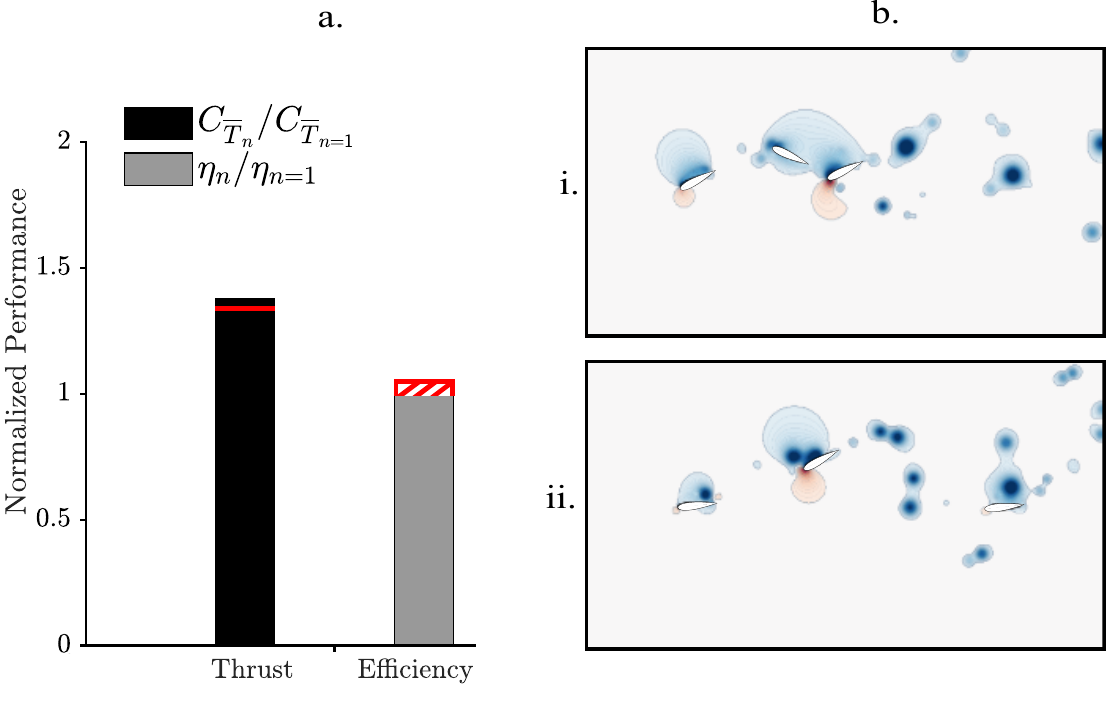}
  \caption{Three-fin system with variable spacing and phase between fins optimized for thrust and for efficiency. Performance is normalized by the single-fin baseline. Red reference lines indicate the optimal three-fin fixed-spacing case from Chapter~\ref{ch5:3_fins}. Representative flow snapshots for each optimized solution are shown in (i--ii).}
  \label{fig:bayesian_all_out}
\end{figure}

Figure \ref{fig:bayesian_all_out} shows that allowing variable spacing provided a marginal improvement in maximum thrust ($\sim5\%$) relative to the fixed-spacing three-fin optimum. The thrust optimized solution favored tighter downstream spacing, with an optimum near $S_{1,2}=1.9$ and $S_{2,3}=0.6$. In contrast, efficiency optimization favored spreading fins farther apart downstream, however it was only able to achieve a efficiency essentially at baseline of a singe fin. Most likely meaning that the optimizer was not able to locate an optimal case within the twenty runs given the increase in parameters. It settled on an optimum near $S_{1,2}=2.9$ and $S_{2,3}=3.8$. 

Overall, variable spacing was most beneficial for maximizing thrust which saw the largest increase in performance by the addition of this parameter, while efficiency was comparatively insensitive to spacing variability within the bounds considered. This result reinforced the broader observation that the principal performance gains in multi-fin systems were controlled primarily by wake timing and phase alignment, while spacing served as a secondary parameter that could enhance or weaken the interaction strength.

\section{Conclusion}

In this chapter we applied Bayesian optimization to explore larger multi-fin propulsion systems beyond the parameter spaces that could be efficiently mapped using traditional sweeps. As the number of fins and control variables increased, the dimensionality of the parameter space quickly became prohibitively large for exhaustive exploration. Bayesian optimization provided an efficient framework for identifying high-performing configurations while simultaneously constructing a surrogate model of the performance landscape. Validation against the fully mapped three-fin phase space confirmed that the optimizer was able to locate the global maximum thrust with significantly fewer simulations than required for a complete parameter sweep.

Using the physical insights developed in chapter \ref{ch5:3_fins}, we implemented a sequential optimization strategy in which each additional downstream fin was optimized primarily relative to its nearest upstream neighbor. This approach leveraged the directional nature of wake interactions, where downstream fins were governed by upstream wakes while upstream fins were largely unaffected by downstream motion. Sequential optimization therefore allowed larger fin systems to be explored efficiently without requiring simultaneous optimization across the full $N$-dimensional phase space.

Applying this strategy to systems ranging from $N=2$ to $N=6$ fins revealed that performance improvements plateau as additional fins are added. Thrust increased modestly as fins were added up to approximately $N=3$, after which the benefit diminished as the wake structure became increasingly complex and chaotic. A similar trend was observed for efficiency, with peak efficiency occurring for systems with approximately three to four fins. Beyond this point additional fins were less able to extract useful energy from the wake.

Interestingly, the optimized fin kinematics formed coherent downstream traveling wave structures. The wavelength of this wave varied depending on the optimization objective: thrust optimized systems exhibited longer wavelengths, while efficiency optimized systems produced shorter wavelengths. This observation suggests that the optimizer naturally identified kinematic patterns that organize the wake in a manner beneficial for downstream propulsion.

Finally, we explored whether allowing non-uniform spacing between fins could further improve performance. While variable spacing produced a small improvement in thrust relative to fixed-spacing systems, the impact on efficiency was comparatively minor. In general, the dominant factor controlling multi-fin performance remained the phase relationship between neighboring fins and the timing of vortex interactions in the wake.

Taken together, these results demonstrate that Bayesian optimization provides a powerful tool for exploring large parameter spaces in multi-fin propulsion systems. More importantly, the results reinforce the physical design heuristic developed in the previous chapters: optimal performance occurs when downstream fins encounter upstream vortices at a favorable phase. This interaction is largely local to neighboring fins, larger fin arrays can be effectively designed through sequential optimization rather than full-system searches. This insight provides a practical pathway for scaling oscillatory propulsion systems to larger numbers of fins while maintaining efficient computational exploration.    
\chapter{Future Work}
\label{ch7:future_work}

The work presented in this dissertation focused on understanding the hydrodynamic interactions within multi-fin oscillatory propulsion systems using computational simulations. Through systematic exploration of two-fin and three-fin systems, along with optimization studies using Bayesian methods, we identified several key physical mechanisms governing thrust generation and efficiency. In particular, we showed that downstream fins are primarily influenced by the wakes of the upstream fins, that optimal performance occurs when fins encounter upstream vortices at favorable phases, and that larger multi-fin systems naturally organize into traveling-wave-like motions.

While these findings provide important insights into the design and optimization of oscillatory propulsion systems, several directions remain for future investigation. In particular, the present work focused primarily on rigid fins operating in uniform freestream flow. Extending the framework to more biologically realistic and operationally relevant scenarios offers promising avenues for future research.

\section{Flexible fins and passive deformation}

Many of the fin systems considered in this study were rigid bodies with prescribed kinematics. However, many biological swimmers utilize flexible fins that deform in response to fluid forces. Flexibility can significantly alter the local angle of attack, the formation of leading-edge vortices, and the timing of vortex shedding. As a result, flexible fins often achieve higher propulsive efficiency than rigid foils.

Future work should explore the role of flexibility in multi-fin propulsion systems. We briefly touched on emulating flexibility in chapter \ref{ch2:single_fin}, by implementing a leading edge spring. One approach would be to incorporate passive structural models into the immersed boundary framework so that fin deformation emerges naturally from fluid–structure interactions. Passive bending or torsional compliance along the body of the fin could allow the fins to adapt their effective kinematics to the surrounding flow, potentially enhancing wake interactions between neighboring fins.

In addition to passive deformation, prescribed flexible kinematics could also be investigated. For example, bending modes to create beneficial camber could greatly increase the performance of single or multi-fin systems. Comparing prescribed flexibility with passive deformation would provide insight into whether optimal kinematics arise naturally from fluid–structure coupling or must be actively controlled.

Understanding how flexibility modifies wake interactions between multiple fins would also help determine whether the phase heuristics identified for rigid fins remain valid when deformation is introduced.

\section{Oscillatory propulsion in quiescent environments}

The simulations in this dissertation assumed a uniform freestream velocity. While this configuration represents steady swimming conditions, many biological and robotic systems must operate in quiescent or near-zero flow environments. In such cases propulsion is not used solely for forward motion but also for hovering, maneuvering, or station-keeping.

Future studies should therefore examine single or multi-fin kinematics in zero or low freestream environments. In these conditions the wake dynamics differ substantially from those observed in forward swimming. Rather than interacting with a convecting wake, the fins interact with vortices that remain in close proximity to the body or near walls and boundaries. This may fundamentally change the kinematics for fins to achieve optimal outcomes. For example, the interactions between multiple oscillating fins may generate recirculating flow structures capable of producing sustained lift or thrust without forward motion. Understanding these mechanisms would be particularly relevant for bio-inspired underwater vehicles operating in or around equipment, reefs, hulls, where slow precise maneuvers or hovering is beneficial for tasks. 

\section{Vortex impingement}

Another possible direction for future study would be to fully understand the benefits of vortex impingement on an oscillating fin by examining the pressure and velocity fields across a range of single-fin systems. This could be accomplished by introducing vortices of varying strength and size into the freestream upstream of the fin. The fin would then encounter these vortices at different points in its oscillation cycle depending on their release timing and vertical position. Such an approach would allow for a detailed investigation of how vortex interactions influence the system, using both time-varying force data and overall performance metrics to quantify their effects.

\section{Experimental validation}

While computational simulations provide detailed insight into the underlying fluid dynamics, experimental validation remains essential for confirming the predicted interaction mechanisms. Future work should therefore include experimental studies of multi-fin propulsion systems.

A controlled experimental setup could consist of a multi-fin oscillatory rig operating in a water channel. Such a system would allow systematic variation of fin phase, spacing, and kinematics while directly measuring thrust and power consumption. Flow visualization techniques such as particle image velocimetry (PIV) could then be used to observe vortex formation and wake interactions.

Experimental measurements would provide valuable confirmation of several key findings from this work. In particular, they could validate the predicted phase relationships that maximize thrust, the directional nature of wake interactions between fins, and the traveling-wave structures observed in optimized fin arrays. Additionally, experiments would help quantify how real world factors such as three-dimensional effects, structural compliance, and measurement uncertainties influence system performance.

Ultimately, combining numerical simulations, optimization techniques, and experimental validation would provide a more complete understanding of multi-fin propulsion systems and help guide the design of efficient bio-inspired underwater vehicles.

\section{Closing remarks}

The results presented in this dissertation represent a step toward understanding how oscillating fins can be arranged and coordinated to produce efficient propulsion. However, the design space for such systems remains vast. Incorporating flexibility, exploring new operating environments, and validating the computational predictions experimentally will help further uncover the principles governing multi-fin swimming systems and enable the development of practical propulsion technologies inspired by biological swimmers.    
                   
                                      
%

\bibliography{sample}
\bibliographystyle{plain}
                   
                   
\appendix{Reduced-Order Model for a Leading-Edge Spring Fin}
\label{app:LEspring_model}

This appendix presents the derivation of the reduced-order dynamical model used to describe the passive pitching motion of a fin with a leading-edge torsional spring. The formulation follows the same scaling arguments used in the prescribed pitching model, while allowing the pitch motion to arise from fluid--structure interaction with the spring.

\section{Leading-Edge Spring Formulation}

We consider a rigid fin undergoing prescribed heave $h(t)$ and passive pitch $\theta(t)$ resisted by a torsional spring of stiffness $k$ located at the leading edge. Using a quarter-chord closure for the hydrodynamic moment about the leading edge,

\begin{equation}
M_z(t)\approx \ell\,F_y(t),
\qquad
\ell \equiv \frac{c}{4},
\end{equation}

\noindent the spring balance becomes

\begin{equation}
k\,\theta(t) = -\,M_z(t)\approx -\frac{c}{4}\,F_y(t).
\label{eq:LEspring_torque_balance}
\end{equation}

\noindent The lateral force is decomposed into lift-like and added-mass-like components

\begin{equation}
F_y(t)=F_{y,L}(t)+F_{y,AM}(t).
\end{equation}

\noindent Using the same scaling closure applied to the prescribed pitch model,

\begin{align}
\frac{F_{y,L}(t)}{\rho s c}
&\sim
\alpha(t)\,U_\infty\,U_{\mathrm{eff}}(t)
+
c\,U_\infty\,\dot{\alpha}(t),
\\
\frac{F_{y,AM}(t)}{\rho s c}
&\sim
\frac{c}{2}\Big(
c\,\ddot{\theta}(t)
+\ddot{h}(t)
+\dot{h}(t)\,\dot{\theta}(t)\,\theta(t)
+U_\infty\,\dot{\theta}(t)
+c\,\dot{\theta}(t)^2\,\theta(t)
\Big).
\end{align}

\noindent The effective inflow velocity and instantaneous angle of attack are

\begin{equation}
U_{\mathrm{eff}}(t)=\sqrt{U_\infty^2+\dot{h}(t)^2},
\qquad
\alpha(t)=-\theta(t)-\arctan\!\Big(\frac{\dot{h}(t)}{U_\infty}\Big).
\end{equation}

\section{Quarter-Chord Moment Closure}

Multiplying by the moment arm $\ell=c/4$ introduces the prefactors

\begin{equation}
Z_L \equiv \frac{\rho s c^2}{4},
\qquad
Z_{AM}\equiv \frac{\rho s c^3}{4}=c\,Z_L.
\end{equation}

\noindent Substituting these definitions into the spring balance yields

\begin{equation}
k\,\theta
=
-\,Z_L\Big(\alpha U_\infty U_{\mathrm{eff}}+cU_\infty\dot{\alpha}\Big)
-\,Z_{AM}\Big(\ddot{\theta}+\ddot{h}+\dot{h}\dot{\theta}\theta+U_\infty\dot{\theta}+c\dot{\theta}^2\theta\Big).
\label{eq:LEspring_balance_Z}
\end{equation}

\section{Pitch Dynamics}

Rearranging Eq.~\ref{eq:LEspring_balance_Z} gives a nonlinear second-order equation governing the pitch dynamics,

\begin{equation}
\ddot{\theta}
=
-\frac{k}{Z_{AM}}\theta
-\frac{Z_L}{Z_{AM}}\Big(\alpha U_\infty U_{\mathrm{eff}}+cU_\infty\dot{\alpha}\Big)
-\ddot{h}
-\dot{h}\dot{\theta}\theta
-U_\infty\dot{\theta}
-c\dot{\theta}^2\theta.
\end{equation}

Using $Z_L/Z_{AM}=1/c$,

\begin{equation}
\ddot{\theta}
=
-\frac{k}{Z_{AM}}\theta
-\frac{U_\infty U_{\mathrm{eff}}}{c}\alpha
-U_\infty\dot{\alpha}
-\ddot{h}
-\dot{h}\dot{\theta}\theta
-U_\infty\dot{\theta}
-c\dot{\theta}^2\theta.
\label{eq:theta_ode_pre_sub}
\end{equation}

\noindent The time derivative of the angle of attack is

\begin{equation}
\dot{\alpha}(t)
=
-\dot{\theta}(t)
-\frac{\ddot{h}(t)/U_\infty}{1+\left(\dot{h}(t)/U_\infty\right)^2}.
\label{eq:alphadot}
\end{equation}

\noindent Substituting $\alpha$ and $\dot{\alpha}$ into Eq.~\ref{eq:theta_ode_pre_sub} yields the compact governing equation

\begin{equation}
\begin{aligned}
\ddot{\theta}
&=
\left(-\frac{k}{Z_{AM}}+\frac{U_\infty U_{\mathrm{eff}}}{c}\right)\theta
+\frac{U_\infty U_{\mathrm{eff}}}{c}
\arctan\!\Big(\frac{\dot{h}(t)}{U_\infty}\Big)
\\
&\quad
-\ddot{h}(t)
\frac{\left(\dot{h}(t)/U_\infty\right)^2}{1+\left(\dot{h}(t)/U_\infty\right)^2}
-\dot{h}(t)\dot{\theta}(t)\theta(t)
-c\dot{\theta}(t)^2\theta(t).
\end{aligned}
\label{eq:theta_closed_compact}
\end{equation}

\noindent where

\[
U_{\mathrm{eff}}(t)=\sqrt{U_\infty^2+\dot{h}(t)^2}.
\]

\section{State-Space Representation}

For numerical integration, the system is written in first-order form. Define the state variables

\begin{equation}
x_1(t)\equiv \theta(t),
\qquad
x_2(t)\equiv \dot{\theta}(t).
\end{equation}

\noindent Then

\begin{equation}
\begin{aligned}
\dot{x}_1 &= x_2,
\\
\dot{x}_2 &=
\left(-\frac{k}{Z_{AM}}+\frac{U_\infty U_{\mathrm{eff}}}{c}\right)x_1
+\frac{U_\infty U_{\mathrm{eff}}}{c}
\arctan\!\Big(\frac{\dot{h}(t)}{U_\infty}\Big)
\\
&\quad
-\ddot{h}(t)
\frac{\left(\dot{h}(t)/U_\infty\right)^2}{1+\left(\dot{h}(t)/U_\infty\right)^2}
-\dot{h}(t)x_2x_1
-cx_2^2x_1.
\end{aligned}
\label{eq:LEspring_state_space}
\end{equation}

\noindent The heave kinematics are prescribed. For the baseline sinusoidal case,

\begin{equation}
h(t)=h_0\sin(\omega t),\qquad
\dot{h}(t)=h_0\omega\cos(\omega t),\qquad
\ddot{h}(t)=-h_0\omega^2\sin(\omega t).
\end{equation}

\section{Force Reconstruction}

After integrating the ODE to obtain $\theta(t)$ and $\dot{\theta}(t)$, the lateral force is reconstructed using the same scaling decomposition. The lift-like component is

\begin{equation}
F_{y,L}(t)=\rho s c\Big(\alpha(t)U_\infty U_{\mathrm{eff}}(t)+cU_\infty\dot{\alpha}(t)\Big),
\end{equation}

\noindent and the added-mass component is

\begin{equation}
F_{y,AM}(t)=\rho s c^2\Big(
b_1\ddot{\theta}(t)
+\ddot{h}(t)
+\dot{h}(t)\dot{\theta}(t)\theta(t)
+U_\infty\dot{\theta}(t)
+c\dot{\theta}(t)^2\theta(t)
\Big).
\end{equation}

\noindent The reconstructed lateral force therefore becomes

\begin{equation}
F_y^{(\mathrm{ODE})}(t)=F_{y,L}(t)+F_{y,AM}(t).
\end{equation}

\begin{equation}
\begin{aligned}
F_y^{(\mathrm{ODE})}(t)
&=\rho s c\Big(\alpha U_\infty U_{\mathrm{eff}}+cU_\infty\dot{\alpha}\Big)
\\
&\quad
+\rho s c^2\Big(
b_1\ddot{\theta}
+\ddot{h}
+\dot{h}\dot{\theta}\theta
+U_\infty\dot{\theta}
+c\dot{\theta}^2\theta
\Big).
\end{aligned}
\end{equation}     
                   
\appendix{Three-Fin Performance Data}
\label{app:three_fin_data}

This appendix summarizes representative cases extracted from the three-fin parameter sweep discussed in Chapter~5 and the Bayesian optimization results presented in Chapter~6. The cases shown correspond to configurations that produced the highest thrust, lowest thrust, highest efficiency, and lowest efficiency within the tested parameter space.

For each configuration we report the system-averaged thrust coefficient $C_T$, the propulsive efficiency $\eta$, and ratios relative to the lead fin. The phase offsets $\phi$ are given in radians and correspond to the phase of each downstream fin relative to the upstream fin.

\begin{table}[H]
\centering
\caption{Performance metrics for selected three-fin configurations identified during the parameter sweep.}
\label{tab:three_fin_performance_cases}

\begin{tabular}{lccccc}
\toprule
Configuration & $C_T$ & $\eta$ & $C_{T,n}/C_{T,1}$ & $\eta_n/\eta_1$ & $\phi$ list \\
\midrule
3 fins (highest thrust) & 0.61 & 0.37 & 1.30 & 1.01 & [0, 5.84, 4.94] \\
3 fins (lowest thrust)  & 0.15 & 0.04 & 0.32 & 0.11 & [0, 3.14, 0.45] \\
3 fins (high efficiency) & 0.40 & 0.37 & 0.85 & 1.01 & [0, 4.05, 2.69] \\
3 fins (low efficiency)  & 0.15 & 0.04 & 0.32 & 0.11 & [0, 3.14, 0.45] \\
\bottomrule
\end{tabular}

\end{table}

The phase configurations illustrate the strong dependence of system performance on the relative timing between fins. In particular, constructive wake interactions can significantly increase the thrust generated by downstream fins, while unfavorable phase relationships can reduce both thrust and efficiency.     
       
\end{document}